\documentclass[12pt]{article}
\usepackage{eqsection,subeqnarray,indent,amsfonts,amssymb}
\usepackage{amsfonts}
\usepackage{epic,eepic}
\usepackage{graphicx}

\begin{document}

\bibliographystyle{plain}

\date{November 27, 2002 \\[1mm] revised April 29, 2003}

\title{\vspace*{-2.5cm}
 Exact Potts Model Partition Functions for Strips of the 
 Triangular Lattice} 

\author{
  \\
  {\small    Shu-Chiuan Chang}                              \\[-0.2cm]
  {\small\it C.~N.~Yang Institute for Theoretical Physics}  \\[-0.2cm]
  {\small\it State University of New York}                  \\[-0.2cm]
  {\small\it Stony Brook, N.~Y.~11794-3840}                 \\[-0.2cm]
  {\small\it USA }                                          \\[-0.2cm]
  {\small\it After 9/10/2002: Department of Applied Physics:} \\[-0.2cm]
  {\small\it Faculty of Science}                            \\[-0.2cm]
  {\small\it Tokyo University of Science}                   \\[-0.2cm]
  {\small\it 1-3 Kagurazaka, Shinjuku-ku}                   \\[-0.2cm]
  {\small\it Tokyo 162-8601}                                \\[-0.2cm]
  {\small\it JAPAN}                                         \\[-0.2cm]
  {\small\tt chang@rs.kagu.tus.ac.jp}                       \\[5mm]
  {\small Jesper Lykke Jacobsen}                            \\[-0.2cm]
  {\small\it Laboratoire de Physique Th\'eorique et Mod\`eles Statistiques}
                                                            \\[-0.2cm]
  {\small\it Universit\'e Paris-Sud}                        \\[-0.2cm]
  {\small\it B\^atiment 100}                                \\[-0.2cm]
  {\small\it F-91405 Orsay}                                 \\[-0.2cm]
  {\small\it FRANCE }                                       \\[-0.2cm]
  {\small\tt jacobsen@ipno.in2p3.fr}                        \\[5mm]
  {\small Jes\'us Salas}                                    \\[-0.2cm]
  {\small\it Departamento de F\'{\i}sica Te\'orica}         \\[-0.2cm]
  {\small\it Facultad de Ciencias, Universidad de Zaragoza} \\[-0.2cm]
  {\small\it Zaragoza 50009}                                \\[-0.2cm]
  {\small\it SPAIN}                                         \\[-0.2cm]
  {\small\it After 1 Oct.\  2003: Departmento de Matem\'aticas} \\[-0.2cm]
  {\small\it Universidad Carlos III de Madrid}              \\[-0.2cm]
  {\small\it Avda.\  de la Universidad, 38}                 \\[-0.2cm]
  {\small\it 28911 Legan\'es, SPAIN}                        \\[-0.2cm]
  {\small\tt jsalas@math.uc3m.es}                           \\[5mm]
  {\small Robert Shrock}                  \\[-0.2cm]
  {\small\it C.~N.~Yang Institute for Theoretical Physics}  \\[-0.2cm]
  {\small\it State University of New York}                  \\[-0.2cm]
  {\small\it Stony Brook, N.~Y.~11794-3840}                 \\[-0.2cm]
  {\small\it USA }                                          \\[-0.2cm]
  {\small\tt robert.shrock@sunysb.edu}                      \\[-0.2cm]
  {\protect\makebox[5in]{\quad}}  
  \\
}

\maketitle

\thispagestyle{empty}   

\newpage

\begin{abstract}

We present exact calculations of the Potts model partition function $Z(G,q,v)$
for arbitrary $q$ and temperature-like variable $v$ on $n$-vertex strip graphs
$G$ of the triangular lattice for a variety of transverse widths equal to $L$
vertices and for arbitrarily great length equal to $m$ vertices, with free
longitudinal boundary conditions and free and periodic transverse boundary
conditions.  These partition functions have the form
$Z(G,q,v)=\sum_{j=1}^{N_{Z,G,\lambda}} c_{Z,G,j}(\lambda_{Z,G,j})^{m-1}$.  We
give general formulas for $N_{Z,G,j}$ and its specialization to $v=-1$ for
arbitrary $L$. The free energy is calculated exactly for the infinite-length
limit of the graphs, and the thermodynamics is discussed.  It is shown how the
internal energy calculated for the case of cylindrical boundary conditions is
connected with critical quantities for the Potts model on the infinite
triangular lattice.  Considering the full generalization to arbitrary complex
$q$ and $v$, we determine the singular locus ${\cal B}$, arising as the
accumulation set of partition function zeros as $m \to \infty$, in the $q$
plane for fixed $v$ and in the $v$ plane for fixed $q$.  Explicit results for
partition functions are given in the text for $L=3$ (free) and $L=3,4$
(cylindrical), and plots of partition function zeros and their asymptotic
accumulation sets are given for $L$ up to 5.  A new estimate for the phase
transition temperature of the $q=3$ Potts antiferromagnet on the 2D triangular
lattice is given.  

\end{abstract}


\clearpage

%
%
\newcommand{\beq}{\begin{equation}}
\newcommand{\eeq}{\end{equation}}
\newcommand{\beqs}{\begin{eqnarray}}
\newcommand{\eeqs}{\end{eqnarray}}
\newcommand{\lsim}{\mathrel{\raisebox{-.6ex}{$\stackrel{\textstyle<}{\sim}$}}}
\newcommand{\gsim}{\mathrel{\raisebox{-.6ex}{$\stackrel{\textstyle>}{\sim}$}}}
\newtheorem{theorem}{Theorem}[section]
\newtheorem{corollary}{Corollary}[section]
\newtheorem{conjecture}{Conjecture}[section]
\newcommand{\be}{\begin{equation}}
\newcommand{\ee}{\end{equation}}
\newcommand{\widebar}{\overline}
\def\reff#1{(\protect\ref{#1})}
\def\spose#1{\hbox to 0pt{#1\hss}}
\def\ltapprox{\mathrel{\spose{\lower 3pt\hbox{$\mathchar"218$}}
 \raise 2.0pt\hbox{$\mathchar"13C$}}}
\def\gtapprox{\mathrel{\spose{\lower 3pt\hbox{$\mathchar"218$}}
 \raise 2.0pt\hbox{$\mathchar"13E$}}}
\def\textprime{${}^\prime$}
\def\proof{\par\medskip\noindent{\sc Proof.\ }}
\def\qed{\hbox{\hskip 6pt\vrule width6pt height7pt depth1pt \hskip1pt}\bigskip}
\def\proofof#1{\bigskip\noindent{\sc Proof of #1.\ }}
\def\half{ {1 \over 2} }
\def\third{ {1 \over 3} }
\def\twothird{ {2 \over 3} }
\def\smfrac#1#2{\textstyle{#1\over #2}}
\def\smhalf{ \smfrac{1}{2} }
\newcommand{\real}{\mathop{\rm Re}\nolimits}
\renewcommand{\Re}{\mathop{\rm Re}\nolimits}
\newcommand{\imag}{\mathop{\rm Im}\nolimits}
\renewcommand{\Im}{\mathop{\rm Im}\nolimits}
\newcommand{\sgn}{\mathop{\rm sgn}\nolimits}
\newcommand{\tr}{\mathop{\rm tr}\nolimits}
\newcommand{\diag}{\mathop{\rm diag}\nolimits}
\newcommand{\Gal}{\mathop{\rm Gal}\nolimits}
\newcommand{\mycup}{\mathop{\cup}}
\newcommand{\Arg}{\mathop{\rm Arg}\nolimits}
\def\hboxscript#1{ {\hbox{\scriptsize\em #1}} }
\def\zhat{ {\widehat{Z}} }
\def\phat{ {\widehat{P}} }
\def\qtilde{ {\widetilde{q}} }
\newcommand{\mod}{\mathop{\rm mod}\nolimits}
\renewcommand{\emptyset}{\varnothing}

\def\scra{\mathcal{A}}
\def\scrb{\mathcal{B}}
\def\scrc{\mathcal{C}}
\def\scrd{\mathcal{D}}
\def\scrf{\mathcal{F}}
\def\scrg{\mathcal{G}}
\def\scrl{\mathcal{L}}
\def\scro{\mathcal{O}}
\def\scrp{\mathcal{P}}
\def\scrq{\mathcal{Q}}
\def\scrr{\mathcal{R}}
\def\scrs{\mathcal{S}}
\def\scrt{\mathcal{T}}
\def\scrv{\mathcal{V}}
\def\scrz{\mathcal{Z}}

\def\q{{\sf q}}

\def\Z{{\mathbb Z}}
\def\R{{\mathbb R}}
\def\C{{\mathbb C}}
\def\Q{{\mathbb Q}}

\def\T{{\mathsf T}}
\def\H{{\mathsf H}}
\def\V{{\mathsf V}}
\def\D{{\mathsf D}}
\def\J{{\mathsf J}}
\def\P{{\mathsf P}}
\def\QQ{{\mathsf Q}}
\def\RR{{\mathsf R}}

\def\bsigma{\mbox{\protect\boldmath $\sigma$}}
\def\bone{{\mathbf 1}}
\def\vv{{\bf v}}
\def\uu{{\bf u}}
\def\w{{\bf w}}

%
%
\section{Introduction} \label{sec.intro}

The $q$-state Potts model has served as a valuable model for the study of phase
transitions and critical phenomena \cite{potts,wurev}.  In this paper we 
present some theorems on structural properties of Potts model partition
functions on triangular-lattice strips of arbitrary width equal to $L$ vertices
and arbitrarily great length equal to $m$ vertices.  We also
report exact calculations of Potts model partition functions for a number of
triangular-lattice strips of various widths and arbitrarily great lengths.
Using these results, we consider the limit of infinite length.  For this
limit we calculate thermodynamic functions and determine the loci
in the complex $q$ and temperature planes where the free energy is 
non-analytic. These loci arise as the continuous accumulation sets of
partition-function zeros. This work is an extension to the triangular lattice
of our earlier study for the square lattice \cite{ts}.  

Consider a graph $G=(V,E)$, defined by its vertex set $V$ and edge set $E$.
Denote the number of vertices and edges as $|V| \equiv n$ and $|E|$,
respectively.  For technical simplicity, we restrict to connected loopless
graphs.  On this graph $G$, at temperature $T$, the Potts model is defined by
the partition function
\beq
Z(G,q,v) = \sum_{ \{ \sigma_n \} } e^{-\beta {\cal H}}
\label{zfun}
\eeq
with the (zero-field) Hamiltonian
\beq
{\cal H} = -J \sum_{\langle i j \rangle} \delta_{\sigma_i \sigma_j}
\label{ham}
\eeq
where $\sigma_i=1,\ldots,q$ are the spin variables on each vertex $i \in V$;
$\beta = (k_BT)^{-1}$; and $\langle i j \rangle \in E$ denotes pairs of 
adjacent vertices.  We use the notation
\beq
K = \beta J \ , \quad a = e^K \ , \quad v = a-1
\label{kdef}
\eeq
so that the physical ranges are (i) $a \ge 1$, i.e., $v \ge 0$ corresponding to
$\infty \ge T \ge 0$ for the Potts ferromagnet, and (ii) $0 \le a \le 1$,
i.e., $-1 \le v \le 0$, corresponding to $0 \le T \le \infty$ for the Potts
antiferromagnet.  One defines the (reduced) free energy per site $f=-\beta F$,
where $F$ is the actual free energy, via
\beq
f(\{G\},q,v) = \lim_{n \to \infty} \ln [ Z(G,q,v)^{1/n}]  
\label{ef}
\eeq
where we use the symbol $\{G\}$ to denote $\lim_{n \to \infty}G$ for a given
family of graphs $G$.

For our results in this paper we shall consider two types of boundary
conditions: free and cylindrical.  Here, free boundary conditions mean free in
both the transverse and longitudinal directions (the latter being the one that
is varied for a fixed width), while cylindrical boundary conditions mean
periodic in the transverse direction and free in the longitudinal direction.
Exact partition functions for arbitrary $q$ and $v$ have previously been
presented for strips of the triangular lattice for width $L=2$ in \cite{ta} and
results have been given for $L=3,4$ in \cite{ks}.  Here we shall give new
explicit results using the transfer matrix method (in the Fortuin--Kasteleyn
representation \cite{kf,fk}) for strips with free and cylindrical boundary
conditions with width $ L=3$ (free) in section 3.2 and $L=3,4$ (cylindrical) in
sections 4.2 and 4.3.  We have obtained new algebraic computer results for
$4\leq L\leq 6$ (free) and $5 \leq L \leq 9$ (cylindrical); these are too
lengthy to include here but are included in the {\sc mathematica} computer file
{\tt transfer\_Tutte\_tri.m} which is available with the electronic version of
this paper in the {\tt cond-mat} archive at {\tt http://arXiv.org}. We shall
also present plots of partition function zeros and their asymptotic
accumulation sets in the limit of infinite length, for widths $2 \le L \le 5$.
Where our results overlap with those for $L=2,3,4$ given before in
\cite{ta,ks}, they are in agreement.  As noted, we shall also give exact
results valid for these strips with arbitrary width and length, namely the
structural theorems of section 2.

There are several motivations for this work.  Clearly, new exact calculations
of Potts model partition functions are of value in their own right.  In
addition, these calculations can give insight into the complex-temperature
phase diagram of the two-dimensional (2D) Potts model on a particular lattice.
This is useful, since the 2D Potts model has never been solved except
in the $q=2$ Ising case. From a mathematical point of view, the partition
function of the Potts model on a graph $G$ is equivalent to the Tutte
polynomial on the same graph $G$ (see below). Thus, we can extract very useful
combinatorial information on the graph $G$.

Let $G^\prime=(V,E^\prime)$ be a spanning subgraph of $G$, i.e. a subgraph
having the same vertex set $V$ and an edge set $E^\prime \subseteq E$. Then
$Z(G,q,v)$ can be written as the sum \cite{birk,kf,fk}
\beq
Z(G,q,v) = \sum_{G^\prime \subseteq G} q^{k(G^\prime)}v^{|E^\prime|}
\label{cluster}
\eeq
where $k(G^\prime)$ denotes the number of connected components of $G^\prime$.
The formula (\ref{cluster}) enables one to generalize $q$ from
${\mathbb Z}_+$ to ${\mathbb R}_+$ (keeping $v$ in its physical range), and  
it also shows that $Z(G,q,v)$ is a polynomial in $q$ and $v$
(equivalently, $a$). 

The Potts model partition function on a graph $G$ is essentially equivalent to
the Tutte polynomial \cite{tutte1,tutte2,tutte3} and Whitney rank polynomial
\cite{whit,wurev,wu87,bbook,welsh,boll}.  Here the Tutte polynomial of an
arbitrary graph $G=(V,E)$ is
\beq
T(G,x,y)=\sum_{G^\prime \subseteq G} (x-1)^{k(G^\prime)-k(G)}
(y-1)^{c(G^\prime)}
\label{t}
\eeq
where $G^\prime$ again denotes a spanning subgraph of $G$ and $c(G^\prime)$
denotes the number of independent circuits in $G^\prime$, satisfying
$c(G^\prime) = |E^\prime|+k(G^\prime)-|V|$.  Since we only consider connected
graphs $G$, we have $k(G)=1$. From (\ref{cluster}) and (\ref{t}), it follows
that the Potts model partition function $Z(G,q,v)$ is related to the Tutte
polynomial $T(G,x,y)$ according to
\beq
Z(G,q,v)=(x-1)^{k(G)}(y-1)^{|V|}T(G,x,y)
\label{zt}
\eeq
where
\begin{subeqnarray}
x   &=&  1+\frac{q}{v} \slabel{xdef} \\
y   &=&  a = v+1       \slabel{ydef}
\end{subeqnarray}
so that
\beq
q=(x-1)(y-1) 
\label{qxy}
\eeq
In addition to the works \cite{ta,ks}, previous exact calculations of Potts
model partition functions for arbitrary $q$ and $v$ on lattice strips and/or
studies of their properties include \cite{ts,bcc,a,ss00,cf,hca,s3a,kc,dg,sdg};
a related early study of chromatic and Tutte
polynomials for recursive families of graphs is \cite{bds}.

Various special cases of the Potts model partition function are of interest.
One special case is the zero-temperature limit of the Potts antiferromagnet,
i.e., $v=-1$.  For sufficiently large $q$, on a given lattice or graph $G$,
this exhibits nonzero ground state entropy $S_0$ (without frustration).  This
is equivalent to a ground state degeneracy per site (vertex), $W > 1$, since
$S_0 = k_B \ln W$.  The $T=0$ partition function of the $q$-state Potts
antiferromagnet on $G$ satisfies
\beq 
Z(G,q,-1)=P(G,q)
\label{zp}
\eeq
where $P(G,q)$ is the chromatic polynomial (in $q$) expressing the number
of ways of coloring the vertices of the graph $G$ with $q$ colors such that no
two adjacent vertices have the same color \cite{birk,bbook,rrev,rtrev}. The
minimum number of colors necessary for this coloring is the chromatic number
of $G$, denoted $\chi(G)$.  We have
\beq
W(\{G\},q)= \lim_{n \to \infty}P(G,q)^{1/n}
\label{w}
\eeq
In the context of our current work we recall that the chromatic number for the
2D triangular lattice is $\chi({\rm tri})=3$.
This chromatic number also applies to 
strips of the triangular lattice with free longitudinal boundary conditions and
free transverse boundary conditions.  For the triangular-lattice strips with
cylindrical (i.e., free longitudinal and periodic transverse) boundary
conditions, $\chi=3$ if the width $L=0$ mod 3 and $\chi=4$ if $L=1$ or 2 mod 3.
References to papers on the special case $v=-1$ are given, e.g., in
\cite{a,dg,transfer3,ts}.

Using the formula (\ref{cluster}) for $Z(G,q,v)$, one can generalize $q$ from
${\mathbb Z}_+$ not just to ${\mathbb R}_+$ but to ${\mathbb C}$ and $v$ from
its physical ferromagnetic and antiferromagnetic ranges $0 \le v \le \infty$
and $-1 \le v \le 0$ to $v \in {\mathbb C}$.  A subset of the zeros of $Z$ in
the two-complex dimensional space ${\mathbb C}^2$ defined by the pair of
variables $(q,v)$ can form an accumulation set in the $n \to \infty$ limit,
denoted ${\cal B}$, which is the continuous locus of points where the free
energy is nonanalytic.  This locus is determined as the solution to a certain
$\{G\}$-dependent equation.  For a given value of $v$, one can consider this
locus in the $q$ plane, and we denote it as ${\cal B}_q(\{G\},v)$.  In the
special case $v=-1$ where the partition function is equal to the chromatic
polynomial, the zeros in $q$ are the chromatic zeros, and ${\cal
B}_q(\{G\},v=-1)$ is their continuous accumulation set in the $n \to \infty$
limit.  With the exact Potts partition function for arbitrary temperature, one
can study ${\cal B}_q$ for $v \ne -1$ and, for a given value of $q$, one can
study the continuous accumulation set of the zeros of $Z(G,q,v)$ in the $v$
plane (complex-temperature or Fisher zeros \cite{fisher} - other early
references include \cite{abe,katsura,suzuki}).  This set will be denoted
${\cal B}_v(\{G\},q)$. 

%
%
\section{General Results for Recursive Families of Graphs} \label{sec.2} 

A recursive family of graphs is one in which one constructs successive members
of the family in a recursive manner starting from an initial member. Recursive
families of graphs that are of particular interest here are strips of regular
lattices of a given width $L$ vertices (with free or cylindrical boundary
conditions) and arbitrarily great length $m$ vertices (with free boundary
conditions).

A general form for the Potts model partition function for the strip graphs
considered here is \cite{a}
\beq
Z(G,q,v) = \sum_{j=1}^{N_{Z,G,\lambda}} c_{G,j}(\lambda_{G,j})^{m-1}
\label{zgsum}
\eeq
where the coefficients $c_{G,j}$ and corresponding terms $\lambda_{G,j}$, as
well as the total number $N_{Z,G,\lambda}$ of these terms, depend on the type
of recursive graph $G$ (width and boundary conditions) but not on its length.
(In \cite{a}, a slightly different labelling convention was used so that
$\lambda_{G,j}^m$ rather than $\lambda_{G,j}^{m-1}$ appeared in the summand of
eq.~(\ref{zgsum}).) In the special case $v=-1$ where $Z$ reduces to the
chromatic polynomial (zero-temperature Potts antiferromagnet), eq.~\reff{zgsum}
reduces to the form \cite{bkw}
\beq
P(G,q) = \sum_{j=1}^{N_{P,G,\lambda}} c_{G,j}(\lambda_{P,G,j})^{m-1} 
\label{pgsum}
\eeq
For the lattice strips of interest here, we define the following explicit
notation.  Let $N_{Z,{\rm sq},{\rm BC}_t \ {\rm BC}_\ell,L,\lambda}$ denote 
the total number of $\lambda$'s for the square-lattice strip with the 
transverse ($t$) and longitudinal ($\ell$) boundary conditions 
${\rm BC}_t$ and ${\rm BC}_\ell$ of width $L$.
Henceforth where no confusion will result, we shall suppress the $\lambda$
subscript.  The explicit labels are $N_{Z,{\rm sq},{\rm FF},L}$ and 
$N_{Z,{\rm tri},{\rm FF},L}$ for the strips of the square and triangular 
lattices with free boundary conditions, and 
$N_{Z,{\rm sq},{\rm PF},L}$ and $N_{Z,{\rm tri},{\rm PF},L}$ for the strips of
these respective lattices with cylindrical boundary conditions.

For the lattice strip graphs of interest here we can express the partition
function via a transfer matrix $\T$ (in the Fortuin-Kasteleyn representation)
of fixed size $M\times M$:
\beq
Z(G,q,v) = \tr \left[ A(q,v) \cdot \T(q,v)^{m-1} \right]
\label{ztransfer}
\eeq
which then yields the form (\ref{zgsum}).  Since the transfer matrix $\T$ and
the boundary-condition matrix $A$ are polynomials in $q$ and $v$, it follows
that the eigenvalues $\{\lambda_k\}$ of the transfer matrix and the
coefficients $c_{G,j}$ are algebraic functions of $q$ and $v$.  

One of the basic structural properties of the Potts model partition function on
a given strip is the number of different eigenvalues of the transfer matrix (in
the Fortuin-Kasteleyn representation), $N_{Z,G,{\rm BC},L}$, in
eq.~\reff{zgsum}.  In \cite{ts}, in addition to proving various formulas for
these numbers for certain strip graphs, we presented a conjecture (denoted
Conjecture 3 in \cite{ts} and given as Theorem~4.3.5 by one of us (S.-C.C.) in
\cite{thesis}).  We now give the proof of this result. 

\bigskip

\begin{theorem} \label{theorem1} For arbitrary $L$,
\beq
2N_{Z,{\rm sq},{\rm PF},L}-N_{Z,{\rm tri},{\rm PF},L} = 
       \cases{ N_{Z,{\rm sqtri},{\rm FP},\frac{L}{2}}   & for even $L$ \cr
    \frac{1}{2}N_{Z,{\rm sqtri},{\rm FP},\frac{L+1}{2}} & for odd $L$ } 
\label{2nzsqpf-nztpf}
\eeq
where the quantity $N_{Z,{\rm sqtri},{\rm FP},L}$ is given by \cite{cf}:
\be
N_{Z,{\rm sqtri},{\rm FP},L} = {2L \choose L}
\label{nzsqtrifp}
\ee
\end{theorem}

{\sl Proof} \quad We recall first that the quantity
$2N_{Z,{\rm sq},{\rm FF},L}-N_{Z,{\rm tri},{\rm FF},L}$ discussed in 
\cite{ts} gives the number of
non-crossing partitions for a transverse slice of these two respective strips
(which is the path graph $T_L$) such that these partitions are symmetric under
reflection about the longitudinal axis. The quantity
$2N_{Z,{\rm sq},{\rm PF},L}-N_{Z,{\rm tri},{\rm PF},L}$ 
gives the corresponding number of non-crossing
partitions for a transverse slice (which is the circuit graph $C_L$) of the two
respective cylindrical strips such that these partitions are symmetric under
reflections about the longitudinal axis and rotations around this axis (the
latter being included since there is no special azimuthal direction). 
We shall prove eq.~\reff{2nzsqpf-nztpf} for odd $L$ first and then
for even $L$.

For odd $L$, let $n=(L+1)/2$, and denote
$2N_{Z,{\rm sq},{\rm PF},L=2n-1}-N_{Z,{\rm tri},{\rm PF},L=2n-1}$
as $X_n$ for simplicity.  Consider a
transverse slice with periodic boundary conditions. Since this is topologically
invariant under rotations around the longitudinal axis, we can label one vertex
as 1 and other vertices 
$2, 3,\ldots, n, n^\prime,(n-1)^\prime,\ldots, 2^\prime$,
in a counterclockwise manner (relative to a specified longitudinal direction),
and consider the reflection symmetry with respect to the longitudinal axis
passing through vertex 1. In order to classify the types of colorings of the
vertices, we shall introduce diagrams consisting of the $L$ vertices on this
transverse slice.  In this context, we shall refer to two vertices as being
``connected'' if these have the same color and shall denote this by using the
Kronecker delta function.

The sets ${\bf P}_{X_n}$ of partitions for $n=1,2,3$ that are invariant 
under this reflection symmetry are: ${\bf P}_{X_1} = \{ 1\}$, 
${\bf P}_{X_2} = \{ 1, \delta_{2,2^\prime}, 
\delta_{1,2}\delta_{1,2'}=\delta_{1,2,2^\prime}\}$, and
\beqs 
{\bf P}_{X_3} & = & \{ 1, \delta_{2,2^\prime}, \  \delta_{3,3^\prime}, \ 
\delta_{1,2,2^\prime}, \ 
\delta_{1,3,3^\prime}, \ 
\delta_{2,3}\delta_{2^\prime,3^\prime}, \ 
\delta_{2,2^\prime}\delta_{3,3^\prime}, \cr & & \qquad  
\delta_{2,3,2^\prime,3^\prime}, \ 
\delta_{1,2,2^\prime}\delta_{3,3^\prime}, \ 
\delta_{1,2,3,2^\prime,3^\prime}\} 
\label{px3}
\eeqs
We can classify these partitions into cases that have $m$ vertices on one side
of the slice (including vertex 1) connected to at least one other vertex (on
the same side or the other side, indicated by the primes above), where $0 \le m
\le n$. For $m=0$, this is the partition 1, that is the identity partition,
defined as the one in which all blocks are ``singletons'', i.e., there are no
connections in the sense given above.  For $m=1$, there is only one
possibility: $\delta_{x_1,x_1^\prime}$, where $2 \le x_1 \le n$. For $m=2$, let
us denote the connected vertices as $x_1$, $x_2$ and, with no loss of
generality, take $x_1 < x_2$; then there are the partitions $\delta_{x_1,x_2}$,
$\delta_{x_1,x_2,x_1^\prime}$, and
$\delta_{x_1,x_1^\prime}\delta_{x_2,x_2^\prime}$. For the first case, $x_1$ can
be vertex 1, but for the second and third cases, $2 \le x_1 \le n$ since $x_1$
and $x_1^\prime$ are different vertices. The corresponding partitions
$\delta_{x_1^\prime,x_2^\prime}$ for the first and second cases can be obtained
by reflection symmetry from the ones that we have listed and hence, for
simplicity, are not shown. With no loss of generality, we take 
$x_1 < x_2 < \ldots < x_m$.  For $m=3$, the partitions are
$\delta_{x_1,x_2}\delta_{x_3,x_3^\prime}$, $\delta_{x_1,x_2,x_3}$,
$\delta_{x_1,x_2,x_1^\prime}\delta_{x_3,x_3^\prime}$,
$\delta_{x_1,x_2,x_3,x_1^\prime}$, $\delta_{x_1,x_1^\prime}\delta_{x_2,x_3}$,
$\delta_{x_1,x_1^\prime}\delta_{x_2,x_3,x_2^\prime}$,
$\delta_{x_1,x_1^\prime}\delta_{x_2,x_2^\prime}\delta_{x_3,x_3^\prime}$.
Having given these illustrations of the specific partitions for $0 \le m \le
3$, we next proceed to the general case.

The partitions that have $m$ vertices on one side of the slice connected
to as least one other vertex (on the same or opposite side) can be
classified further. Let us denote $a_m$ as the number of cases where the
vertex $x_1$ has only the connection $\delta_{x_1,x_1^\prime}$ for $1 \le
m$. In these cases, $x_1$ cannot be vertex 1, and the number of the
partitions for each $m \le n-1$ is ${n-1 \choose m}$.  The last three
partitions of $m=3$ given above are examples of these cases. We denote
$b_m$ as the number of cases where the vertex $x_1$ has connection to at
least one unprimed vertex with or without $\delta_{x_1,x_1^\prime}$ for $2
\le m$. The first and second partitions for $m=3$ are examples of these
cases without $\delta_{x_1,x_1^\prime}$, and the third and fourth
partitions for $m=3$ are examples of these cases with
$\delta_{x_1,x_1^\prime}$. Notice that while $x_1$ cannot be vertex 1 for
the cases with $\delta_{x_1,x_1^\prime}$, which have the number of
partitions ${n-1 \choose m}$ for each $m \le n-1$, $x_1$ can be vertex 1
for the cases without $\delta_{x_1,x_1^\prime}$, and the number of
partitions for each $m \le n$ is ${n \choose m}$. Therefore,
\beq 
X_n = 1 + \sum_{m=1}^{n-1}a_m{n-1 \choose m} + \sum_{m=2}^{n-1}b_m {n-1
\choose m} + \sum_{m=2}^{n} b_m {n \choose m} 
\label{xn} 
\eeq 
Next, we shall obtain expressions for $a_m$ and $b_m$. The cases for $a_m$
can be obtained from all the cases with $m-1$ vertices by changing $x_i$
to $x_{i+1}$ for $1 \le i \le m-1$ and adding $\delta_{x_1,x_1^\prime}$,
i.e., $a_m$ is the same as the total number of the cases with $m-1$
vertices, 
\beq 
a_m = a_{m-1} + 2b_{m-1} 
\label{am} 
\eeq 
Now consider the cases for $b_m$ without $\delta_{x_1,x_1^\prime}$. These can
be further divided into two possibilities: the cases with
$\delta_{x_1,x_2}$ and the cases with $\delta_{x_1,x_i}$ where $2 < i \le
m$. Denote the numbers of these two possibilities as $d_m$ and $e_m$,
respectively. Clearly, 
\beq 
b_m = d_m + e_m 
\label{bm} 
\eeq 
The cases for $d_m$ can be obtained from the cases for $b_{m-1}$ by
changing $x_i$ to $x_{i+1}$ for $1 \le i \le m-1$ and adding
$\delta_{x_1,x_2}$, and from all the cases with $m-2$ vertices, where the
number is $a_{m-1}$, by changing $x_i$ to $x_{i+2}$ for $1 \le i \le m-2$
and adding $\delta_{x_1,x_2}$. Therefore,
\beq
d_m = a_{m-1} + b_{m-1} 
\label{dm}
\eeq
The cases for $e_m$ can be obtained from the cases for $d_{m-2}$ by
changing $x_i$ to $x_{i+2}$ for $2 \le i \le m-2$ and adding
$\delta_{x_2,x_3}$, from the cases for $d_{m-3}$ by changing $x_i$ to
$x_{i+3}$ for $2 \le i \le m-3$ and adding $\delta_{x_2,x_3,x_4}$, etc.
In general, the cases for $e_m$ are obtained from the cases for
$d_{m-\ell}$ by changing $x_i$ to $x_{i+\ell}$ for $2 \le i \le m-\ell$
and $2 \le \ell \le m-2$, and adding a set of $\ell$ vertices, where all
the vertices in this set must have at least one connection to at least one
other vertex in this set without reflection or rotation symmetry. The number 
of all possible connections of this set with $\ell$ vertices is the Riordan 
number (given as sequence A005043 in \cite{sl}), which will be denoted as 
$r_\ell$ and may be defined via the generating function \cite{stanley}
\beq 
R(z) = \frac{1+z-(1-2z-3z^2)^{1/2}}{2z(1+z)} = \sum_{\ell=0}^\infty r_\ell
z^\ell
\label{rz}
\eeq
Therefore,
\beq
e_m = \sum_{\ell=2}^{m-2}d_{m-\ell} \ r_\ell 
\label{em}
\eeq

In terms of generating functions
\beqs
A(z) & = & a_1z + a_2z^2 + a_3z^3       + \ldots \cr\cr
B(z) & = & b_2z^2 + b_3z^3 + b_4z^4     + \ldots \cr\cr
D(z) & = & d_2z^2 + d_3z^3 + d_4z^4     + \ldots \cr\cr
E(z) & = & e_4z^4 + e_5z^5 + e_6z^6     + \ldots \cr\cr
R(z) & = & 1 + r_2z^2 + r_3z^3 + r_4z^4 + \ldots 
\label{abedrgf}
\eeqs
we can re-express eqs.~\reff{am} to~\reff{em} as
\beqs
\frac{A(z)-a_1z}{z} & = & A(z) + 2B(z) \cr\cr
B(z) & = & D(z) + E(z) \cr\cr
\frac{D(z)}{z} & = & A(z) + B(z) \cr\cr
E(z) & = & D(z)(R(z)-1) 
\label{abdeeq}
\eeqs
We find
\beq
A(z) = \frac{z}{\sqrt{1-2z-3z^2}} 
\label{agf}
\eeq
The coefficients in the expansion of $A(z)$ in eq.~(\ref{abedrgf}) are, up to a
shift, the central trinomial coefficients (given as sequence
A002426 in \cite{sl}), i.e., for each value $1 \le m$ the 
largest coefficient of $(1+z+z^2)^{m-1}$.  Next, 
\beq
B(z) = \frac{1}{2} \biggl [\frac{1-z}{\sqrt{1-2z-3z^2}} - 1 \biggr ] 
\label{bgf}
\eeq 
The coefficients in the expansion of $B(z)$ in (\ref{abedrgf}) are given by the
coefficients for the next-to-central column in the expansion of 
$(1+z+z^2)^{m-1}$ for $1 \le m$ (listed as sequence A005717 in \cite{sl}).  For
$D(z)$ we have the closed form 
\beq
D(z) = \frac{z}{2} \biggl [\sqrt{\frac{1+z}{1-3z}} - 1 \biggr ] 
\label{dgf}
\eeq
The coefficients in the expansion of $D(z)$ in (\ref{abedrgf}) are the 
numbers of directed animals of size $m-1$ on the square
lattice for $1 \le m$ (given as sequence A005773 in \cite{sl}).  Finally, we
have 
\beq   
E(z) =
z\frac{1-3z+(1-2z)\sqrt{1-2z-3z^2}}{1-2z-3z^2+(3z-1)\sqrt{1-2z-3z^2}} - 1
+ z 
\label{egf}
\eeq
The coefficients in the expansion of $E(z)$ in (\ref{abedrgf}) are the
coefficients forming the second column from the center, in a tabular format, in
the expansion of $(1+z+z^2)^{m-2}$ for $2 \le m$ (given as sequence A014531 in
\cite{sl}).

Recall the binomial transformation for two sequences of numbers
$[s_0, s_1, s_2, \ldots]$, $[t_0, t_1, t_2, \ldots]$ with generating functions
$S(z) = \sum_{n=0}^\infty s_nz^n$ and $T(z) = \sum_{n=0}^\infty t_nz^n$.
If these sequences have the relation
\beq
t_n = \sum_{m=0}^n {n \choose m}s_m \ ,
\label{bt}
\eeq
then \cite{bs} 
\beq
T(z) = \frac{1}{1-z}S(\frac{z}{1-z}) 
\label{btgf}
\eeq
In our case, if the generating function of $X_n$ is $X(z) = 
\sum_{n=1}^\infty X_nz^n$, then we can combine eqs.~(\ref{xn}),
(\ref{agf}) and (\ref{bgf}) to get 
\beqs
X(z) & = & \frac{1}{1-z} - 1 + \frac{z}{1-z}A \left(\frac{z}{1-z} \right) + 
\frac{z}{1-z}B\left(\frac{z}{1-z}\right) + \frac{1}{1-z}
B\left(\frac{z}{1-z}\right) \cr\cr
& = & \frac{1}{2} \biggl [ \frac{1}{\sqrt{1-4z}} - 1 \biggr ] 
\label{xgf}
\eeqs
This has the expansion 
\beq
X(z) = \sum_{n=1}^\infty {2n-1 \choose n} z^n 
\label{Xzexpansion}
\eeq
Note that ${2n-1 \choose n} = \frac{1}{2}{2n \choose n} = \frac{1}{2}{L+1
\choose (L+1)/2}$ (given as sequence A001700 in \cite{sl}).

Let us next proceed to consider the number of non-crossing partitions for a
slice of the transverse vertices which has periodic boundary conditions and
reflection symmetry for even $L$. Denote $n=L/2$, and
$2N_{Z,{\rm sq},{\rm PF},L=2n}-N_{Z,{\rm tri},{\rm PF},L=2n}$ 
by $Y_n$ for simplicity. There are two
possibilities: the reflection axis does not go through any vertex or goes
through two vertices. These possibilities will be denoted as type I and and
type II partitions, respectively, and the number of partitions of each of these
two types as $Y_n^{\rm I}$ and $Y_n^{\rm II}$. For type I partitions, 
label the vertices
on one side of the reflection axis as 1, 2,\ldots, $n$ and the corresponding
reflected vertices as $1^\prime, 2^\prime, \ldots, n^\prime$ on the other side.
For type II partitions, label the vertices as 1, 2, \ldots, $n$, $n+1$,
$n^\prime$, \ldots, $2^\prime$, where vertices 1 and $n+1$ are on 
the reflection axis.  
The sets ${\bf P}_{Y_n^{\rm I}}$ of type I partitions for $n=1,2$ having this
reflection symmetry are: 
${\bf P}_{Y_1^{\rm I}} = \{ 1, \delta_{1,1^\prime}\}$ and
${\bf P}_{Y_2^{\rm I}} = \{ 1, \delta_{1,1^\prime}, \delta_{2,2^\prime},
\delta_{1,2}\delta_{1^\prime,2^\prime}, \delta_{1,2,1^\prime},
\delta_{1,1^\prime}\delta_{2,2^\prime}\}$. 
The sets ${\bf P}_{Y_n^{\rm II}}$ of
type II partitions for $n=1,2$ having this reflection symmetry are: 
${\bf P}_{Y_1^{\rm II}} = \{ 1, \delta_{1,2}\}$ and 
${\bf P}_{Y_2^{\rm II}} = \{ 1,
\delta_{2,2^\prime}, \delta_{1,3}, \delta_{1,2,2^\prime},
\delta_{2,3,2^\prime}, \delta_{1,2,3,2^\prime,3^\prime}\}$.  We
notice that partitions 1, i.e. identity and $\delta_{1,2,\ldots,L}$ (i.e., a
unique block) are both contained in the type I and type II classes of
partitions.  Since we have rotational symmetry, the partitions which are not
symmetric with respect to the central axis perpendicular to the reflection axis
are counted twice in either type I or type II classes of partitions. A similar
statement applies to the partitions which are symmetric with respect to the
axis perpendicular to the reflection axis if $L$ is a multiple of 4. If $L$ is
not a multiple of 4, the partitions that are symmetric with respect to the 
perpendicular axis are counted once in both type I and type II classes. 
Therefore, $2Y_n$ is the sum of all possible partitions in these two classes of
partitions.

We again classify these partitions into cases which have $m$ vertices on one
side of the slice connected to at least one other vertex (including vertex 1
and vertex $n+1$ for type II partitions). It is clear that $0 \le m \le n$ for
type I partitions and $0 \le m \le n+1$ for type II partitions. Let us consider
type I first. For $m=0$, there is the partition 1 (identity). For $m=1$, there
is only one possibility: $\delta_{x_1,x_1^\prime}$, where $1 \le x_1 \le
n$. For $m=2$, there are $\delta_{x_1,x_2}$, $\delta_{x_1,x_2,x_1^\prime}$ and
$\delta_{x_1,x_1^\prime}\delta_{x_2,x_2^\prime}$, where $x_1 < x_2$.  The
corresponding $\delta_{x_1^\prime,x_2^\prime}$ for the first and second cases
are not shown for simplicity.  We shall again take $x_1 < x_2 < \ldots < x_m$.
The number of the cases with $m$ vertices connected is $a_m+2b_m = a_{m+1}$ as
discussed above eq.~(\ref{am}), and the number of the partitions is 
${n \choose m}$ for each $0 \le m \le n$. Therefore,
\beq
Y_n^{\rm I} = \sum_{m=0}^{n}a_{m+1}{n \choose m} \quad {\rm for} \ 1 \le n 
\label{yin}
\eeq
Let the generating function of $Y_n^{\rm I}$ be 
$Y^{\rm I}(z) = \sum_{n=1}^\infty Y_n^{\rm I} z^n$.  
Using eq.~(\ref{agf}) and modifying eq.~(\ref{btgf}), we have
\beq
Y^{\rm I}(z) = \frac{1}{1-z}\left [ 
        \frac{A(\frac{z}{1-z})}{\frac{z}{1-z}}\right ]
 - a_1 = \frac{1}{\sqrt{1-4z}} - 1 
\label{yigf} 
\eeq

For type II partitions, neither vertex 1 nor vertex $n+1$ has a corresponding
symmetric partner, $1^\prime$ and $(n+1)^\prime$, respectively. We have to
classify the cases where $x_1$ is only connected to at least one other unprimed
vertex (in these cases, $x_1$ can be vertex 1), the number of which cases was
denoted $b_m$ earlier, into two possibilities: the cases where $x_m$ is only
connected to at least one other unprimed vertex (so that $x_m$ can be vertex
$n+1$), and the cases where $x_m$ is connected to $x_m^\prime$ (so that $x_m$
cannot be vertex $n+1$) among other possible connections.  The number of
partitions for these two possibilities will be denoted as $f_m$ for $2 \le m$
and $h_m$ for $3 \le m$, respectively. The partitions where $x_1$ is connected
to $x_1^\prime$, i.e., $\delta_{x_1,x_1^\prime}$, among other possible
connections can also be classified into two possibilities: the cases where
$x_m$ is only connected to other unprimed vertices, and the cases where $x_m$
is connected to $x_m^\prime$ among other possible connections. The number of
partitions for these two possibilities will be denoted as $h_m^\prime$ for $3
\le m$ and $i_m$ for $1 \le m$, respectively. Notice that $h_m^\prime = h_m$
because of the reflection symmetry. We shall need another set of partitions
where $x_1$ is connected to $x_m$ with possible connection to other unprimed
vertices, and all the other vertices are connected to at least one other
unprimed vertex; we denote the number of partitions for them as $j_m$ for $2
\le m$. If we add an additional connection $\delta_{x_1,x_1^\prime}$ to these
partitions, and denote the number of partitions as $k_m$, then $k_1=1$ and
$k_m=j_m$ for $2 \le m$.

For the cases for $f_m$, $x_1$ and $x_m$ can be vertices 1 and $n+1$, 
respectively, so the number of the partitions is ${n+1 \choose m}$ for 
each $m$. For the cases for $h_m$, $x_1$ can be vertex 1 but $x_m$ cannot be 
vertex $n+1$, so the number of the partitions is ${n \choose m}$ for each 
$m$. For the cases for $i_m$, neither $x_1$ can be vertex 1 nor $x_m$ can 
be vertex $n+1$, so the number of the partitions is ${n-1 \choose m}$ for 
each $m$. Therefore,
\beq
Y_n^{\rm II} = 1 + \sum_{m=2}^{n+1}f_m {n+1 \choose m} + 2\sum_{m=3}^nh_m {n
\choose m} + \sum_{m=1}^{n-1}i_m {n-1 \choose m} 
\label{yiin}
\eeq
We next obtain expressions for $f_m$, $h_m$ and $i_m$. From the definitions, 
\beq
f_m + h_m = b_m
\label{fhb}
\eeq
and
\beq
h_m^\prime + i_m = a_m + b_m = d_{m+1} \ ,
\label{hiab}
\eeq
where we use eq.~(\ref{dm}). The cases for $f_m$ include all the cases for
$j_m$, plus the cases for both $f_{m-2}$ and $h_{m-2}$ with the additional
connection $\delta_{x_{m-1},x_m}$, plus the cases for both $f_{m-3}$ and
$h_{m-3}$ with the additional connection $\delta_{x_{m-2},x_{m-1},x_m}$,
etc. We have
\beq
f_m = j_m + \sum_{\ell=2}^{m-2}j_\ell \ [f_{m-\ell}+h_{m-\ell}] 
= j_m + \sum_{\ell=2}^{m-2}j_\ell \ b_{m-\ell} 
\label{fm}
\eeq
The cases for $h_m$ can be obtained by adding $\delta_{x_m,x_m^\prime}$ to
the cases for $f_{m-1}$ and $h_{m-1}$, and adding
$\delta_{x_{m-1},x_m,x_m^\prime}$ to the cases for $f_{m-2}$ and
$h_{m-2}$, etc. We have
\beq
h_m = \sum_{\ell=1}^{m-2}k_\ell \ [f_{m-\ell}+h_{m-\ell}] 
= \sum_{\ell=1}^{m-2}k_\ell \ b_{m-\ell} 
\label{hm}
\eeq
The cases for $h_m^\prime$ can be obtained by adding
$\delta_{x_{m-1},x_m}$ to the cases for $h_{m-2}^\prime$ and $i_{m-2}$,
and adding  $\delta_{x_{m-2},x_{m-1},x_m}$ to the cases for
$h_{m-3}^\prime$ and $i_{m-3}$, etc. We have
\beq
h_m^\prime = \sum_{\ell=2}^{m-1}j_\ell \ [h_{m-\ell}^\prime+i_{m-\ell}]
= \sum_{\ell=2}^{m-1}j_\ell \ d_{m-\ell+1} \ ,
\label{hprimem}
\eeq
which should be equal to $h_m$, as mentioned before. The cases for $i_m$
include all the cases for $k_m$, plus the cases for both $h_{m-1}^\prime$ and
$i_{m-1}$ with additional connection $\delta_{x_m,x_m^\prime}$, plus the cases
for both $h_{m-2}^\prime$ and $i_{m-2}$ with additional connection
$\delta_{x_{m-1},x_m,x_m^\prime}$, etc. We have
\beq
i_m = k_m + \sum_{\ell=1}^{m-1}k_\ell \ [h_{m-\ell}^\prime+i_{m-\ell}]
= k_m + \sum_{\ell=1}^{m-1}k_\ell \ d_{m-\ell+1} 
\label{im} 
\eeq
The cases for $j_m$ can be obtained from the cases for $r_m$ by removing
the cases for $r_{m-2}$ with additional $\delta_{x_{m-1},x_m}$, removing
the cases for $r_{m-3}$ with additional $\delta_{x_{m-2},x_{m-1},x_m}$,
etc. We have
\beq
j_m = r_m - \sum_{\ell=2}^{m-2}j_\ell \ r_{m-\ell} 
\label{jm}
\eeq

In terms of the generating functions
\beqs
F(z) & = & f_2z^2 + f_3z^3 + f_4z^4 + \ldots \cr\cr
H(z) & = & H(z)^\prime = h_3z^3 + h_4z^4 + h_5z^5 + \ldots \cr\cr
I(z) & = & i_1z + i_2z^2 + i_3z^3   + \ldots \cr\cr
J(z) & = & j_2z^2 + j_3z^3 + j_4z^4 + \ldots \cr\cr
K(z) & = & k_1z + k_2z^2 + k_3z^3   + \ldots
\label{fhijkgf}
\eeqs
we can reexpress eqs.~(\ref{fhb}) to~(\ref{jm}) as
\beqs
F(z) + H(z) & = & B(z) \cr\cr
H(z) + I(z) & = & \frac{D(z)}{z} \cr\cr
F(z) & = & J(z) + B(z)J(z) \cr\cr
H(z) & = & B(z)K(z) = \frac{D(z)}{z}J(z) \cr\cr
I(z) & = & K(z) + \frac{D(z)}{z}K(z) \cr\cr
J(z) & = & R(z)-1 - (R(z)-1)J(z) = (R(z)-1)(1-J(z)) 
\label{ghijgf}
\eeqs
We find
\beq 
J(z) = \frac{R(z)-1}{R(z)} = \frac{1}{2}(1-z-\sqrt{1-2z-3z^2}) \ ,
\label{jgf}
\eeq
which is essentially the Motzkin number \cite{motzkin,donaghey} (given as
sequence A001006 in \cite{sl}).  This can be understood since the cases for
$j_m$ are in one-to-one correspondence with the non-crossing,
non-nearest-neighbor partitions of $m-1$ vertices with free boundary condition,
as given in eq.~(2.1.1) of \cite{ts}, i.e., $j_m$ = $M_{m-2}$. We also have
\beq
K(z) = z + J(z) = \frac{1}{2}(1+z-\sqrt{1-2z-3z^2}) 
\label{kgf}
\eeq
The first five equations in eq.~(\ref{ghijgf}) are redundant, and can be 
solved to have
\beq 
F(z) = \frac{z^2}{\sqrt{1-2z-3z^2}} = zA(z) \ ,
\label{fgf}
\eeq
i.e., $f_m = a_{m-1}$, which is the largest coefficient of 
$(1+z+z^2)^{m-2}$ for $2 \le m$. Next, we have
\beq
H(z) = \frac{1}{2} \biggl [\left ( \frac{1-2z}{1-3z} \right ) \sqrt{1-2z-3z^2}
 - 1 \biggr ] 
\label{hgf}
\eeq
The coefficients in the expansion of $H(z)$ are the numbers of directed 
animals of size $m-2$ on the square lattice with the first two quadrants 
(which can grow in right, left and up directions) for $2 \le m$ (given as 
sequence A005774 in \cite{sl}). Finally,
\beq
I(z) = z \sqrt{\frac{1+z}{1-3z}} = 2D(z) + z \ ,
\label{igf}
\eeq
which is given as sequence A0025565 in \cite{sl}.

Denote the generating function of $Y_n^{\rm II}$ as 
$Y^{\rm II}(z) = \sum_{n=1}^\infty Y_n^{\rm II}z^n$.  
Combining eqs.~(\ref{yiin}), (\ref{fgf}), (\ref{hgf}), 
(\ref{igf}), and using the binomial transformation (\ref{btgf}), we obtain the
relation 
\beqs
Y^{\rm II}(z) & = & \frac{1}{1-z} -1 + \frac{1}{z(1-z)} F(\frac{z}{1-z}) +
\frac{1}{1-z} H(\frac{z}{1-z}) + \frac{z}{1-z} I(\frac{z}{1-z}) \cr\cr
& = & \frac{1}{1-z} -1 + \frac{z}{(1-z)^2\sqrt{1-4z}} +
\frac{1}{2(1-z)^2} \biggl [ \frac{1-3z}{\sqrt{1-4z}} +z-1 \biggr ] 
\nonumber \\
&   &  \qquad + \frac{z^2}{(1-z)^2\sqrt{1-4z}} \nonumber \\ 
& = & \frac{1}{\sqrt{1-4z}} - 1 \ , 
\label{yiigf}
\eeqs
which is the same as the $Y^{\rm I}(z)$ in eq.~(\ref{yigf}). We finally have
\beq
Y(z) = \frac{Y^{\rm I}(z) + Y^{\rm II}(z)}{2} = \frac{1}{\sqrt{1-4z}} - 1 \ , 
\label{ygf}
\eeq
which is the generating function of ${2n \choose n} = {L \choose L/2}$, 
i.e., the central binomial coefficients given as sequence A000984 in 
\cite{sl}. 

Eqs.~(\ref{xgf}) and~(\ref{ygf}) prove the theorem for odd and even 
$L$, respectively.  
\ $\Box$
 
\bigskip

Another structural result was given as Conjecture 4 in \cite{ts} and
Theorem~4.3.7 in \cite{thesis} for the number of $\lambda$'s for the triangular
lattice with cylindrical boundary condition for the Potts model partition
function.  We state this and give the proof. 

\bigskip

\begin{theorem} \label{theorem2} For arbitrary $L$,
\beq
N_{Z,{\rm tri},{\rm PF},L}=\frac{1}{L}\biggl [ C_L + \sum_{d|L; \ 1 \le d < L} 
\phi(L/d){2d \choose d} \biggr ]  
\label{nztripfgen}
\eeq
where $d|L$ means that $d$ divides $L$ and $\phi(n)$ is the Euler 
function, equal to the number of positive integers not exceeding the
positive integer $n$ and relatively prime to $n$. 
\end{theorem}

{\sl Proof} \quad As shown in \cite{ss00}, $N_{Z,{\rm tri},{\rm FF},L}=C_L$ 
is the
number of non-crossing partitions of a set of $L$ vertices with free
boundary conditions. 
$L N_{Z,{\rm tri},{\rm PF},L}-N_{Z,{\rm tri},{\rm FF},L}$ is equal to $2(L-1)$
for prime $L$, as shown in \cite{ts,thesis}, since all the partitions of
$N_{Z,{\rm tri},{\rm PF},L}$ have periodicity $L$ for prime $L$ except for the
partitions $1$ (i.e., all blocks being singletons) and
$\delta_{1,2,\ldots,L}$ (i.e., a unique block) which have periodicity 1. 
Consider a general $L$ and assume that $L$ has the factor $d$ (so that for 
prime $L$, $d$ can only be 1 or $L$).  Denote the number of partitions
which have periodicity $d$ modulo rotations as $2\alpha_d$.  
Then 
\beq
L N_{Z,{\rm tri},{\rm PF},L}-N_{Z,{\rm tri},{\rm FF},L} = 
                      \sum_{d|L} 2\alpha_d(L-d)
\label{nztripf}
\eeq
where $d$ are all the positive integers that divide $L$.

There are two kinds of partitions among $2\alpha_d$, each of which contains
$\alpha_d$ specific partitions. The first kind of partitions is defined by the
condition that each set of $d$ adjacent vertices does not have connection with
any other vertex in the complementary subset of vertices. Let us label the
vertices in one set of $d$ vertices as 1,2,\ldots,$d$. For $d=1$, the set of
partition is $\{ 1\}$. For $d=2$, the set of partition is $\{
\delta_{1,2}\}$. For $d=3$, the set of partitions is $\{ \delta_{1,2},
\delta_{1,3}, \delta_{1,2,3}\}$. The second kind of partitions is defined by
the condition that all of the first vertices of each of the sets of $d$
vertices are connected to each other. For $d=1$, the set of partition is $\{
\delta_{1,1^\prime}\}$, where $1^\prime$ is the first vertex of the adjacent
set of $d$ vertices. For $d=2$, the set of partition is $\{
\delta_{1,1^\prime}\}$. For $d=3$, the set of partitions is $\{
\delta_{1,1^\prime}, \delta_{1,1^\prime}\delta_{2,3},
\delta_{1,1^\prime,3}\}$. Since there is a one-to-one correspondence between
these two kinds of partitions, let us only consider the second kind of the
partitions which always have the connection $\delta_{1,1^\prime}$. Denote
$\alpha^\prime_d$ as the number of partitions which has periodicity less than
or equal to $d$ modulo rotations. This can be written in term of
$\alpha_d$ as 
\beq
\alpha^\prime_d = \sum_{d^\prime|d} \alpha _{d^\prime} 
\label{alphaprimealpha}
\eeq
For the set of $d$ vertices, if vertex 1 does not connect to any other vertex
except $1^\prime$, the first vertex of an adjacent set, then the number of
partitions is equal to $N_{Z,{\rm tri},{\rm FF},d-1}=C_{d-1}$. 
If vertex 1 has a connection
to vertex $b$ for $1 < b < d$ in addition to vertex $1^\prime$, then the number
of partitions is $C_{b-2}C_{d-b}$. The connection of vertex 1 to vertex $d-b+2$
is equivalent to the connection of vertex 1 to vertex $b$ under rotation
symmetry, and should not be considered again. Of course, vertex 1 can have
connection to more than one vertices in addition to
$\delta_{1,1^\prime}$. Therefore, to calculate $\alpha^\prime_d$, we have to
partition $d$ first. For each partition of $d$, e.g., $[x_1, x_2,\ldots, x_i]$
with $x_1+x_2+\ldots+x_i=d$, where $i$ is the numbers of vertices in the set of
$d$ vertices connected to vertex 1 (including vertex 1), we multiply all the
corresponding $C_{x_i-1}$'s by the number of different combinations of this
partition modulo rotations (thus [2,2,1,1] is the same as [2,1,1,2] and
[1,1,2,2], but different from [2,1,2,1]). As an example, for $d = 5$, we have
the partitions $5 = 4+1 = 3+2 = 3+1+1 = 2+2+1 = 2+1+1+1 = 1+1+1+1+1$, and
therefore
\beqs 
\alpha^\prime _5 & = & C_4 + \frac{1}{2}C_3C_0{2 \choose 1} +
\frac{1}{2}C_2C_1{2 \choose 1} + \frac{1}{3}C_2C_0^2{3 \choose 1} +
\frac{1}{3}C_1^2C_0{3 \choose 2} \nonumber \\
&   & \qquad  + \frac{1}{4}C_1C_0^3{4 \choose 1} + C_0^5{5\choose 5} \cr\cr
& = & 14 + 5 \times 1 + 2 \times 1 + 2 \times 1^2 + 1^2 \times 1 + 1
\times 1^3 + 1^5 =  26 
\label{alpha_example}
\eeqs
The factors $\frac{1}{2}$, $\frac{1}{3}$, and $\frac{1}{4}$ are included
because of the equivalence under rotations.  

We find that $\alpha^\prime_d$ is the same as the rooted planar trees with $d$
edges with the property that rotations about the root vertex yield equivalent
trees, given as sequence A003239 in \cite{sl}. The reason can be explained as
follows.  Define a ``planted'' tree as a rooted tree with the property that the
root vertex has degree one (where the degree of a vertex is the number of edges
connected to it).  Now we can construct planted trees from subtrees of the
rooted planar tree that are each connected to the root vertex. Note that the
number of planted tree with $d$ edges is $C_{d-1}$ \cite{stanley}.  The
sequence A003239 also gives the number of necklaces with a total of $2d$ beads
where $d$ beads have one color and the other $d$ beads have another
color. The explicit formula is
\beq
\alpha^\prime_d = \frac{1}{2d} \sum_{d^\prime|d} \phi (d/d^\prime)
{2d^\prime \choose d^\prime} \ ,
\label{alphaprime}
\eeq  
where $\phi(n)$ was defined above after eq.~(\ref{nztripfgen}). 

Now the number of partitions that have periodicity equal to $d$ modulo
rotations, $\alpha_d$, is the M\"obius transformation \cite{bs} 
of $\alpha^\prime_d$, given by 
\beq
\alpha_d = \sum_{d^\prime|d} \mu (d/d^\prime) \alpha^\prime _{d^\prime}
\label{mobiusalpha1}
\eeq
where $\mu(n)$ is the M\"obius function, defined as $-1$ if $n$ is prime, $0$
if $n$ has a square factor, and $1$ for other $n$. We find that $\alpha_d$ is
listed as sequence A022553 in \cite{sl}, which is the Lyndon words containing a
total of $2d$ letters with $d$ letters of one type, and the other $d$ letters
of another. One has the explicit formula
\beq
\alpha_d = \frac{1}{2d} \sum_{d^\prime|d} \mu (d/d^\prime) {2d^\prime 
\choose d^\prime} \ ,
\label{mobiusalpha2}
\eeq
i.e., $2d\alpha_d$, is the M\"obius transformation of 
${2d \choose d}$, or equivalently,
\beq
{2d \choose d} = \sum_{d^\prime|d} 2d^\prime \alpha _{d^\prime} 
\label{mobiusalphaprime}
\eeq
Therefore, the total number of partitions that have periodicity $d$
modulo rotations is 
\beq
2\alpha_d = \frac{1}{d} \sum_{d^\prime|d} \mu (d/d^\prime) {2d^\prime 
\choose d^\prime}
\label{mobiusalpha3}
\eeq
This is given as sequence A060165 in \cite{sl}. 

We finally have
\beqs
LN_{Z,{\rm tri},{\rm PF},L}-N_{Z,{\rm tri},{\rm FF},L} & = & 
       \sum_{d|L}2\alpha_d(L-d) \cr\cr
& = & \sum_{d|L} 2L\alpha_d - {2L \choose L} \cr\cr
& = & \sum_{d|L} \frac{L}{d} \sum_{d^\prime|d} \mu (d/d^\prime)
 {2d^\prime \choose d^\prime} - {2L \choose L} \cr\cr
& = & \sum_{d^\prime|L} \ \sum_{d|L;d^\prime|d} \frac{L}{d} \mu
 (d/d^\prime) {2d^\prime \choose d^\prime} - {2L \choose L} \cr\cr
& = &  \sum_{d^\prime|L} \phi (L/d^\prime) {2d^\prime \choose d^\prime} - 
 {2L \choose L} \cr\cr
& = & \sum_{d|L; \ 1 \le d < L} \phi (L/d) {2d \choose d} 
\label{LtNtriPFmNtriFF}
\eeqs
and the theorem follows. \ $\Box$

\bigskip

Theorems~\ref{theorem1} and~\ref{theorem2} imply a simple corollary
which gives an exact formula for $N_{Z,{\rm sq},{\rm PF},L}$:

\begin{corollary} \label{corollary1} For arbitrary $L$,
\beq
N_{Z,{\rm sq},{\rm PF},L} = 
                     \frac{1}{2} \left[ N_{Z,tri,PF,L} +  {L \choose [L/2]} 
\right]
\label{nzsqpf}
\eeq
\end{corollary}

{\sl Proof} \quad For even $L$ is trivial. For odd $L$, we only have to 
notice that
\beq
(1/2){L+1 \choose (L+1)/2} = \frac{L(L-1) \cdots \frac{L+3}{2}}
{(\frac{L-1}{2})!} = {L\choose (L-1)/2} = {L \choose [L/2] } 
\label{notice}
\eeq
This proves the corollary. $\Box$
    
\bigskip

One of us (S.-C.C.) has presented a conjecture, Conjecture~ 4.2.2 in 
\cite{thesis}, which generalizes Theorem~3 in \cite{ts} 
(i.e., Theorem~4.2.6 in \cite{thesis}). We restate this conjecture here. 
Consider the number of 
$\lambda$'s in the chromatic polynomial for a strip of the triangular lattice 
with cylindrical boundary conditions: 

\bigskip

\begin{conjecture} \label{conjecture1} For arbitrary $L$,
\beq
N_{P,{\rm tri},{\rm PF},L}=\frac{1}{L}\biggl [ d_L + \sum_{d|L; \ 1 \le d < L} 
\phi(L/d)t_d \biggr ]
\label{nptripfgen}
\eeq
where $t_d = a_{d+1}$ is the largest coefficient in the expansion of 
$(1+z+z^2)^d$, i.e., the central trinomial coefficient (given as sequence 
A002426 in \cite{sl}), and $d_L$ is essentially the Riordan number $r_L$ 
(given as sequence A005043 in \cite{sl})
\be
d_L = \left\{ \begin{array}{ll}
              1   & \qquad L=1 \\
              r_L & \qquad L\geq 2
              \end{array} \right. 
\ee
\end{conjecture}
where the reader is cautioned not to confuse the $d_L$ in
eq.~(\ref{nptripfgen}) with the different quantity $d_m$ in eqs.~(\ref{bm}) and
(\ref{dm}).  We can motivate this conjecture as follows. We know that $d_L$ is
the number of non-crossing non-nearest-neighbor partitions of a set of $L$
vertices with periodic boundary conditions, as shown in \cite{ss00}. The number
$L N_{P,{\rm tri},{\rm PF},L}-d_L$ is equal to $L-1$ for prime $L$, as shown in
\cite{ts,thesis}, since all the partitions of $N_{P,{\rm tri},{\rm PF},L}$ have
periodicity $L$ for prime $L$ except for the partitions $1$ (i.e., all blocks
being singletons) which have periodicity 1. Consider a general $L$ having the
factor $d$, and denote the number of non-nearest-neighbor partitions that have
periodicity $d$ modulo rotations as $\beta _d$.  Then
\beq
L N_{P,{\rm tri},{\rm PF},L}-d_L = \sum_{d|L} \beta _d(L-d) 
\label{nptripf}
\eeq

There are two kinds of partitions among $\beta_d$; we denote the number of
these as $\beta^{\rm I}_d$ and $\beta^{\rm II}_d$, respectively. 
The first kind of
partitions is defined by the condition that each set of $d$ adjacent vertices
does not have connection to any other vertices in other sets. Label the $d$
vertices as 1,2,\ldots,$d$. 
For $d=1$, the set of partitions consists of just $\{1\}$. 
There is no non-nearest-neighbor partition for $d=2$. For $d=3$, the set
of partitions is $\{ \delta_{1,3}\}$. For $d=4$, the set of partitions is $\{
\delta_{1,4}, \delta_{1,3}\}$. The second kind of partitions is defined by the
condition that all of the first vertices of each set of $d$ vertices are
connected to each other. There is no non-nearest-neighbor partition for
$d=1$. For $d=2$, the set of partitions is comprised of $\{
\delta_{1,1^\prime}\}$, where $1^\prime$ is the first vertex of an adjacent set
of $d$ vertices. For $d=3$, the set of partitions is
$\{\delta_{1,1^\prime}\}$. For $d=4$, the set of partitions is $\{
\delta_{1,1^\prime}, \delta_{1,1^\prime}\delta_{2,4}\}$. Notice that for $d=4$,
$\{\delta_{1,1^\prime,3}\}$ is the same as $\{\delta_{1,1^\prime}\}$ for $d=2$,
and should not be included to avoid double-counting. 

Consider the first kind of partitions. If the first vertex 1 is connected
to the last vertex $d$ among other possible connections with other vertices
in the set of $d$ vertices, the number of these partitions is $r_{d-1}$, the
Riordan number (given as sequence A005043 in \cite{sl}). This can be seen
by identifying the vertices 1 and $d$ to form a circuit with $d-1$ 
vertices \cite{ss00}. In addition, the set of $d$ vertices can also be 
partitioned into several parts where the first and the last vertices of 
each part are connected, but parts are not connected to each other. To 
calculate $\beta^{\rm I}_d$, we partition $d$, and apply the same 
transformation as illustrated in Theorem~\ref{theorem2} 
on $r_{d-1}$, then apply a M\"obius transformation. The first few values of $\beta^{\rm I}_d$ are 1, 0, 1, 2, 5, 11, 28, 68, 174, 445 for $1 \le d \le 10$.

Consider the second kind of partitions.  By an argument similar to that given
in the proof of Theorem~\ref{theorem2}, if vertex 1 is not connected to any
other vertex except $1^\prime$ among the set of $d$ vertices, then the number
of partitions is just $N_{P,{\rm tri},{\rm FF},d-1}=M_{d-2}$. 
In addition, the vertex 1 can
also be connected to other vertices among the set of $d$ vertices. To calculate
$\beta^{\rm II}_d$, we again partition $d$, and apply the same transformation, 
as illustrated in Theorem~\ref{theorem2} on $M_{d-2}$.  
We then apply a M\"obius
transformation. The first few values of $\beta^{\rm II}_d$ are 0, 1, 1, 2, 5, 11, 28, 68, 174, 445 for $1 \le d \le 10$. We find that $\beta^{\rm I}_d$ and $\beta^{\rm II}_d$ 
are the
same up to $d=10$, except for $d=1$ and 2.  
Although $\beta_d$ is not listed in \cite{sl}, we find that the following
relation holds up to $d=10$, and is similar to eq.~(\ref{mobiusalpha3}):
\beq
\beta _d = \frac{1}{d} \sum_{d^\prime|d} \mu (d/d^\prime) t_d 
\label{beta}
\eeq
If this is correct for arbitrary $d$, then the conjecture is proved.

Combining Conjecture~\ref{conjecture1} and Conjecture~1 in \cite{ts}, we
have 

\bigskip

\begin{conjecture} \label{conjecture2} For arbitrary $L$,
\beq
N_{P,{\rm sq},{\rm PF},L} = \cases{ \frac{1}{2}\biggl [ 
 N_{P,{\rm tri},{\rm PF},L} + 
\frac{1}{2}N_{P,{\rm sqtri},{\rm FP},\frac{L}{2}} \biggr ] & for even $L$ \cr
\frac{1}{2}\biggl [ N_{P,{\rm tri},{\rm PF},L} +
\frac{1}{4}N_{P,{\rm sqtri},{\rm FP},\frac{L+1}{2}} - \frac{1}{2}
r_{\frac{L-1}{2}} \biggr ] & for odd $L \ge 3$ }
\label{npsqpfgen}
\eeq
where $N_{P,{\rm sqtri},{\rm FP},L}$ was given in \cite{cf}, and $r_n$ is the 
Riordan number (given as sequence A005043 in \cite{sl}).
\end{conjecture}

%
%
\section{Potts Model Partition Functions for Triangular-lattice Strips 
         with Free Boundary Conditions} \label{sec.free_bc}

The Potts model partition function $Z(G,q,v)$ for a triangular-lattice strip 
of width $L$ and length $m$ with free boundary conditions is given by
\begin{subeqnarray}
Z(L_{\rm F} \times m_{\rm F},q,v) &=&
             \vv^{\rm T} \cdot \H \cdot \T^{m-1} \cdot \uu_{\rm id} \\
        & = & \w^{\rm T} \cdot \T^{m-1} \cdot \uu_{\rm id} 
\label{def_Z_free}
\end{subeqnarray}
where $\w^{\rm T} = \vv^{\rm T} \cdot \H$.  Hereafter we shall follow
the notation and the computational methods developed in
\cite{ts,ss00,transfer2,transfer3} (For chromatic polynomials, a related matrix
formulation has been discussed in \cite{b,matmeth,baxter_86,baxter_87}.)
Concerning notation, no confusion should result between the vertex set $V$, the
variable $v$, and the vector $\vv$.  Here $\T=\V \cdot \H$ is the transfer
matrix, and $\H$ (resp.\ $\V$) corresponds to adding one more layer of
horizontal (resp.\ vertical) bonds. The matrices $\T$, $\V$, and $\H$ act on
the space connectivities of sites on the top layer, whose basis elements are
indexed $\vv_{\cal P}$ are indexed by partitions ${\cal P}$ of the single-layer
vertex set $\{1,\ldots,L\}$. In particular, $\uu_{\rm id} =
\vv_{\{\{1\},\{2\},\ldots,\{L\}\}}$.

To simplify the notation, we shall denote the elements of the basis 
$\vv_{\cal P}$ by a shorthand using Kronecker delta functions: for instance
$\vv_{\{\{1,3\},\{2\},\{4,5\}\}}$ will be written $\delta_{1,3}\delta_{4,5}$. 
We denote the set of basis elements for a given strip as 
${\bf P} = \{ \vv_{\cal P} \}$. 

As we are dealing with planar lattices, only non-crossing 
partitions occur. The number of such partitions is given by the 
Catalan numbers
\be
C_n = {1\over n+1} \left( \begin{array}{c} 
                            2n\\
                            n
                           \end{array} \right)
\label{def_catalan}
\ee
In the triangular-lattice strip with free boundary conditions there is no
additional symmetry that allows us to reduce the number of partitions; 
thus the dimension of the transfer matrix for width $L$ is \cite{ss00} 
\be
N_{Z,{\rm tri},{\rm FF},L} = C_L 
\label{dimension_free}
\ee

We have obtained the transfer matrices $\T(L_{\rm F})$ and the vectors
$\w$ and $\uu_{\rm id}$ using symbolic computation with {\sc mathematica}
as in \cite{ss00,ts}. We have double-checked these results using a different
program written in {\sc perl}. 

An equivalent way to present a general formula for the partition function is
via a generating function.  Labelling a lattice strip of a given type 
and width as $G_m$, with $m$ the length, one has
\beq
\Gamma(G,q,v,z) = \sum_{m=0}^\infty z^m Z(G_m,q,v)
\label{gamma}
\eeq
where $\Gamma(G,q,v,z)$ is a rational function
\beq
\Gamma(G,q,v,z) = \frac{{\cal N}(G,q,v,z)}{{\cal D}(G,q,v,z)}
\label{gammaform}
\eeq
with
\beq
{\cal N}(G,q,v,z)=\sum_{j=0}^{\deg_z({\cal N})} A_{G,j}z^j
\label{numgamma}
\eeq
\beq
{\cal D}(G,q,v,z) = 1+\sum_{j=1}^{N_{Z,{\rm tri},{\rm BC},L}} b_{G,j}z^j 
                  = \prod_{j=1}^{N_{Z,{\rm tri},{\rm BC},L}}
                        (1- \lambda_{G,j}z)
\label{dengamma}
\eeq
where the subscript ${\rm BC}$ denotes the boundary conditions.  
In the transfer-matrix formalism, the $\lambda_{G,j}$'s in the denominator of
the generating function, eq.~(\ref{dengamma}), are the eigenvalues of $\T$.

Strips of the triangular lattice are well-defined for widths $L \ge 2$.  The
partition function $Z(G,q,v)$ has been calculated (for arbitrary $q$, $v$, and
$m$) for the strip with $L=2_{\rm F}$ in \cite{ta} by two of us using a
systematic iterative application of the deletion-contraction theorem.
$Z(G,q,v)$ was also studied for arbitrary $q$ and $v$ and zeros calculated for
$L=3,4$ and various lengths in \cite{ks}.  Here after reviewing the $L_{\rm
F}=2_{\rm F}$ case in our formalism, we shall report explicit results for the
partition function for $L_{\rm F} = 3_{\rm F}$.  For $4_{\rm F} \le L_{\rm F}
\le 6_{\rm F}$, the expressions for $\T(L_{\rm F})$, $\w(L_{\rm F})$
and $\uu_{\rm id}(L_{\rm F})$ are so lengthy that we cannot include them here.
They are
available from the authors on request and in the {\sc mathematica} file {\tt
transfer\_Tutte\_tri.m} which is available with the electronic version of this
paper in the {\tt cond-mat} archive at {\tt http://arXiv.org}. 

%
%
\subsection{$L=2_{\rm F}$} \label{sec.2F} 

Although this partition function was given earlier in \cite{ta}, it is useful
to review the calculation from the point of view of the (spin-representation)
transfer matrix. The number of elements in the basis is equal to $C_2=2$:
${\bf P} = \{ 1, \delta_{1,2} \}$. In this basis, the transfer matrices and
the other relevant quantities are given by
\begin{subeqnarray}
\T &=& \left( \begin{array}{cc}
             q^2 + 4 q v + 5 v^2 + v^3  & (1+v)(q + 3 v + v^2) \\ 
             v^2 (q + 3 v + v^2)        & v^2 (1+v) (2+v)  
             \end{array} \right) \\
\w^{\rm T}           &=&  q \left( q + v, 1 + v \right) \\  
\uu_{\rm id}^{\rm T} &=&    \left( 1,     0     \right)
\label{Tt2ff}  
\end{subeqnarray}
Because certain expressions recur in transfer matrices for wider strips, it
is convenient to re-express (\ref{Tt2ff}) in terms of these expressions; we
have
\begin{subeqnarray}
\T &=& \left( \begin{array}{cc}
              T_0 &  D_1E_3 \\
           v^2E_3 &  v^2D_1D_2 
           \end{array} \right) \\
\w^{\rm T} &=& q\left( F_1,D_1 \right)
\label{Tt2ffcompact}
\end{subeqnarray}
where
\begin{subeqnarray}
\slabel{def_Dk}
D_k &=& v+k \\
\slabel{def_Fk}
F_k &=& q+kv \\
\slabel{def_Ek}
E_k &=& q^2 + kv + v^2 \\
T_0 &=& q^2+4qv+5v^2+v^3
\end{subeqnarray}
In terms of this transfer matrix and these vectors one calculates the partition
function $Z(G_m,q,v)$ for the strip with a given length $m$ via
eq.~(\ref{def_Z_free}).  Equivalently, one can calculate the partition function
using a generating function, and this was the way in which the results were
presented in \cite{ta}, with 
\beq
{\cal N} = \prod_{j=1}^2(1-\lambda_{tf2,j}z)
\eeq
and
\beq
\lambda_{tf2,(1,2)} = \frac{1}{2} \biggl [ T_{S12} \pm (q+3v+v^2)\sqrt{R_{S12}}
 \ \biggr ]
\label{lams}
\eeq
where
\beq
T_{S12}=v^4+4v^3+7v^2+4qv+q^2
\label{t12}
\eeq
and
\beq
R_{S12} = q^2+2qv-2qv^2+5v^2+2v^3+v^4 
\label{rs12}
\eeq
The product of these eigenvalues, which is the determinant of $\T$, is 
\beq
{\rm det}(\T) = v^2(1+v)^2(v+q)^2 = v^2D_1^2F_1^2
\label{det_tff2}
\eeq
The vanishing of this determinant at $v=-1$ and $v=-q$ occur because in each
case one of the two eigenvalues is absent for, respectively, the chromatic and
flow polynomials \cite{f}.  Analogous formulas can be given for
${\rm det}(\T)$ for higher values of $L$; we omit these for brevity.
%
%
%
\subsection{$L=3_{\rm F}$} \label{sec.3F} 

The number of elements in the basis is equal to $C_3=5$:
${\bf P} = \{ 1, \delta_{1,2}, \delta_{1,3},\delta_{2,3},\delta_{1,2,3}
\}$. 
In this basis, the transfer matrices and the other relevant quantities 
are given by
\begin{subeqnarray}
\T &=& \left( \begin{array}{ccccc}
    T_1     &     D_1 F_3 E_3 &     T_2 &     D_1 T_3     & D_1^2 T_4      \\
v^2 F_3 E_3 & v^2 D_1 D_2 F_2 & v^2 T_5 & v^2 D_1 T_4     & v^2 D_1^2 D_2  \\
v^3 E_3     & v^2 D_1     E_3 & v^2 T_5 & v^3 D_1 D_2     & v^2 D_1^2 D_2  \\
v^2 E_3 E_4 & v^2 D_1 D_2 E_3 & v^2 T_6 & v^2 D_1 D_2 E_4 & v^2 D_1^2 D_2^2 \\ 
v^4 D_3 E_3 & v^3 D_1 D_2 E_3 & v^3 T_7 & v^4 D_1 D_2 D_3 & v^3 D_1^2 D_2^2 \\
             \end{array} \right) \\
\w^{\rm T}         &=& q \left( F_1^2, D_1 F_1, E_2, D_1 F_1, D_1^2  \right) \\
\uu_{\rm id}^{\rm T} &=&  \left( 1, 0, 0, 0, 0 \right)  
\end{subeqnarray}
where the $T_k$ are shorthand notations used in this section, defined as 
\begin{subeqnarray}
T_1 &=& q^3 + 7 q^2 v + 19 q v^2 + 19 v^3 + 2 q v^3 + 5 v^4 \\
T_2 &=& q^2 + 7 q v + 16 v^2 + q v^2 + 9 v^3 + 2 v^4 \\
T_3 &=& q^2 + 6 q v + 11 v^2 + q v^2 + 4 v^3 \\
T_4 &=& q + 5 v + 2 v^2 \\
T_5 &=& q + 6 v + 4 v^2 + v^3\\
T_6 &=& 2 q + 13 v + 13 v^2 + 6 v^3 +v^4\\
T_7 &=& q + 12 v + 13 v^2 + 6 v^3 + v^4
\end{subeqnarray}
%
%
%
\section{Potts Model Partition Functions for Triangular-lattice Strips 
         with Cylindrical Boundary Conditions}

The Potts model partition function $Z(G,q,v)$ 
for a triangular-lattice strip of width $L$ vertices and length $m$ vertices
with cylindrical boundary conditions can be written as 
\begin{subeqnarray}
Z(L_{\rm P} \times m_{\rm F},q,v) &=&
             \vv^{\rm T} \cdot \H \cdot \T^{m-1} \cdot \uu_{\rm id} \\
        & = & \w^{\rm T} \cdot \T^{m-1} \cdot \uu_{\rm id}
\label{def_Z_cyl}
\end{subeqnarray}
where again $\w^{\rm T} = \vv^{\rm T} \cdot \H$.  Results on
$Z(G,q,v)$ for $2_{\rm P} \le L_{\rm P} \le 4_{\rm P}$ with cylindrical
boundary conditions were previously given in \cite{ks}.  Here we shall
present explicit results for this partition function for strips of the
triangular lattice with cylindrical boundary conditions and widths $2_{\rm P}
\le L_{\rm P} \le 4_{\rm P}$.  Results for the transfer matrices in
computer-readable files for widths up to $L_{\rm P}=9_{\rm P}$ are also
available upon request; they are too lengthy to present here.

In the computation of the transfer matrix for a triangular-lattice strip 
with cylindrical boundary conditions there is a technical complication 
in order to treat correctly the last diagonal bond joining columns $L$ and 1 
\cite[Section~3]{ss00}. Instead of considering a triangular-lattice strip 
of width $L$ and cylindrical boundary conditions, we start with a 
strip of width $L+1$ and free boundary conditions. The parameter $v$ is 
the same for all edges, except for the vertical edges corresponding to column 
$L+1$, where it takes the value $v=0$. After performing the computation, we 
identify columns 1 and $L+1$. This procedure implies that there are  
double vertical edges (with parameters $v$ and 0) connecting sites on column 1;
but their net contribution is $v$, as expected.

We obtain in this way a transfer matrix of
dimension $C_{L+1}$. This matrix can be simplified by noting that there
are many zero eigenvalues. Let us denote by $\{\vv_j^{(s)}\}$ 
(resp.\ $\{\vv_j^{(n)}\}$ ) the basis elements corresponding to $L+1$ being 
(resp.\  not being) a singleton. The number of elements of $\{\vv_j^{(s)}\}$ is
$C_L$; hence the cardinality of $\{\vv_j^{(n)}\}$ is $C_{L+1}-C_L$. 
The zero eigenvalues are associated with certain 
eigenvectors of the form $\vv_1 - \vv_2$ where $\vv_1 \in \{\vv_j^{(s)}\}$ and
$\vv_2 \in \{\vv_j^{(n)}\}$. Let us make the following change of variable
\be
\vv_i^\prime = \left\{ \begin{array}{ll}
               \vv_i^{(s)} - \vv_j^{(n)} & \qquad i=1,\ldots,C_{L+1}-C_L \\
               \vv_i^{(s)}               & \qquad i=C_{L+1}-C_L+1,\ldots,C_{L+1}
                       \end{array}\right.
\ee
where the first elements corresponds to eigenvectors with zero eigenvalues, 
and the last $C_L$ elements are those of the basis $\{\vv_j^{(s)}\}$. 
Then, the transfer matrix takes the simple form 
\be
\T^\prime = \left( \begin{array}{cc}
                    0  & \T^{(n)} \\
                    0  & \T^{(s)} 
                   \end{array}
            \right)
\ee
where $\T^{(s)}$ (resp.\  $T^{(n)}$) is a matrix of dimension $C_L$ (resp.\ 
$C_{L+1}-C_L$). In this new basis the vectors $\w$ and 
$\uu_{\rm id}$ take the form
\begin{subeqnarray}
\w^\prime            &=& \left( 0, \w^{(s)} \right) \\
\uu_{\rm id}^\prime  &=& \left( 0, \uu_{\rm id}^{(s)} \right) 
\end{subeqnarray}
Thus, we can write the partition function in terms of $\T^{(s)}$, $\w^{(s)}$ 
and $\uu_{\rm id}^{(s)}$ alone
\be
Z(L_{\rm P} \times m_{\rm F},q,v) =
         \w^{(s)\ {\rm T}} \cdot [\T^{(s)}]^{m-1} \cdot \uu_{\rm id}^{(s)}
\ee

So far we have a transfer matrix with the same dimension as for free boundary
conditions (namely, $N_{Z,{\rm tri},{\rm FF},L} = C_L$). 
We can reduce even more the dimension of the transfer matrix by noting that 
cylindrical boundary conditions introduce 
an extra symmetry (i.e. translations along the transverse direction). 
In particular, we can make a further change of basis in the subspace 
$\{\vv_j^{(s)}\}$ so that 
\be
\vv_i^{(s)\prime} = \left\{ \begin{array}{ll}
          \vv_i^{(s,n)} & \qquad i=1,\ldots,C_L- N_{Z,{\rm tri},{\rm PF},L} \\
          \vv_i^{(s,t)} & \qquad i=C_L-N_{Z,{\rm tri},{\rm PF},L}+1,\ldots,C_L
                       \end{array}\right.
\ee
where the last $N_{Z,{\rm tri},{\rm PF},L}$ 
(resp.\ the first $C_L-N_{Z,{\rm tri},{\rm PF},L}$) 
elements $\{\vv_i^{(s,t)}\}$ (resp.\ $\{\vv_i^{(s,n)}\}$) are 
translational-invariant (resp.\  non-translational-invariant) 
combinations of the original vectors $\{\vv_i^{(s)}\}$. In this new basis,
the transfer matrix $\T^{(s)}$ takes a block diagonal form 
\be
\T^{(s)\prime} = \left( \begin{array}{cc}
                    \T^{(s,n)}  & 0 \\
                    0           & \T^{(s,t)}
                   \end{array}
            \right)
\ee
and the vectors $\w^{(s)}$ and $\uu_{\rm id}^{(s)}$ take the form  
\begin{subeqnarray}
\w^{(s)\prime}             &=& \left( 0, \w^{(s,t)} \right) \\
\uu_{\rm id}^{(s)^\prime}  &=& \left( 0, \uu_{\rm id}^{(s,t)} \right) 
\end{subeqnarray}
The partition function can be computed using the transfer matrix $\T^{(s,t)}$ 
\be
Z(L_{\rm P} \times m_{\rm F},q,v) =
                  \w^{(s,t)\ {\rm T}} \cdot [\T^{(s,t)}]^{m-1} 
                                      \cdot \uu_{\rm id}^{(s,t)}
\ee
The dimension of $\T^{(s,t)}$ will be denoted by $N_{Z,{\rm tri},{\rm PF},L}$ 
and is given by Theorem~\ref{theorem2} (see also Table~\ref{Table_dimensions}). 

For $L\geq 6$ there is a further simplification, which is also present in the
chromatic-polynomial case \cite{transfer3}: in the translation-invariant
subspace the transfer matrix does commute with the reflection operation. Thus,
we can pass to a new basis consisting of connectivities that are either even or
odd under reflection. In this new basis, the transfer matrix $\T^{(s,t)}$ has
the block diagonal form 
\be
\T^{(s,t)\prime} = \left( \begin{array}{cc}
                          \T^{(s,t)}_{-}  & 0 \\
                              0           & \T^{(s,t)}_{+}
                          \end{array}
                   \right)
\ee
and the vectors $\w^{(s,t)}$ and $\uu_{\rm id}^{(s,t)}$ take the form
\begin{subeqnarray}
\w^{(s,t)\prime}            &=& \left( 0, \w_{+}^{(s,t)} \right) \\
\uu_{\rm id}^{(s,t)^\prime} &=& \left( 0, \uu^{(s,t)}_{{\rm id},+} \right)
\end{subeqnarray}
Thus, we can compute the partition function using the transfer matrix 
$\T^{(s,t)\prime}$ 
\be
Z(L_{\rm P} \times m_{\rm F},q,v) =
                  \w_{+}^{(s,t)\ {\rm T}} \cdot 
                     [\T^{(s,t)}_{+}]^{m-1} 
                     \cdot \uu^{(s,t)}_{{\rm id},+}
\ee
In what follows, we shall drop the superindices $^{(st)}$ and 
the subindex $_{+}$ to simplify the notation. 
The dimension of the transfer matrix
$\T^{(s,t)}_{+} \equiv \T$ corresponds to the number of partition classes 
of the numbers $\{1,2,\ldots,L\}$ that are invariant under translations and
reflections. Thus, this number is precisely $N_{Z,{\rm sq},{\rm PF},L}$ and
is given in terms of $L$ by eq.~\reff{nzsqpf} (see Table~\ref{Table_dimensions}
for some numerical values).

We have obtained the transfer matrices $\T(L_{\rm P})$ and the vectors 
$\w(L_{\rm P})$ and $\uu_{\rm id}(L_{\rm P})$ using symbolic computation 
with {\sc mathematica} for $L_{\rm P} \leq 5_{\rm P}$.  
For $6_{\rm P} \le L_{\rm P} \le 9_{\rm P}$, we have used two different 
programs (one written in {\sc C} and the
other one in {\sc perl}), which were also used to double-check the results for
$L\leq 5_{\rm P}$.  For $5_{\rm P} \le L \le 9_{\rm P}$, the expressions for
$\T(L_{\rm P})$ and the vectors $\w(L_{\rm P})$ and $\uu_{\rm id}(L_{\rm P})$
can be found in the {\sc mathematica} file {\tt transfer\_Tutte\_tri.m} which
is available with the electronic version of this paper in the {\tt cond-mat}
archive at {\tt http://arXiv.org}. 

%
%
\subsection{$L=2_{\rm P}$} \label{sec.2P} 

The number of elements in the basis is two, with ${\bf P} = \{ 1,
\delta_{1,2}\}$; the transfer matrices and the other relevant quantities are
given by
\begin{subeqnarray}
\T &=& \left( \begin{array}{cc}
             q^2 + 6 q v + 12 v^2 + q v^2 + 8 v^3 + 2 v^4  & 
             D_1^2 (q + 4 v + 2v^2) \\ 
             v^2 (2 q + 12 v + 13 v^2 + 6 v^3 + v^4)       & 
             v^2 D_1^2 D_2^2  
             \end{array} \right) \\
\w^{\rm T}           &=&  q \left( E_2, D_1^2 \right) \\ 
\uu_{\rm id}^{\rm T} &=&    \left( 1, 0 \right)  
\end{subeqnarray}
As before, we have used the shorthand notation introduced above 
\reff{def_Dk}/\reff{def_Ek}.  

%
%
\subsection{$L=3_{\rm P}$} \label{sec.3P}

The number of (translational-invariant) elements in the basis is three: 
${\bf P} = \{ 1, \delta_{1,2}+\delta_{1,3}+\delta_{2,3}, \delta_{1,2,3} \}$.
The transfer matrices and the vectors $\w$ and $\uu_{\rm id}$ are given by
\begin{subeqnarray}
\T &=& \left( \begin{array}{ccc}

              T_{11} & 3     D_1     T_{12} & D_1^3 T_{13} \\
      v^2     T_{21} &   v^2 D_1     T_{22} & v^2 D_1^3 D_2^2 \\
      v^4 D_3 T_{31} & 3 v^3 D_1 D_2 T_{32} & v^3 D_1^3 D_2^3 
             \end{array} \right) \\
\w^{\rm T}           &=&  q \left( V_1, 3 D_1 E_2 , D_1^3 \right) \\ 
\uu_{\rm id}^{\rm T} &=&    \left( 1, 0, 0 \right)  
\end{subeqnarray}
where the factors $D_k, E_k$ are given in \reff{def_Dk}/\reff{def_Ek}, and 
\begin{subeqnarray}
V_1    &=& q^2 + 3 q v + 3 v^2 + 2v^3 \\[2mm]
T_{11} &=& q^3 + 9 q^2 v + 33 q v^2 + 4 q v^3 + 50 v^3 + 21 v^4 + 3 v^5 \\
T_{12} &=& q^2 + 8 q v + 20 v^2 + 2 q v^2 + 14 v^3 + 3 v^4 \\ 
T_{13} &=& q + 6 v + 3 v^2\\ 
T_{21} &=& q^2 + 10 q v + 30 v^2 + 2 q v^2 + 22 v^3 + 7 v^4 + v^5 \\  
T_{22} &=& 6 q + 38 v + 2 q v + 42 v^2 + 18 v^3 + 3 v^4\\ 
T_{31} &=& 3 q + 18 v + 15 v^2 + 6 v^3 + v^4\\ 
T_{32} &=& q + 12 v + 13 v^2 + 6 v^3 + v^4
\end{subeqnarray}
%

%
%
\subsection{$L=4_{\rm P}$} \label{sec.4P}

The number of (translational-invariant) elements in the basis is six:
${\bf P} = \{ 1,
 \delta_{1,2}+\delta_{2,3}+\delta_{3,4}+\delta_{1,4},
 \delta_{1,3}+\delta_{2,4},
 \delta_{1,2,3}+\delta_{1,2,4}+\delta_{1,3,4}+\delta_{2,3,4},
 \delta_{1,2,3,4},
 \delta_{1,4}\delta_{2,3} + \delta_{1,2}\delta_{3,4} \}$.
The transfer matrix is given by
\begin{eqnarray}
& & \T = \\
& &
   \left( \begin{array}{cccccc}
                   T_{11}&     4 D_1         T_{12}&         2 E_4   T_{13} &
     4 D_1^2       T_{14}&       D_1^4 T_{15}      &         2 D_1^2 T_{16} \\
   v^2             T_{21}& 2 v^2 D_1         T_{22}&   v^2           T_{23} &
   v^2 D_1^2       T_{24}&   v^2 D_1^4 D_2^2       & 2 v^2 D_1^2 D_2 T_{26} \\
   v^3             T_{31}& 2 v^2 D_1         T_{32}& 4 v^2         T_{34}^2 &
 4 v^2 D_1^2 D_2   T_{34}&   v^2 D_1^4 D_2^2       &   v^2 D_1^2     T_{36} \\
   v^4 D_3         T_{41}&   v^3 D_1         T_{42}& 2 v^3     T_{34} T_{54}&
 2 v^3 D_1^2 D_2   T_{44}&   v^3 D_1^4 D_2^3       &   v^3 D_1^2 D_2 T_{46} \\
   v^6 D_3^2       T_{51}& 4 v^5 D_1 D_2 D_3 T_{52}& 2 v^4         T_{54}^2 &
 4 v^4 D_1^2 D_2^2 T_{54}&   v^4 D_1^4 D_2^4       &
                                               2 v^5 D_1^2 D_2^2 D_3 T_{56}\\
   v^4           T_{62}^2& 4 v^4 D_1 D_2     T_{62}&   v^4           T_{63} &
 2 v^4 D_1^2 D_2^2       & 0                       & 2 v^4 D_1^2 D_2^2
\end{array} \right)  \nonumber
\end{eqnarray}
where the factors $D_k(v)$ and $E_k$ are defined in
\reff{def_Dk}/\reff{def_Ek}; the $T_{ij}$ are given by
\begin{subeqnarray}
T_{11} &=& q^4 + 12 q^3 v + 62 q^2 v^2 + 164 q v^3 + 4 q^2 v^3 + 192 v^4 +
           29 q v^4 \nonumber \\
       & & \qquad + 72 v^5 + 6 v^6 \\
T_{12} &=& q^3 + 11 q^2 v + 47 q v^2 + q^2 v^2 + 80 v^3 + 11 q v^3 +
           40 v^4 + 5 v^5 \\
T_{13} &=& q^2 + 8 q v + 24 v^2 + q v^2 + 16 v^3 + 4 v^4 \\
T_{14} &=& q^2 + 10 q v + 28 v^2 + 3 q v^2 + 20 v^3 + 4 v^4 \\
T_{15} &=& q + 8 v + 4 v^2 \\
T_{16} &=& q^2 + 10 q v + 32 v^2 + 3 q v^2 + 24 v^3 + 5 v^4 \\
T_{21} &=& q^3 + 12 q^2 v + 54 q v^2 + 2 q^2 v^2 + 96 v^3 + 18 q v^3 +
        59 v^4 + q v^4 \nonumber \\
       & & \qquad + 12 v^5 + v^6 \\
T_{22} &=& 3 q^2 + 28 q v + q^2 v + 80 v^2 + 13 q v^2 + 65 v^3 + q v^3 +
        18 v^4 + 2 v^5 \\
T_{23} &=& 3 q^2 + 34 q v + 96 v^2 + 23 q v^2 + 120 v^3 +
        8 q v^3 + 67 v^4 + q v^4 \nonumber \\
       & & \qquad + 18 v^5 + 2 v^6 \\
T_{24} &=& 10 q + 56 v + 6 q v + 63 v^2 + q v^2 + 26 v^3 + 4 v^4\\
T_{26} &=& q + 8 v + 5 v^2 + v^3\\
T_{31} &=& 2 q^2 + 18 q v + 48 v^2 + 4 q v^2 +  31 v^3 + 8 v^4 + v^5 \\
T_{32} &=& q^2 + 14 q v + 52 v^2 + 4 q v^2 + 43 v^3 + 14 v^4 + 2 v^5 \\
T_{34} &=& q + 6 v + 4 v^2 + v^3 \\
T_{36} &=& 2 q + 24 v + 27 v^2 + 12 v^3 + 2 v^4 \\
T_{41} &=& q^2 + 11 q v + 36 v^2 + 2 q v^2 + 24 v^3 + 7 v^4 + v^5\\
T_{42} &=& 2 q^2 + 43 q v + q^2 v + 216 v^2 + 28 q v^2 + 270 v^3
              + 4 q v^3 + 138 v^4 \nonumber \\
       & & \qquad + 36 v^5 + 4 v^6 \\
T_{44} &=& 3 q + 24 v + q v + 27 v^2 + 12 v^3 + 2 v^4\\
T_{46} &=& q + 24 v + 27 v^2 + 12 v^3 + 2 v^4\\
T_{51} &=& 4 q + 24 v + 17 v^2 + 6 v^3 + v^4\\
T_{52} &=& 2 q + 18 v + 15 v^2 + 6 v^3 + v^4\\
T_{54} &=& q + 12 v + 13 v^2 + 6 v^3 + v^4\\
T_{56} &=& 4 + 3 v + v^2  \\
T_{62} &=& q + 6 v + 2 v^2\\
T_{63} &=& 3 q + 24 v + 26 v^2 + 12 v^3 + 2 v^4
\end{subeqnarray}
The vectors $\w$ and $\uu_{\rm id}$ are given by
\begin{subeqnarray}
\w^{\rm T} &=&  q \left( V_1, 4 D_1 V_2, 2 E_2^2, 4 D_1^2 E_2, D_1^4,
                                  2 D_1^2 E_2 \right) \\ 
\uu_{\rm id}^{\rm T} &=&  \left( 1, 0,0,0,0,0 \right)  
\end{subeqnarray}
where 
\begin{subeqnarray}
V_1 &=& q^3 + 4 q^2 v + 6 q v^2 + 4 v^3 + v^4 \\
V_2 &=& q^2 + 3 q v + 3 v^2 + v^3 
\end{subeqnarray}
%

%
%
\section{Partition Function Zeros in the $\lowercase{q}$ Plane}

In this section we shall present results for zeros and continuous 
accumulation sets ${\cal B}_q$ (in the $q$-plane) for the 
partition function of the Potts antiferromagnet on triangular-lattice
strips of widths $L\leq 5$ with free or cylindrical boundary conditions.

\begin{figure}[hbtp]
\vspace*{-1.5cm}
\centering
\begin{tabular}{cc}
   \includegraphics[width=190pt]{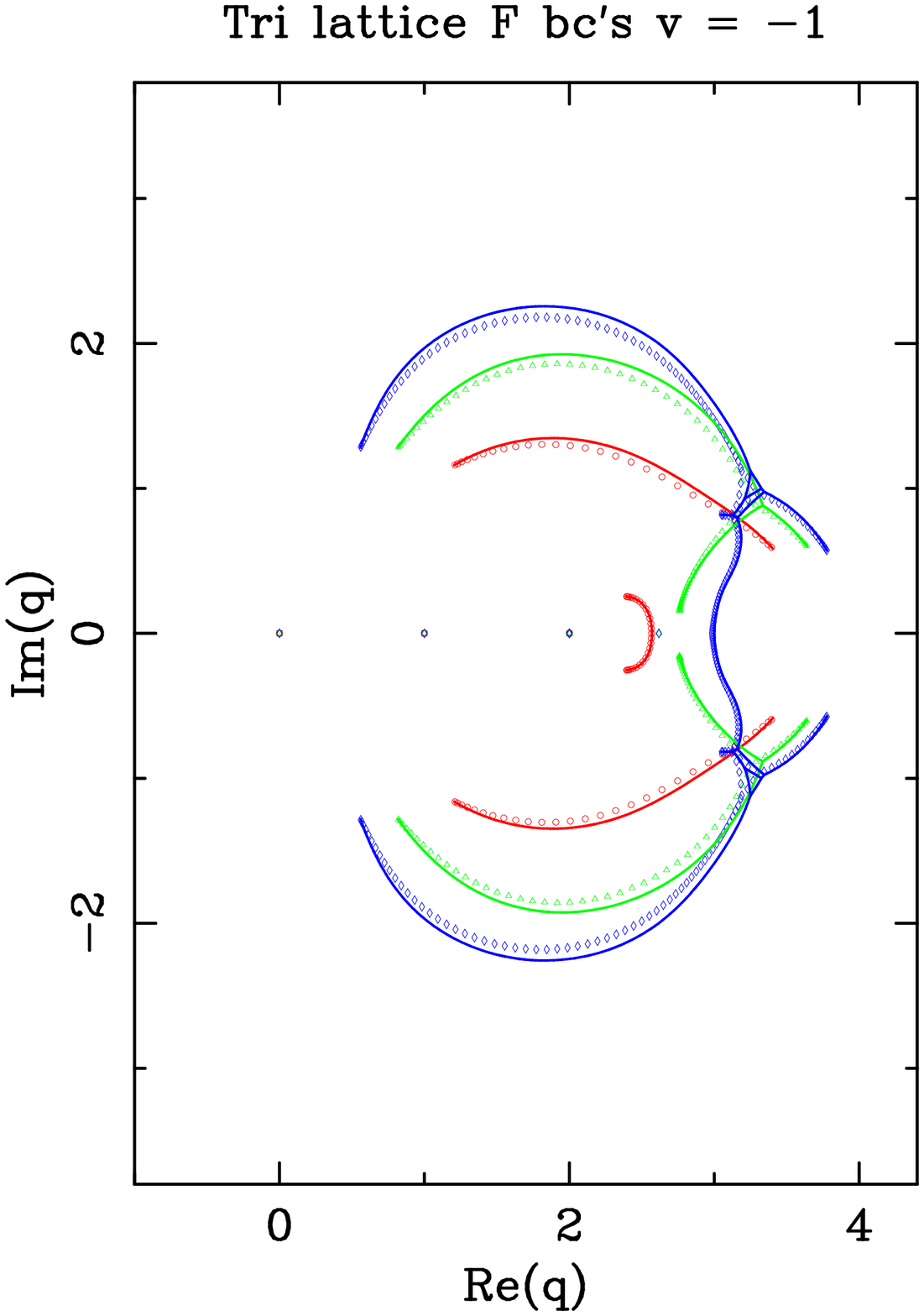} &
   \includegraphics[width=190pt]{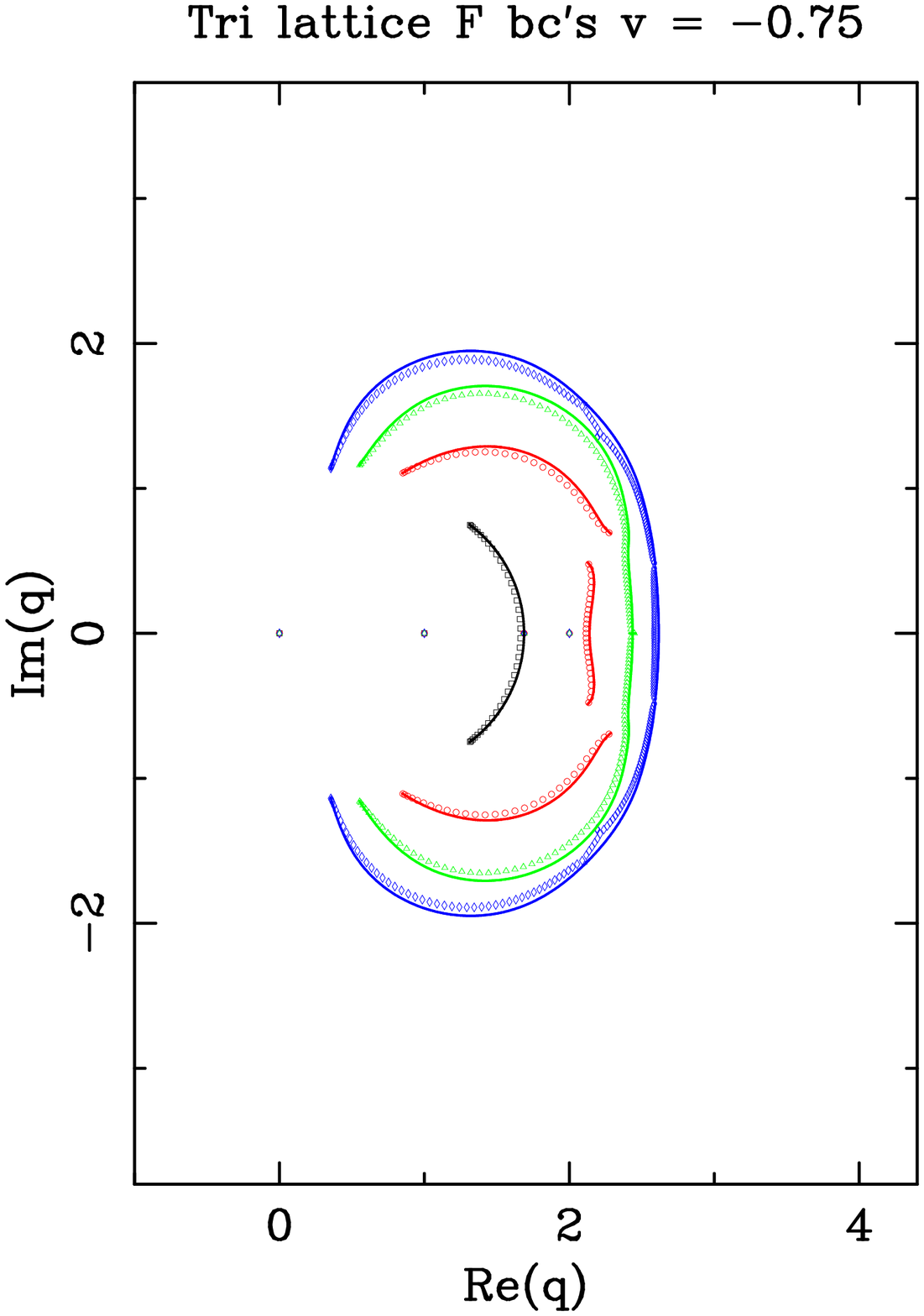} \\[1mm]
   \phantom{(((a)}(a)    & \phantom{(((a)}(b) \\[5mm]
   \includegraphics[width=190pt]{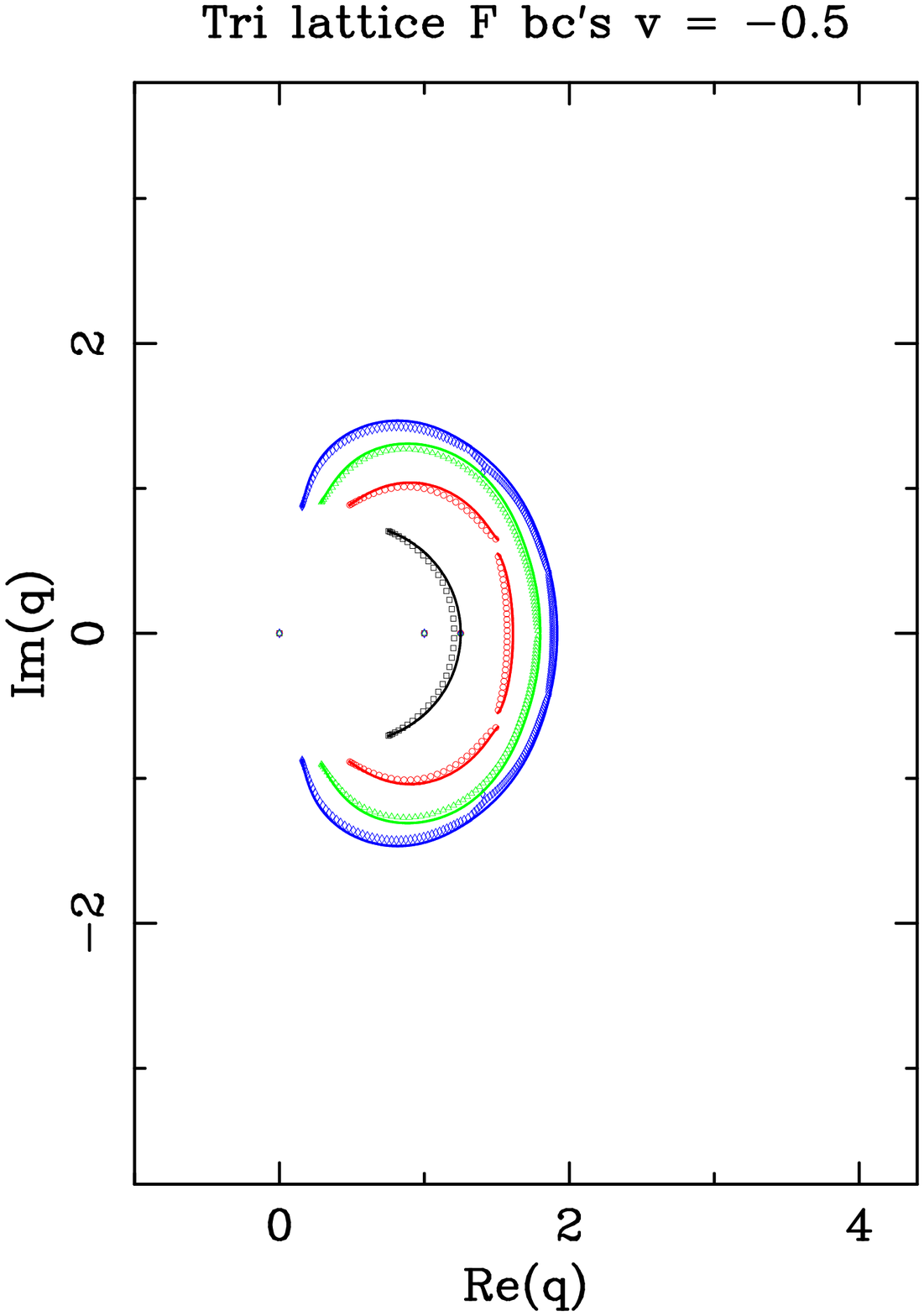} &
   \includegraphics[width=190pt]{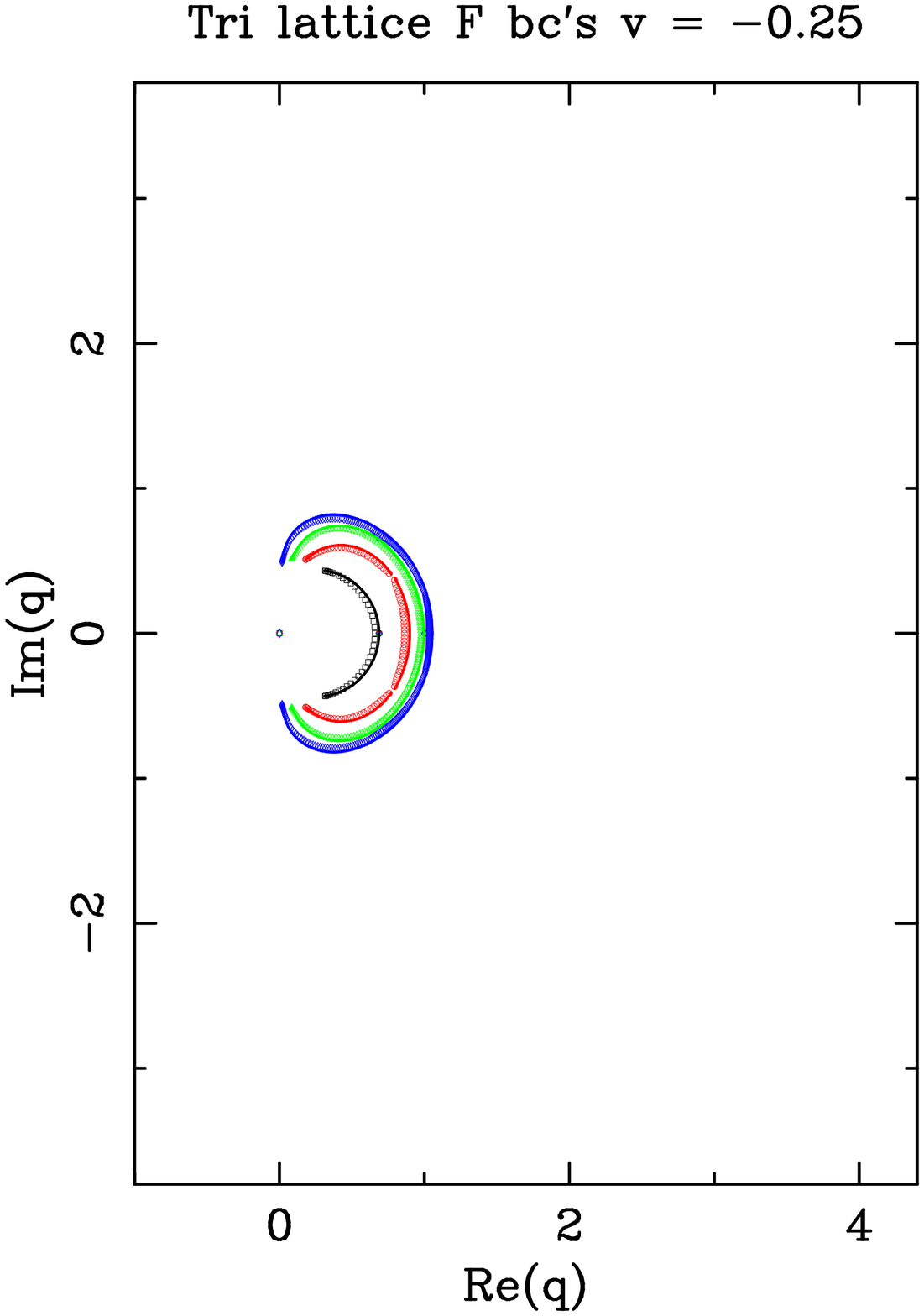} \\[1mm]
   \phantom{(((a)}(c)    & \phantom{(((a)}(d) \\
\end{tabular}
\caption[a]{\protect\label{figures_qplane_F} Limiting curves forming the
singular locus ${\cal B}_q$ for the Potts model free energy for (a) $v=-1$, (b)
$v=-3/4$, (c) $v=-1/2$, and (d) $v=-1/4$ on strips with free boundary
conditions and several widths $L$: 2 (black), 3 (red), 4 (green), and 5 (blue).
We also show the partition-function zeros for the strips $L_{\rm F} \times (10
L)_{\rm F}$ for the same values of $L$: 2 ($\Box$, black), 3 ($\circ$, red), 4
($\triangle$, green), and 5 ($\Diamond$, blue).  Where the results for $2 \le L
\le 4$ overlap with those in \cite{ta,ks} (see also \cite{strip,t,transfer3}
for $v=-1$), they agree and are included here for comparison.  
}
\end{figure}

In Figure~\ref{figures_qplane_F} we show the partition-function zeros in the
$q$-plane for strips of sizes $L_{\rm F} \times (10L)_{\rm F}$ with $2\leq L
\leq 5$ and free boundary conditions. We also show the corresponding limiting
curves ${\cal B}_q(L,v)$ for the infinite-length limit.
Figure~\ref{figures_qplane_F}(a) displays the zeros for $v=-1$ (i.e., the
chromatic zeros \cite{strip,t,transfer3}), and
Figures~\ref{figures_qplane_F}(b)--(d) display the corresponding zeros for the
non-zero temperatures $v=-0.75$ (b), $v=-0.5$ (c), and $v=-0.25$ (d).  The case
$L=2$ was studied in \cite{ta} and the cases $L=3,4$ were studied in
\cite[Figs. 3.15, 3.22]{ks} for arbitrary temperature, and our zeros are
in agreement with these earlier works.

\begin{figure}[hbtp]
\vspace*{-1.5cm}
\centering
\begin{tabular}{cc}
   \includegraphics[width=190pt]{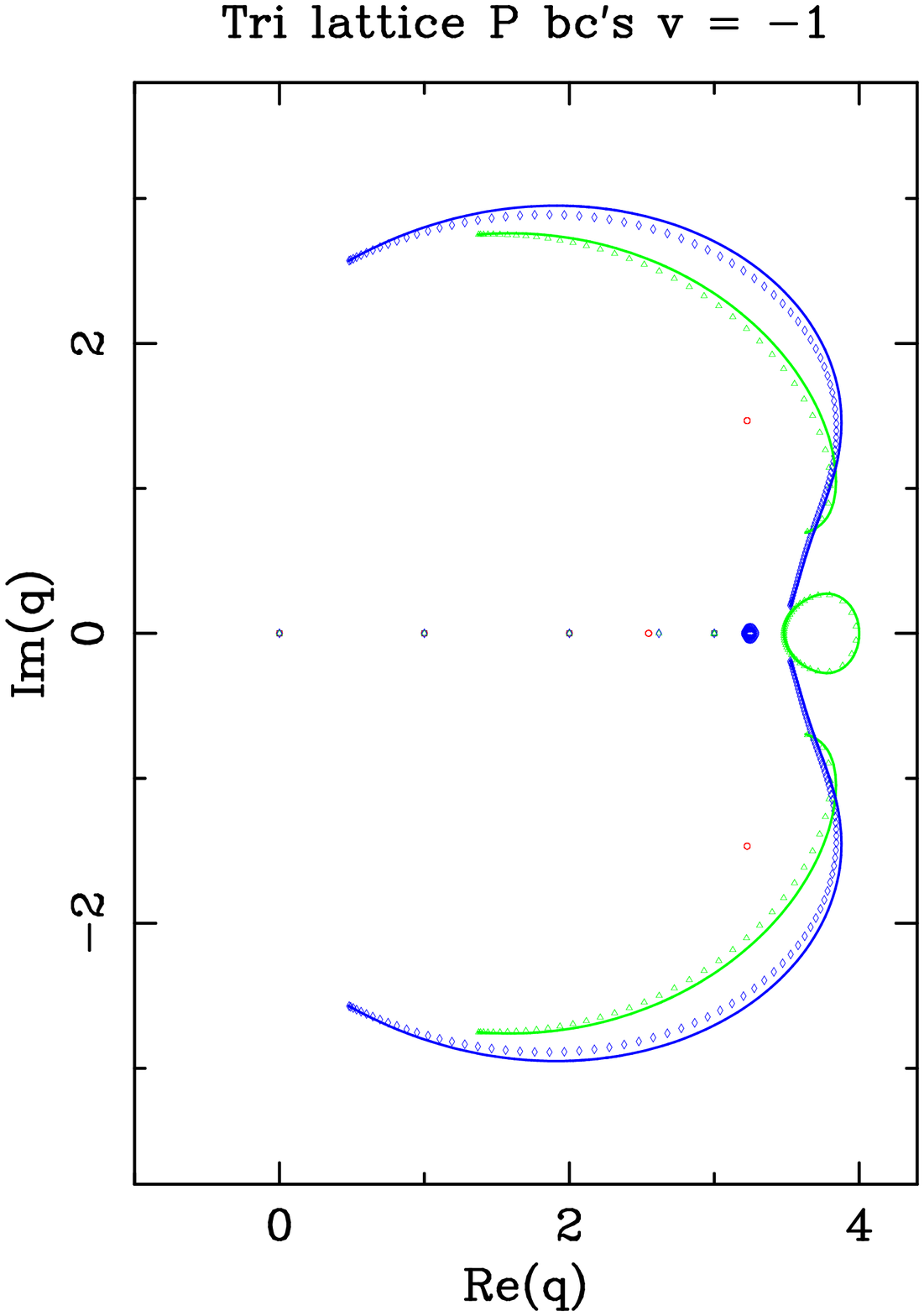} &
   \includegraphics[width=190pt]{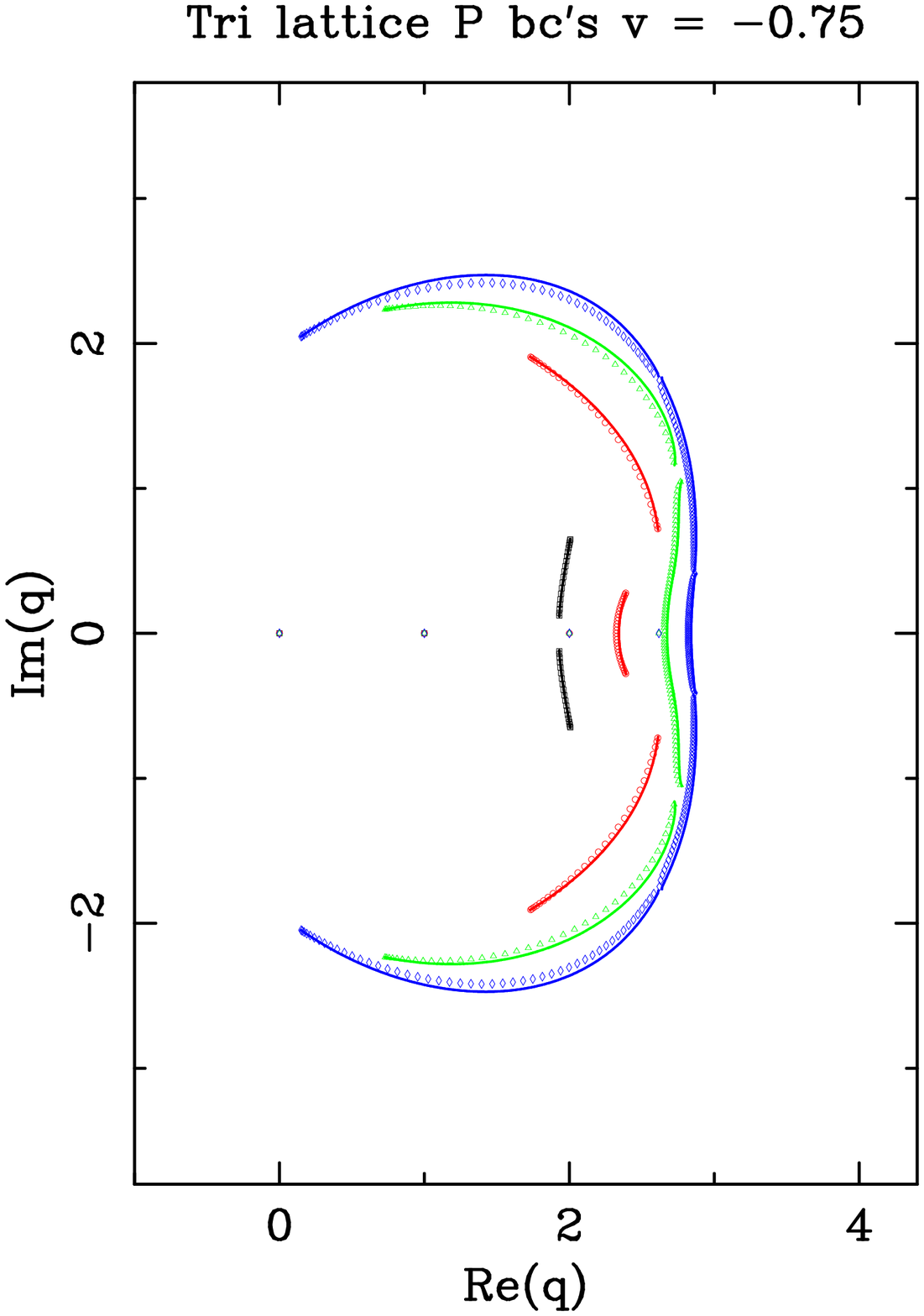} \\[1mm]
   \phantom{(((a)}(a)    & \phantom{(((a)}(b) \\[5mm]
   \includegraphics[width=190pt]{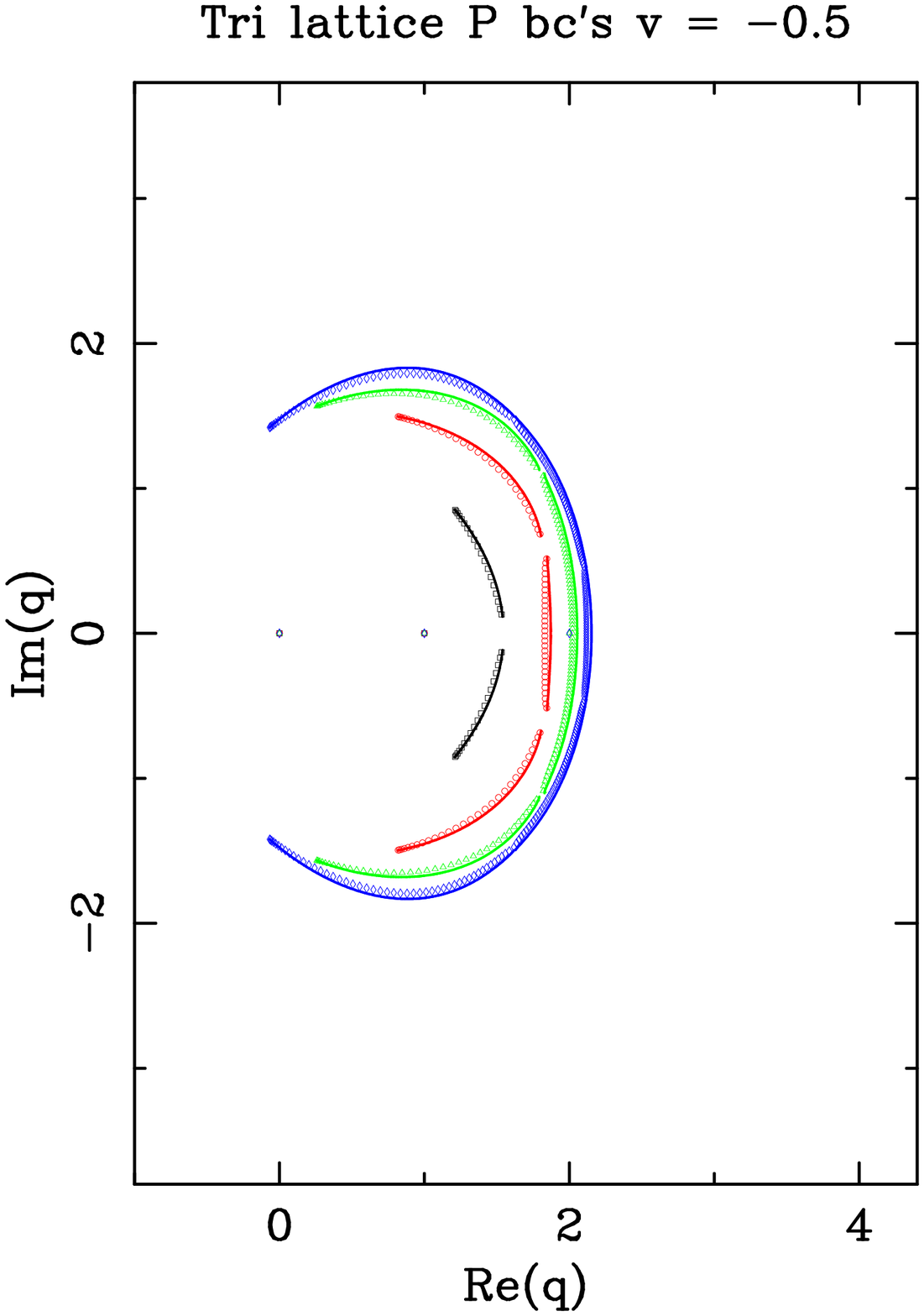} &
   \includegraphics[width=190pt]{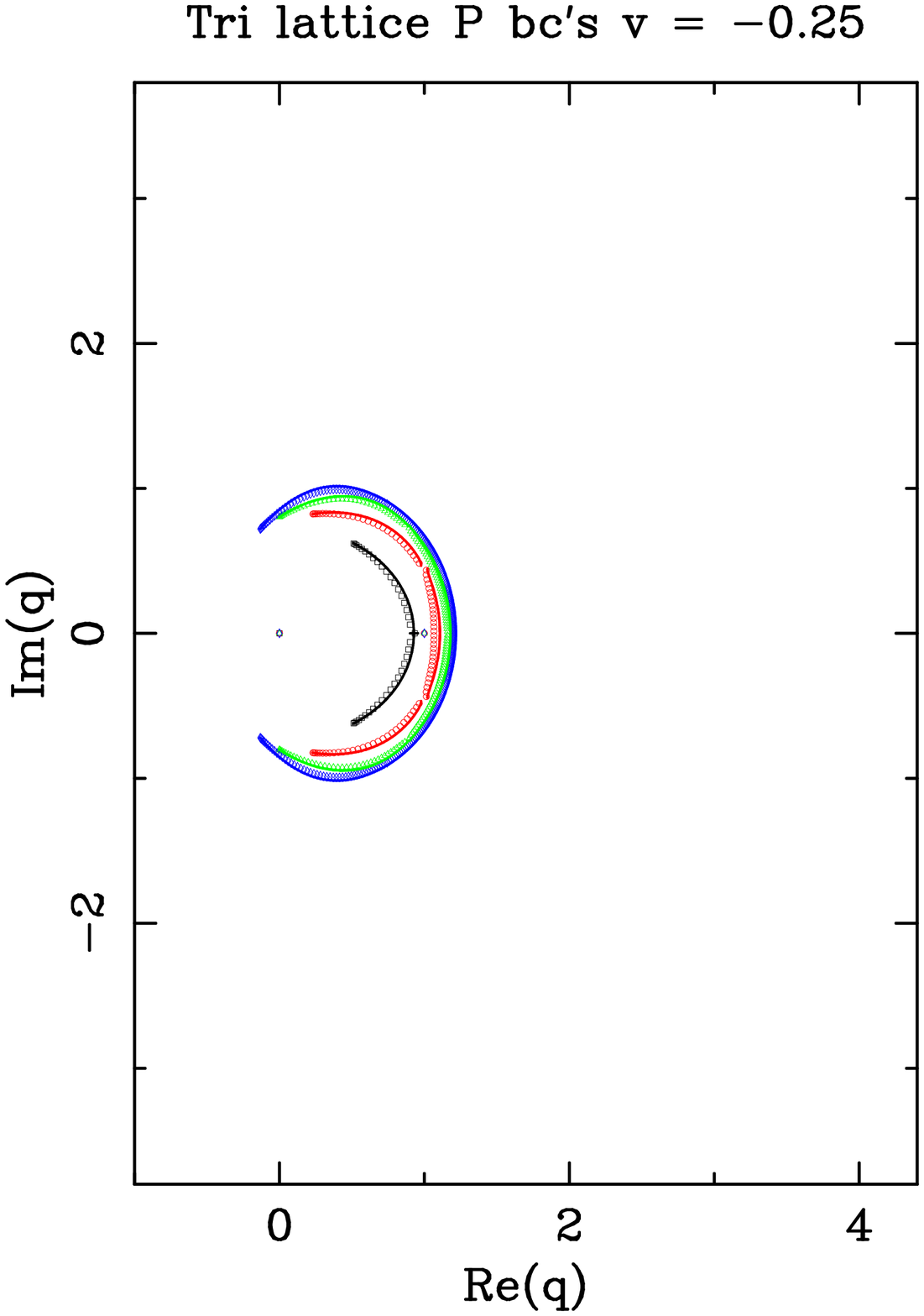} \\[1mm]
   \phantom{(((a)}(c)    & \phantom{(((a)}(d) \\
\end{tabular}
\caption[a]{\protect\label{figures_qplane_P}
Limiting curves forming the singular locus ${\cal B}_q$ for the Potts model
free energy for (a) $v=-1$, (b) $v=-3/4$, (c) $v=-1/2$, and  (d) $v=-1/4$
on strips with cylindrical boundary conditions and
several widths $L$: 2 (black), 3 (red), 4 (green), and 5 (blue).
We also show the partition-function zeros for the strips
$L_{\rm P} \times (10 L)_{\rm F}$ for the same values of $L$.
The symbols are as in Figure~\protect\ref{figures_qplane_F}.
Where the results for
$2 \le L \le 4$ overlap with those in \cite{ta,ks} and for $v=-1$ 
(see also \cite{strip2,t,transfer3} for $3 \le L \le 5$), they agree and are
included here for comparison.
}
\end{figure}

The corresponding partition-function zeros and accumulation sets
for triangular-lattice strips with cylindrical boundary conditions
are shown in Figure~\ref{figures_qplane_P}. Again, 
Figure~\ref{figures_qplane_P}(a)--(d) displays the zeros for $v=-1$ (a),
$v=-0.75$ (b), $v=-0.5$ (c), and $v=-0.25$ (d).  The cases $L=3,4$ were 
studied previously in (Figs. 3.19, 3.25 of) \cite{ks} for arbitrary
temperature, and our zeros are in agreement with this work.

The case $v=-1$, which is the zero-temperature Potts antiferromagnet (chromatic
polynomial) has been previously studied in
\cite{strip,strip2,t,ss00,transfer3} for the free longitudinal boundary
conditions and in \cite{wcyl,wcy,t,tor4,cf} for periodic longitudinal
boundary conditions. For the case of free longitudinal boundary conditions
\cite{transfer3} contains results for $L\leq 9_{\rm F}$ and $L\leq 12_{\rm
P}$. Our Figures~\ref{figures_qplane_F}(a) and~\ref{figures_qplane_P}(a)
include calculations up to $L=5$ for comparison with other values of $v$.
Although some curves, such as those for cylindrical boundary conditions, may
enclose regions, the curves do not enclose regions containing the origin.  One
observes that for either type of transverse boundary condition, as the width
$L$ increases, the left-hand arc endpoints move slowly toward the origin.  When
this was observed in earlier work for several different lattice strips,
\cite{strip}, it motivated the suggestion that in the limit $L \to \infty$ for
strips with free longitudinal boundary conditions, the limiting ${\cal B}_q$
would separate the $q$ plane into regions including a curve passing through
$q=0$ \cite{strip,bcc}.  The specific calculation of ${\cal B}_q(v=-1)$ in the
limit $L \to \infty$ reported by Baxter \cite{baxter_87} has this feature.
(For critical comments on certain features of Baxter's results, see the
detailed discussion in \cite{transfer3}.)  The property that ${\cal B}_q$
separates the $q$ plane into regions with one of the curves on ${\cal B}_q$
passing through the origin is also observed for lattice strips with finite
width $L$ if one imposes periodic longitudinal boundary conditions
\cite{w,pg,wcyl,wcy,pm,tk,t,tor4,s4}.

In making inferences about possible $L \to \infty$ characteristics of the
continuous accumulation set of zeros ${\cal B}(G_s, L \times \infty)$ in the
$q$ or $v$ plane for infinite-length, width-$L$ lattice strip graphs of type
$G_s$, one should recall that, in general, $\lim_{L \to \infty} {\cal B}(G_s,L
\times \infty)$ is different from the continuous accumulation set of the zeros
of the partition function for the usual 2D thermodynamic limit defined by
starting with an $L_y \times L_x$ section of a regular lattice and letting
$L_x$ and $L_y$ both approach infinity with $L_y/L_x$ a finite nonzero
constant.  This type of noncommutativity was encountered in previous studies of
${\cal B}$ for the Potts model free energy on infinite-length, finite-width
strips with periodic longitudinal boundary conditions \cite{a,ta,s3a,hca}; for
these strips, ${\cal B}$ is noncompact in the $v$ plane, reflecting the fact
that the Potts model has a ferromagnetic critical point only at $T=0$
(i.e. $K=\infty$, hence $v=\infty$) for any width $L$, no matter how great,
whereas for the 2D lattice defined in the thermodynamic limit, it has a
ferromagnetic critical point at a finite temperature, so ${\cal B}$ is compact
in the $v$ plane.  Noncommutativity of this type was also found in studies of
${\cal B}_q$.  For example, in calculations of ${\cal B}_q$ for infinite-length
strips of the triangular lattice with cyclic boundary conditions, it was found
that this locus always passes through $q=2$ \cite{wcyl,wcy,t,tor4}, whereas, in
contrast, the locus found in \cite{baxter_87} for the infinite-width limit of
strips with cylindrical boundary conditions does not pass through $q=2$.
Similarly, in calculations of ${\cal B}_q$ for infinite-length strips of the
square lattice with cyclic boundary conditions, it was found that this locus
always passes through $q=2$, whereas in calculations of ${\cal B}_q$ for
infinite-length strips of the square lattice with cylindrical boundary
conditions \cite{strip2,ss00,s4,transfer2}, it was found that ${\cal B}$ does
not pass through $q=2$, strongly suggesting that this difference will persist
in the limit $L_y \to \infty$.

At nonzero temperature for the antiferromagnet, as represented in our figures
by the range $-0.75 \leq v \leq -0.25$, the partition-function zeros in the
$q$-plane have a different shape: as $L$ increases, the limiting curves ${\cal
B}_q(L,v)$ tend to a bean-shaped curve or set of arcs, open on the left,
without substantial protruding branches, in contrast to many of the $v=-1$
curves.  For a given value of $v$ in the range considered, as the width $L$
increases, the curve envelope moves outward somewhat and the arc endpoints on
the left move slowly toward $q=0$.  This behavior is consistent with the
hypotheses that for a given $v$, as $L \to \infty$, (i) ${\cal B}_q$ would
approach a limiting locus as $L \to \infty$ and (ii) this locus would separate
the $q$ plane into different regions, with a curve passing through $q=0$ as
well as a maximal real value, $q_c(v)$.  This is qualitatively the same type of
behavior that was found earlier for the square-lattice strips
\cite{a,s3a,ts}. In particular, our results are consistent with the inference
that as $L\to\infty$, ${\cal B}_q$ for $v=-1$ would pass through $q_c(v=-1)=4$,
corresponding to the property that the $q=4$ Potts antiferromagnet has a
zero-temperature critical point on the (infinite) triangular lattice
\cite{baxter_70}. 

For a given $L$, as $v$ increases from $-1$ to 0, i.e., as the inverse
temperature $\beta$ decreases from infinity to 0 for the antiferromagnet, the
zeros and the limiting curve contract to a point at $q=0$.  This is an
elementary consequence of the fact that these lattice strips have fixed maximal
vertex degree and as the parameter $K=\beta J$ approaches zero, the spin-spin
interaction term in ${\cal H}$, eq.~(\ref{ham}), vanishes, so that the sum over
states just counts all $q$ possible spin states independently at each vertex,
and $Z(G,q,v)$ approaches the value $Z(G,q,0)=q^n$.  An upper bound on the
magnitudes of the zeros is given by the following theorem by Sokal:

\begin{theorem} (Sokal \cite{Sokal_01}) \label{theorem_sokal}
Let $G=\{V,E\}$ be a loopless finite undirected graph of maximum degree $\leq
r$, equipped with complex edge weights $\{v_e\}_{e\in E}$ satisfying
$|1+v_e|\leq 1$ for all $e$. Let $|v|_{\rm max}=\max_{e\in E} |v_e|$.  Then all
the zeros of $Z(G,q,\{v_e\})$ lie in the disc $|q|< C(r) v_{\rm max}$ with
$C(r) \leq 7.963907 r$.
\end{theorem} 

This is a loose bound; for all of the strips with cylindrical
boundary conditions and for the free strips with widths $L \ge 3$, the maximal
degree is $r=6$, so that the above theorem implies that $|q| \lsim 
47.8 |v|_{\rm max}$.  Thus, for example, for $v=-1$, this reads 
$|q| \lsim 47.8$, whereas in fact $|q| \lsim 4$ for free boundary conditions
and $|q| \lsim 4.5$ for cylindrical boundary conditions.  A general feature is
that the limiting curves and associated zeros tend to be located mostly in the
$\real(q) \ge $ half plane. 

As noted, it is evident in Figures~\ref{figures_qplane_F}
and~\ref{figures_qplane_P} that as $L$ increases, the accumulation set ${\cal
B}_q(L,v)$ moves outwards. As expected, the convergence to the limit
$L\to\infty$ seems to be faster with cylindrical boundary conditions, as there
are no surface effects when the length is made infinite.  

One can also plot ${\cal B}_q$ for the ferromagnetic region $0 \le v \le
\infty$ (e.g., Figs. 2 and 13 of \cite{ta}).  Although we have not included
these plots here, we note that an elementary Peierls argument shows that the
$q$-state Potts ferromagnet on infinite-length, finite-width strips has no
finite-temperature phase transition and associated magnetic long range order.
Hence, for this model ${\cal B}_q$ does not cross the positive real $q$ axis
for $0 \le v < \infty$.

%
%
\section{Partition Function Zeros in the $\lowercase{v}$ Plane}

\subsection{General}

In this section we shall present results for zeros and continuous accumulation
sets ${\cal B}_v$ (in the $v$-plane) for the partition function of the Potts
antiferromagnet on triangular-lattice strips of widths $L\leq 5$ and free or
cylindrical boundary conditions.  Our results hold for arbitrarily great length
$L$ and for any real or complex value of $q$; they thus complement calculations
of the Potts model partition function for fixed positive integer values of $q$
on sections of the triangular lattice \cite{mm,hcl,p,p2}.  We shall focus here
on integer values of $q$, since these are the most relevant from a physical
point of view.  We recall the possible noncommutativity in the definition of
the free energy for certain integer values of $q$ (see eqs.~(2.10), (2.11) of
\cite{a} or (1.17) of \cite{ta}):
$\lim_{n \to \infty} \lim_{q \to q_s} Z(G,q,v)^{1/n} \ne \lim_{q \to q_s} 
\lim_{n \to \infty} Z(G,q,v)^{1/n}$
As discussed in \cite{a}, because of this noncommutativity, the formal
definition (\ref{ef}) is, in general, insufficient to define the free energy
$f$ at these special points $q_s$; it is necessary to specify the order of the
limits that one uses in the above equation. We denote the two definitions
using different orders of limits as $f_{qn}$ and $f_{nq}$:
$f_{nq}(\{G\},q,v) = \lim_{n \to \infty} \lim_{q \to q_s} n^{-1} \ln Z(G,q,v)$
and 
$f_{qn}(\{G\},q,v) = \lim_{q \to q_s} \lim_{n \to \infty} n^{-1} \ln Z(G,q,v)$.
 
As a consequence of this noncommutativity, it follows that for the special set
of points $q=q_s$ one must distinguish between (i) $({\cal
B}_v(\{G\},q_s))_{nq}$, the continuous accumulation set of the zeros of
$Z(G,q,v)$ obtained by first setting $q=q_s$ and then taking $n \to \infty$,
and (ii) $({\cal B}_v(\{G\},q_s))_{qn}$, the continuous accumulation set of the
zeros of $Z(G,q,v)$ obtained by first taking $n \to \infty$, and then taking $q
\to q_s$.  For these special points (cf.~eq.~(2.12) of \cite{a}),
\beq
({\cal B}_v(\{G\},q_s))_{nq} \ne ({\cal B}_v(\{G\},q_s))_{qn} 
\label{bnoncom}
\eeq
Here this noncommutativity will be relevant for $q=0$ and $q=1$. 

In Figures~\ref{zeros_tri_allF_q=0}--\ref{zeros_tri_allF_q=4} we show the
partition-functions zeros in the $v$-plane (for a fixed value of $q$) for
strips of sizes $L_{\rm F} \times (10L)_{\rm F}$ with $2\leq L \leq 5$ and free
boundary conditions.  We also show the corresponding limiting curves ${\cal
B}_v(L,q)$ for the limit of infinite strip length. For simplicity, we have
displayed each value of $L$ on a different plot: $L=2$ (a), $L=3$ (b), $L=4$
(c), and $L=5$ (d). The corresponding partition-function zeros and accumulation
sets for triangular-lattice strips with cylindrical boundary conditions are
shown in Figures~\ref{zeros_tri_allP_q=0}--\ref{zeros_tri_allP_q=4} with the
same notation as for the former figures.  Complex-temperature phase diagrams
and associated partition function zeros were given in \cite{ta} for $L=2$ for
free and periodic longitudinal boundary conditions and free transverse boundary
conditions.  Results for $L=4_{\rm F}$ and $L=4_{\rm P}$ at zero and finite
temperatures were given before in \cite{ks}.  Our present calculations are in
agreement with, and extend, this previous work.

On the infinite triangular lattice (defined via the 2D thermodynamic limit as
given above), the phase transition point separating the paramagnetic (PM) and
ferromagnetic (FM) phases is determined as the (unique) real positive solution
of the equation \cite{kimjoseph}
\beq
v^3+3v^2-q=0
\label{tri_eq}
\eeq
In previous studies such as \cite{a,ta}, it has been found that although
infinite-length, finite-width strips are quasi-one-dimensional systems and
hence the Potts model has no physical finite-temperature transition on such
systems, some aspects of the complex-temperature phase diagram have close
connections with those on the (infinite) triangular lattice.  We shall discuss
some of these connections below.  

\subsection{$q=0$} 

 From the cluster representation of $Z(G,q,v)$, eq.~(\ref{cluster}), it follows
that this partition function has an overall factor of $q^{k(G)}$, where $k(G)$
denotes the number of components of $G$, i.e., an overall factor of $q$ for a
connected graph.  Hence, $Z(G,q=0,v)=0$.  In the transfer matrix formalism,
this is evident from the overall factor of $q$ coming from the vector 
$\w$. However, if we first take the limit $n \to \infty$ to define ${\cal B}$
for $q \ne 0$ and then let $q \to 0$ or, equivalently, extract the factor $q$
from the left vector $\w$, we obtain a nontrivial locus, namely 
$({\cal B}_v(\{G\},0)_{qn}$.  This is a consequence of the noncommutativity
(\ref{bnoncom}) for $q=0$. 

\begin{figure}[hbtp]
\centering
\begin{tabular}{cc}
   \includegraphics[width=200pt]{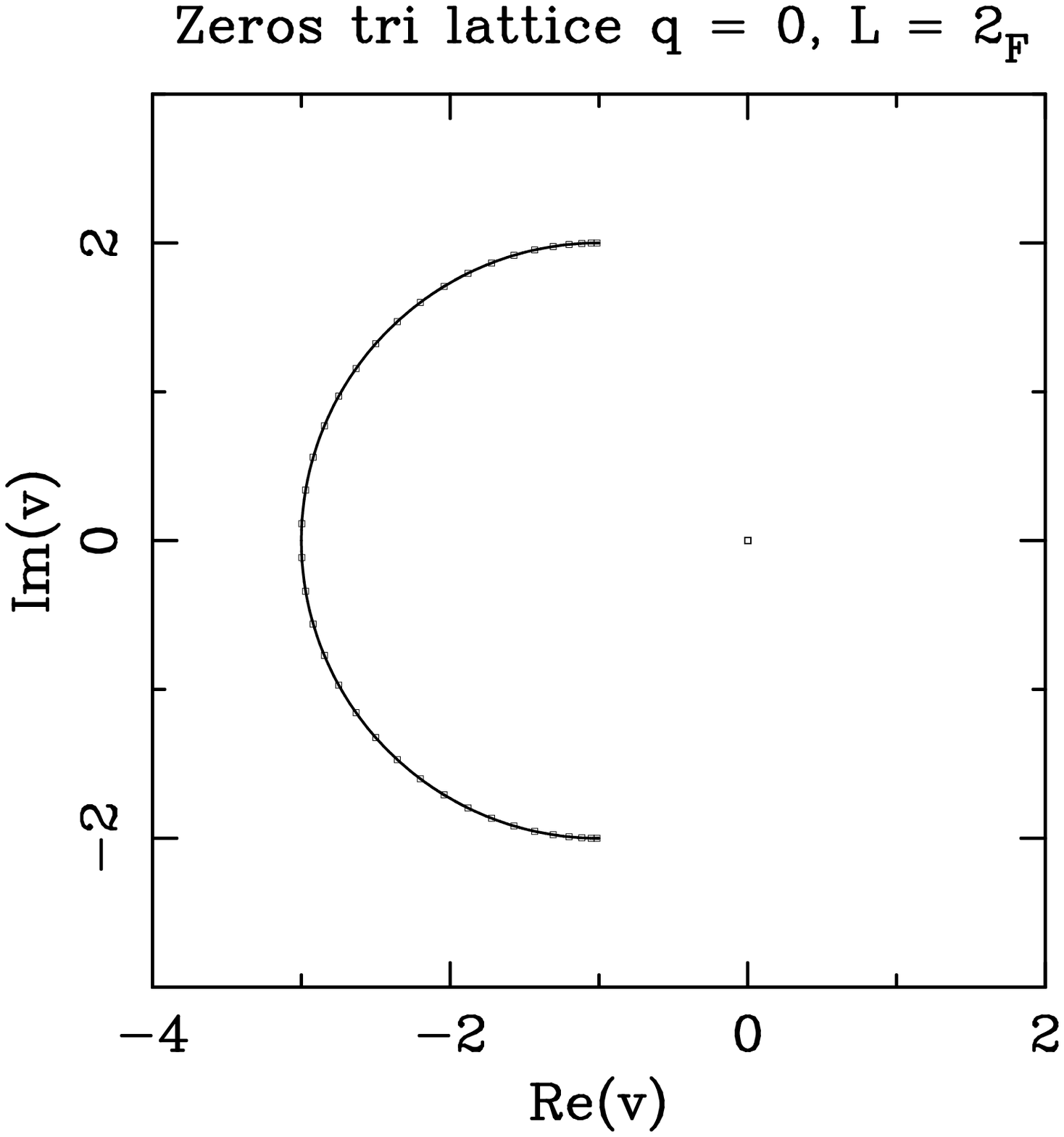} &
   \includegraphics[width=200pt]{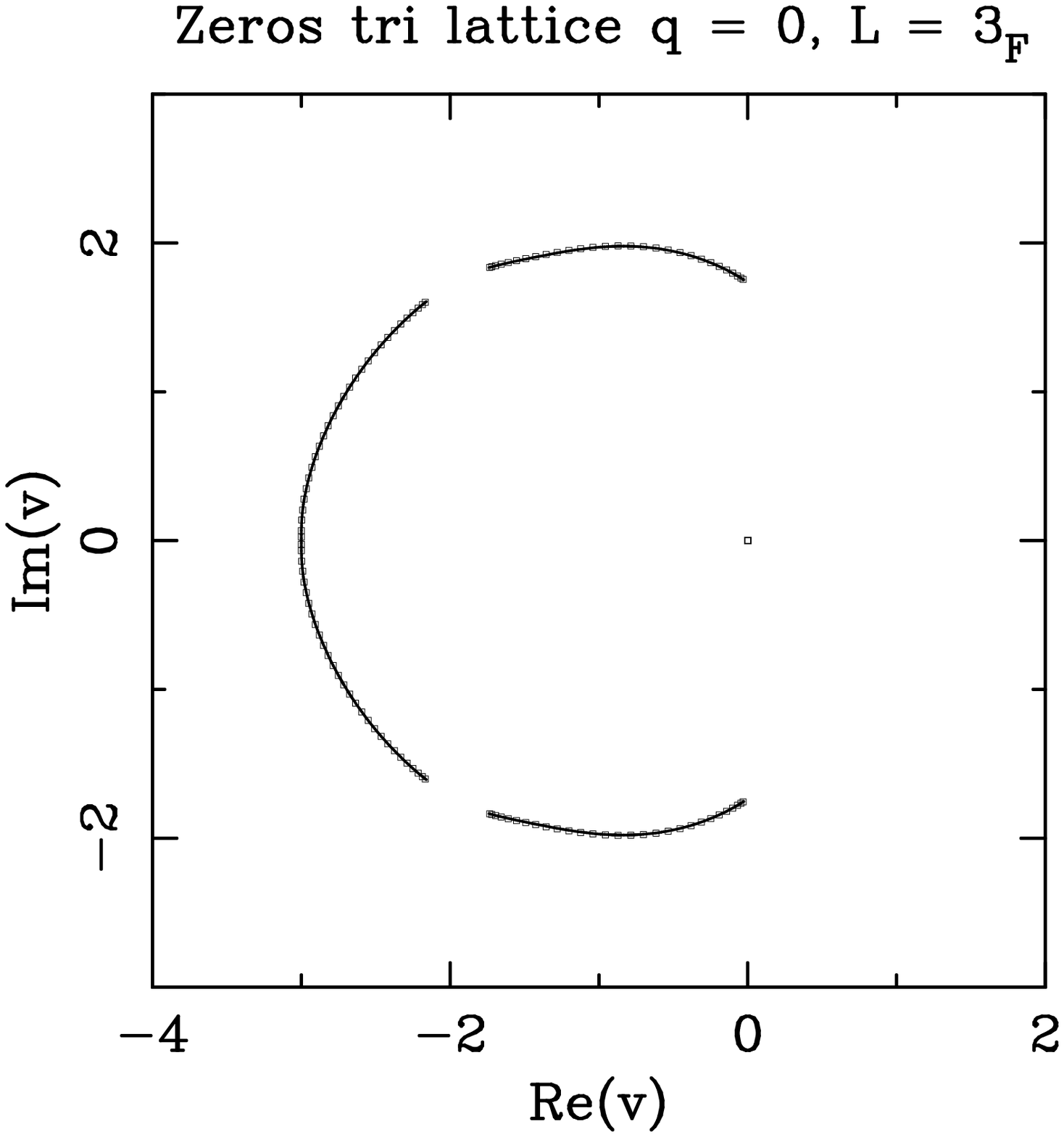} \\[1mm]
   \phantom{(((a)}(a)    & \phantom{(((a)}(b) \\[5mm]
   \includegraphics[width=200pt]{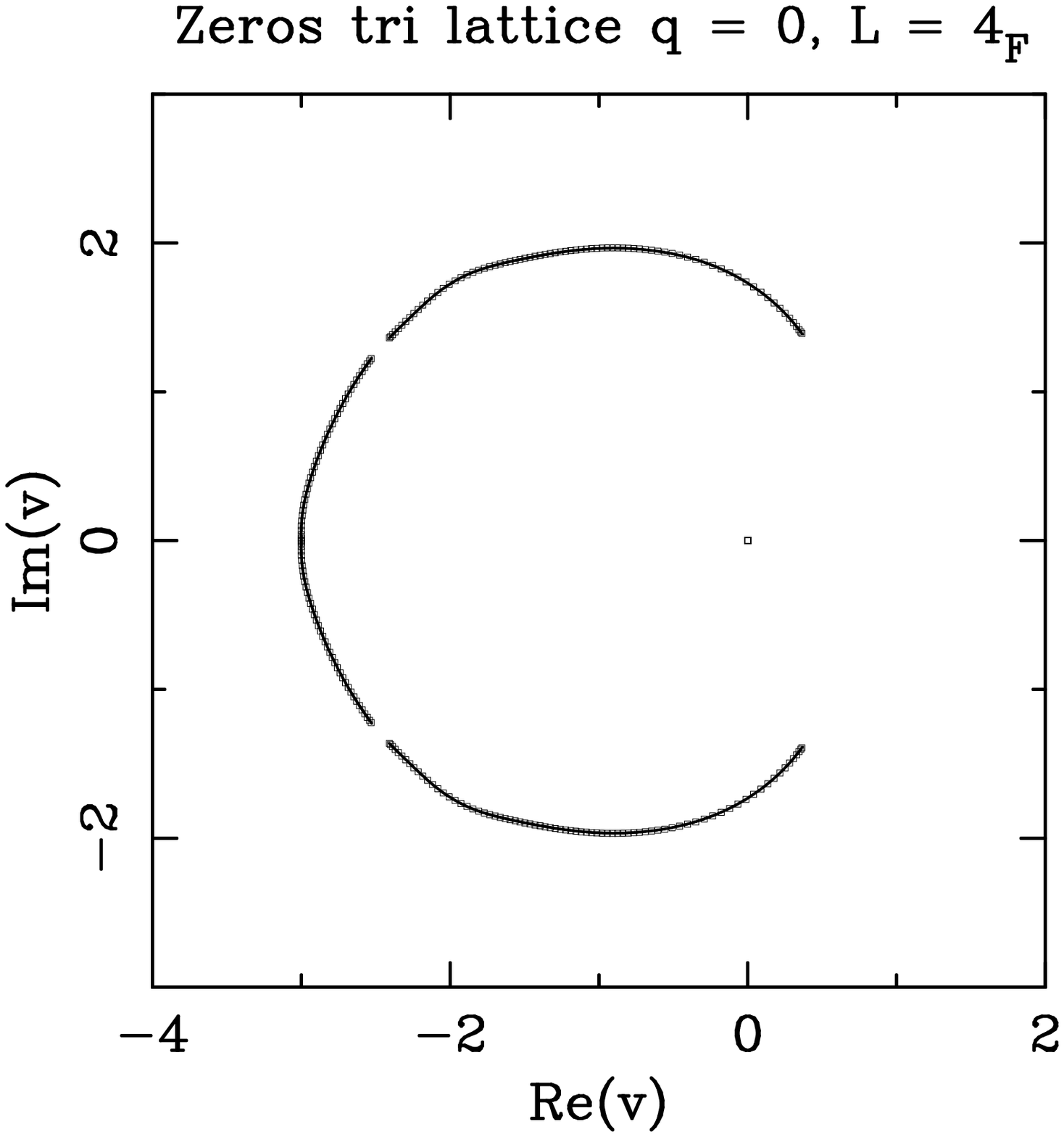} &
   \includegraphics[width=200pt]{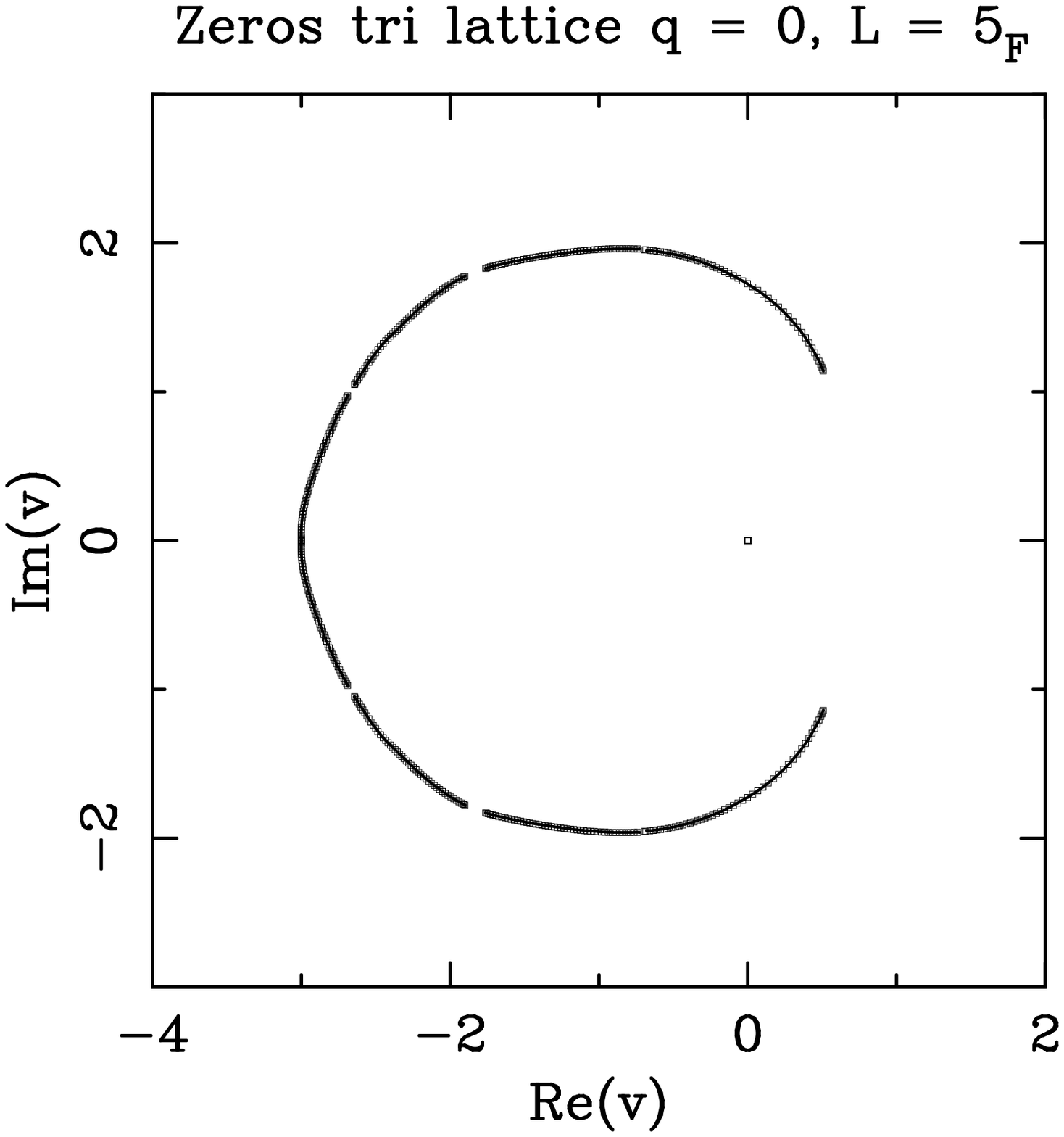} \\[1mm]
   \phantom{(((a)}(c)    & \phantom{(((a)}(d) \\
\end{tabular}
\caption[a]{\protect\label{zeros_tri_allF_q=0} Limiting curves forming the
singular locus ${\cal B}$, in the $v$ plane, for the free energy, defined with
the order $f_{qn}$, of the Potts model for $q=0$ on the $L_{\rm F} \times
\infty_{\rm F}$ triangular-lattice strips with (a) $L=2$, (b) $L=3$, (c) $L=4$,
and (d) $L=5$.  We also show the zeros of $Z(G,0,v)/q$ corresponding to the
strips $L_{\rm F} \times (10L)_{\rm F}$ for each value of $L$.  }
\end{figure}

\begin{figure}[hbtp]
\centering
\begin{tabular}{cc}
   \includegraphics[width=200pt]{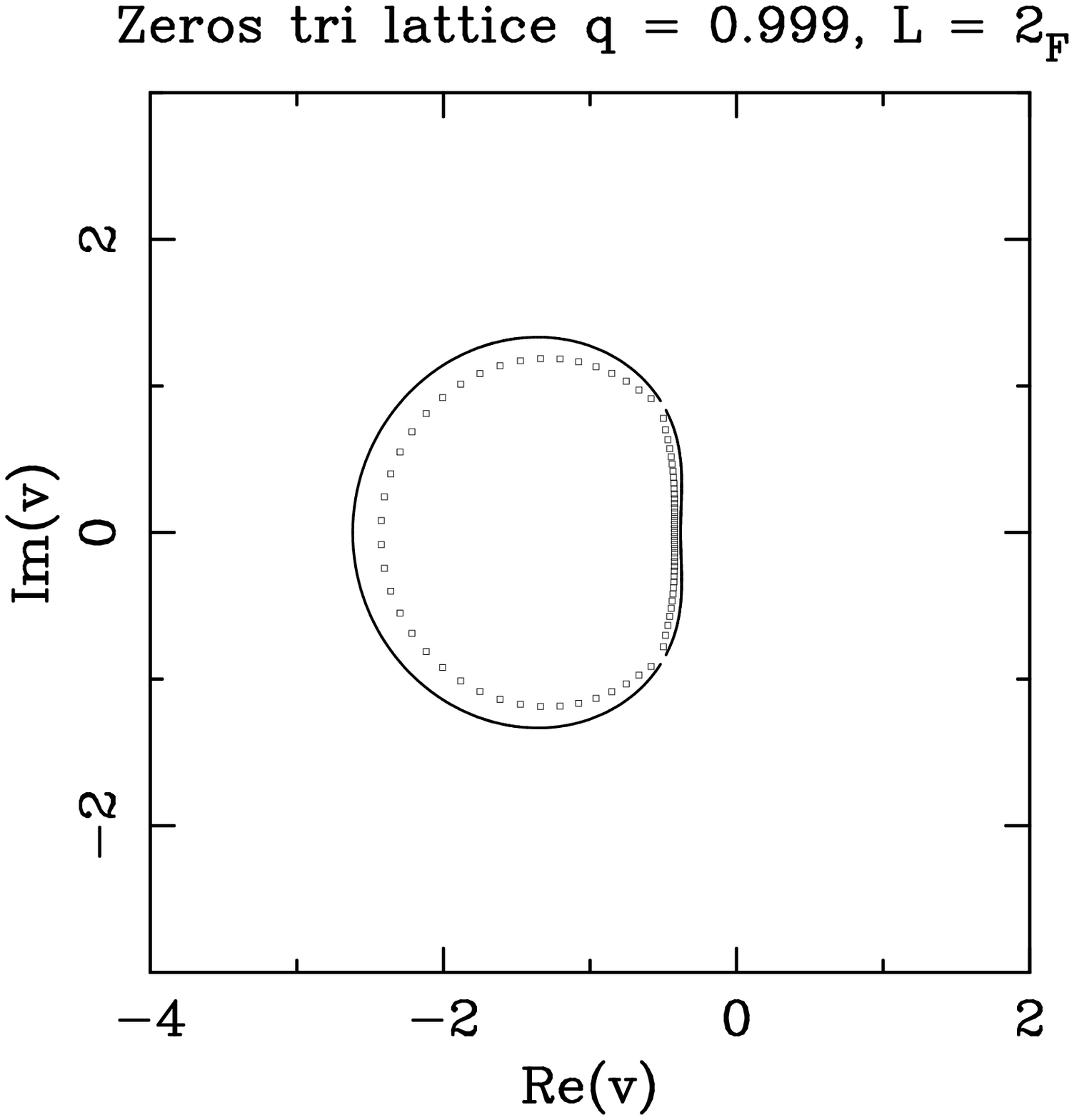} &
   \includegraphics[width=200pt]{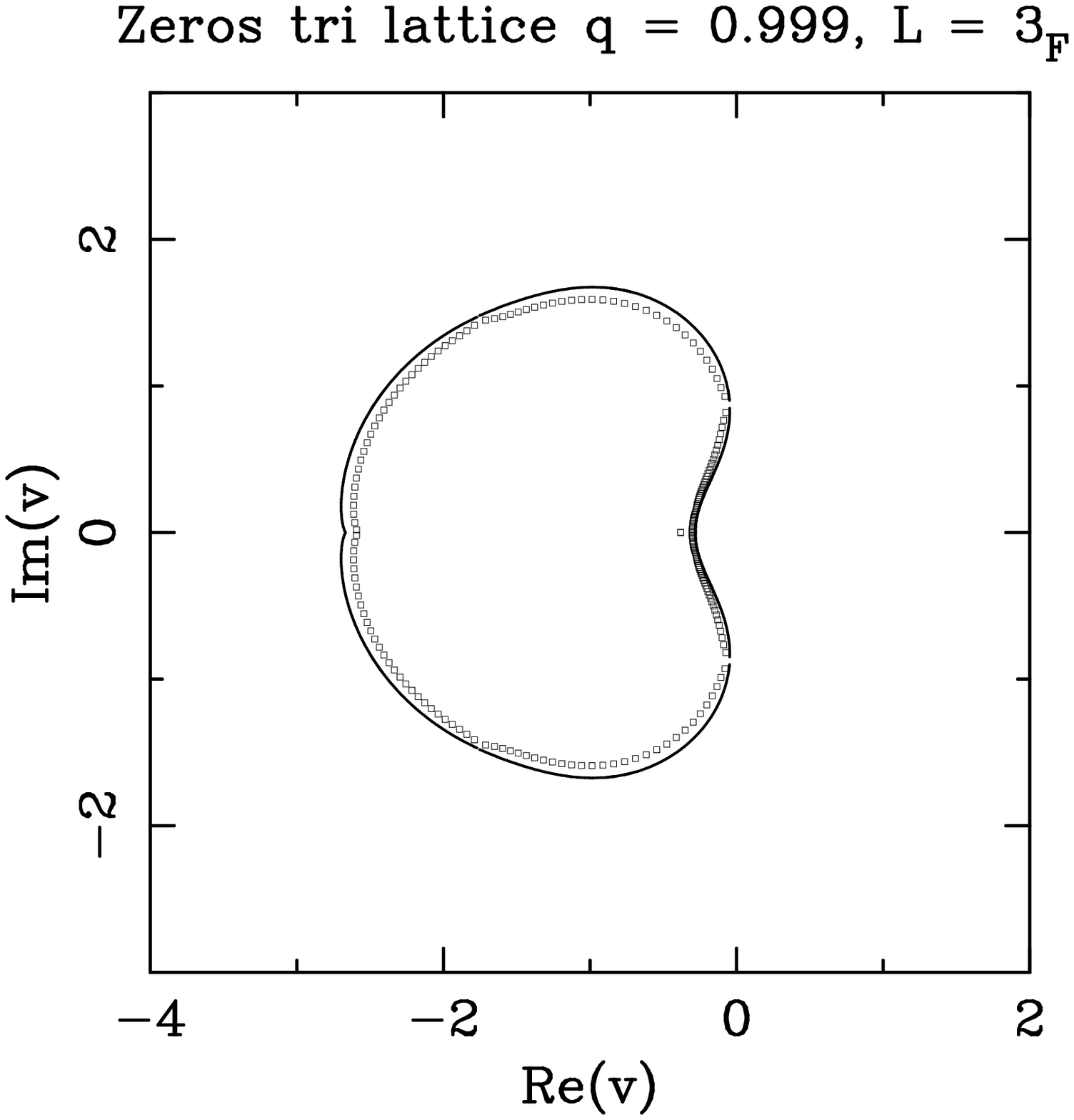} \\[1mm]
   \phantom{(((a)}(a)    & \phantom{(((a)}(b) \\[5mm]
   \includegraphics[width=200pt]{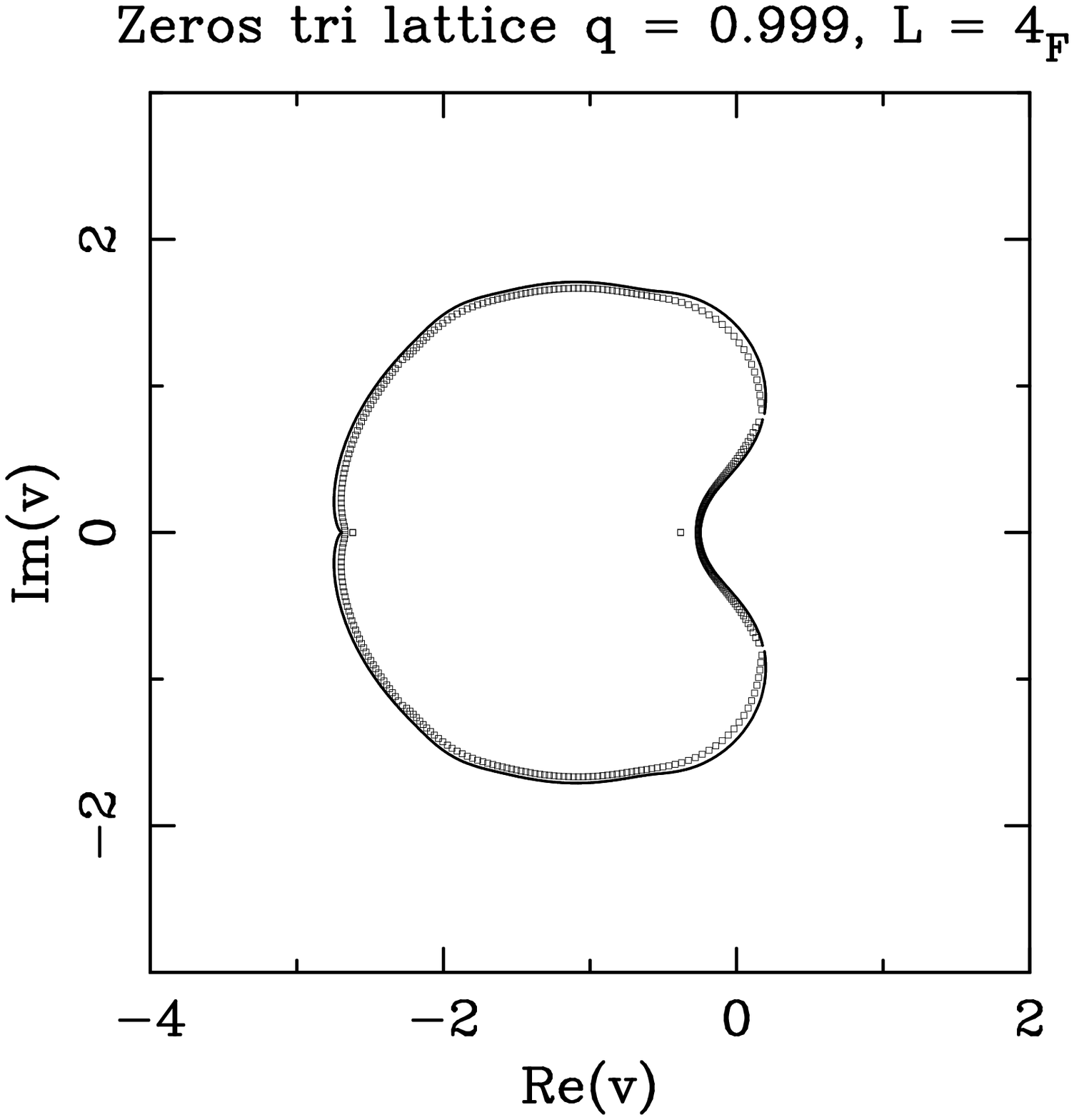} &
   \includegraphics[width=200pt]{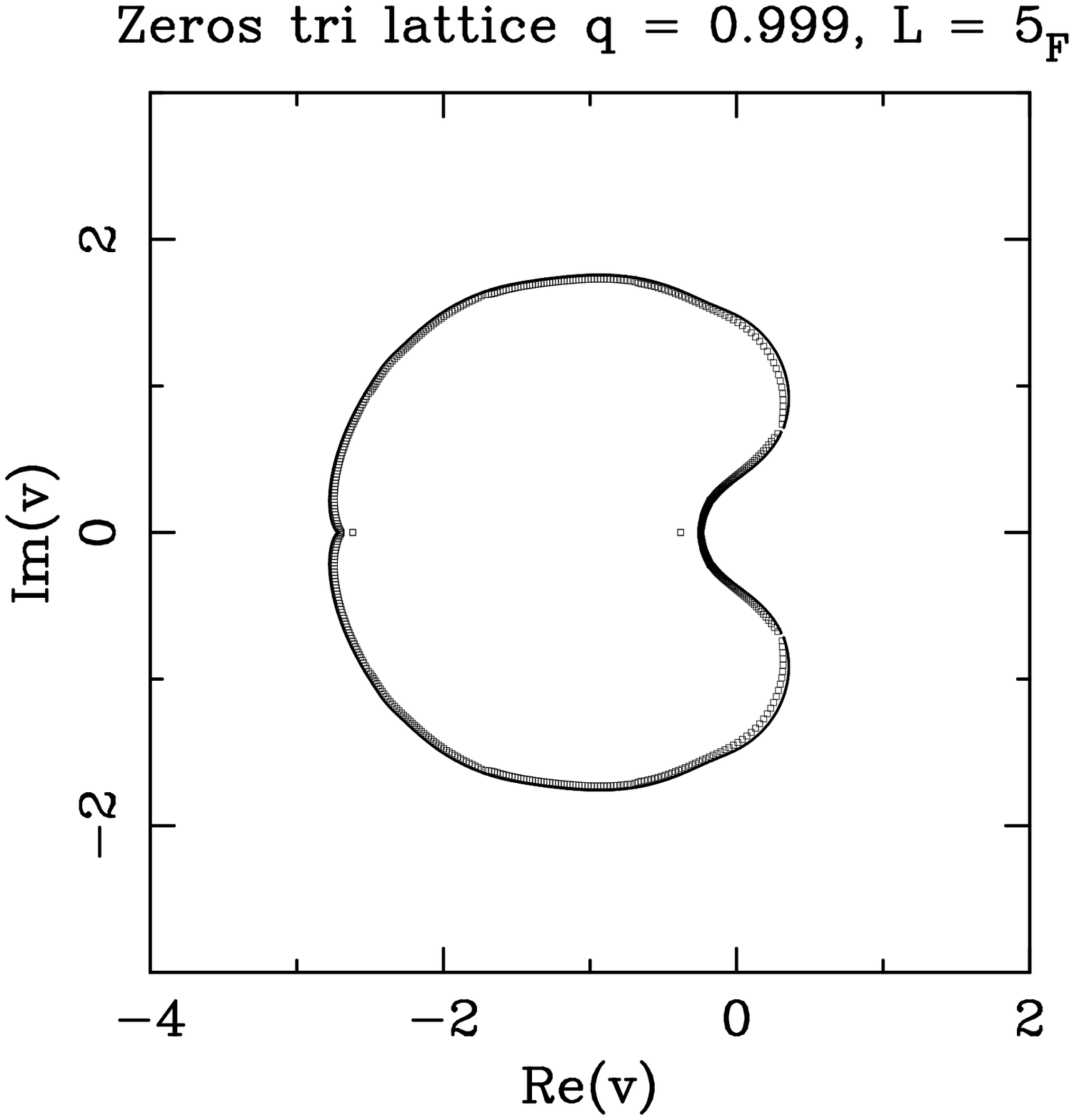} \\[1mm]
   \phantom{(((a)}(c)    & \phantom{(((a)}(d) \\
\end{tabular}
\caption[a]{\protect\label{zeros_tri_allF_q=1}
Limiting curves forming the singular locus ${\cal B}$, in the $v$ plane, for
the free energy of the Potts model for $q=0.999$ on the
$L_{\rm F} \times \infty_{\rm F}$ triangular-lattice strips with
(a) $L=2$, (b) $L=3$, (c) $L=4$, and (d) $L=5$.  This are essentially
equivalent to the limiting curves for $f_{qn}$ at $q=1$.  
We also show the partition-function zeros corresponding to the strips
$L_{\rm F} \times (10L)_{\rm F}$ for each value of $L$.
}
\end{figure}

With the second order of limits or the equivalent removal of the factor of $q$
in $Z$, we obtain the locus ${\cal B}_v(q=0)$ shown in
Figures~\ref{zeros_tri_allF_q=0} (free boundary conditions)
and~\ref{zeros_tri_allP_q=0} (cylindrical boundary conditions).  The
accumulation set ${\cal B}_v(L,q=0)$ seems to converge to a roughly circular
curve. We see in Figures~\ref{zeros_tri_allF_q=0} and~\ref{zeros_tri_allP_q=0}
that the limiting curves cross the real $v$-axis at $v\approx -3$.  We note the
interesting feature that this is a root of eq.~(\ref{tri_eq}) for $q=0$.  For
the case of cylindrical boundary conditions, ${\cal B}_v(q=0)$ includes a small
line segment on the real axis near $v=-3$, with a length that decreases as $L$
increases.  As $L$ increases, the arc endpoints on the upper and lower right
move toward the real axis.  It is possible that these could pinch this axis at
$v=0$ as $L \to \infty$, corresponding to the other root of (\ref{tri_eq}) for
$q=0$.

\subsection{$q=1$} 
 
For $q=1$, the spin-spin interaction in (\ref{ham}) always has the Kronecker
delta function equal to unity, and hence the Potts model partition function 
is trivially given by 
\be 
Z(G,q=1,v) = e^{K|E|} = (1+v)^{|E|} 
\label{Z_q=1}
\ee
where $|E|$ is the number of edges in the graph $G$.  This has a single zero at
$v=-1$.  But again, one encounters the noncommutativity (\ref{bnoncom}) for
$q=1$.  It is interesting to analyze this in terms of the transfer matrix
formalism. At this value of $q$, both the transfer matrix and the left vector 
$\w$ are non-trivial.  There is thus a cancellation of terms that
yields the result \reff{Z_q=1}. The case $L=2_{\rm P}$ is the simplest one to
analyze: the eigenvalues and coefficients for $q=1$ are given by 
\begin{subeqnarray}
\lambda_1(1,v) &=& 2v^2 \ ;   \qquad \quad c_1(1,0) = 0 \\ 
\lambda_2(1,v) &=& (1+v)^6 \ ; \quad       c_2(1,0) = (1+v)^2 
\end{subeqnarray}
Thus, only the second eigenvalue contributes to the partition function, and 
it gives the expected result 
$Z(2_{\rm P}\times m_{\rm F},q=1,v) = (1+v)^{6m-4}$. 
For $L=2_{\rm F}$ we obtain: 
\begin{subeqnarray}
\lambda_1(1,v) &=&  v^2 \ ; \phantom{2}\qquad \quad c_1(1,0) = 0 \\
\lambda_2(1,v) &=& (1+v)^4  \ ;        \quad        c_2(1,0) = (1+v)
\end{subeqnarray}
giving rise to $Z(2_{\rm F}\times m_{\rm F},q=1,v) = (1+v)^{4m-3}$.  The case
$L=3_{\rm P}$ is similar: there is a single eigenvalue $\lambda_1(1,v)=(1+v)^9$
with a non-zero coefficient $c_1(1,v)=(1+v)^3$, and the other two
eigenvalues [which are the roots of $x^2 - xv^2(3v^3+13 v^2 +24v+3) +
v^5(1+v)(6+5v+6v^2)$] have identically zero coefficients 
$c_{2,3}(1,v)=0$. Thus, the partition function takes the form 
$Z(3_{\rm P}\times m_{\rm F},q=1,v) = (1+v)^{9m-6}=(1+v)^{|E|}$.

In general, we conclude that at $q=1$ only the eigenvalue
$\lambda=(1+v)^{3L_{\rm P}}$ (resp.\ $\lambda=(1+v)^{3L_{\rm F}-2}$)
contributes to the partition function for cylindrical (resp.\ free) boundary
conditions, and its coefficient is $c=(1+v)^{L_{\rm P}}$ (resp.\
$c=(1+v)^{L_{\rm F}-1}$).  The other eigenvalues do not contribute as
they have zero coefficients. This is the analogue of what was found for the 
strips with cyclic boundary conditions, where the various $\lambda$'s
fall into sets $\lambda_{G,d,j}$ such that all of the $\lambda$'s with a fixed
$d$ have a unique coefficient which is a polynomial of degree $d$ in $q$ given
by \cite{a,ta,cf} 
\beq
c^{(d)}  =  U_{2d}\Bigl ( \frac{\sqrt{q}}{2}\Bigr ) 
         =  \sum_{j=0}^d (-1)^{j}{2d-j \choose j}q^{d-j}
\label{cd}
\eeq
where $U_\nu(z)$ is the Chebyshev polynomial of the second kind.  These
coefficients vanish at certain values of $q$, which means that if one evaluates
the partition function first at these values and then takes the limit $n \to
\infty$, the corresponding $\lambda$'s will not contribute to $Z$, while if one
takes $n \to \infty$ first, calculates the free energy and the locus ${\cal
B}_{qn}$, and then sets $q$ equal to one of these values, the $\lambda$'s will,
in general contribute.  In particular, we recall (eq.~(2.18) of \cite{cf}) that
if $q=1$, then $c^{(d)}$ vanishes if $d=1$ mod 3.  Thus, we see similar
manifestations of the noncommutativity (\ref{bnoncom}) for strips with free and
periodic boundary conditions.  

In our present case, in order to obtain ${\cal B}_{qn}$, we have computed the
Tutte-polynomial zeros and the corresponding limiting curves for $q=0.999$ (see
Figures~\ref{zeros_tri_allF_q=1} and~\ref{zeros_tri_allP_q=1}).  The
accumulation sets ${\cal B}_v(L,q=0.999)$ for $L = 2_{\rm F}$ to $L = 5_{\rm
F}$ consists of arcs that come close to forming an almost closed bean-shaped
curve, with an involution on the right that deepens as $L$ increases.  Solving
the $q=1$ special case of eq.~(\ref{tri_eq}) yields the roots $v =
-2.879385..$, $v= -0.652703...$, and $v= 0.5320888..$.  The locus ${\cal B}_v $
crosses the real $v$ axis at two points, and our results are consistent with
the inference that as $L \to \infty$, these two crossing points are the first
two roots listed above.  For free boundary conditions with width $L=3,4,5$,
${\cal B}_{qn}$ exhibits a small involution on the left.  A noteworthy feature
of this locus is that it is relatively smooth, without the prongs that tend to
occur for the other values of $q$ discussed here.

\subsection{$q=2$}

\begin{figure}[hbtp]
\centering
\begin{tabular}{cc}
   \includegraphics[width=200pt]{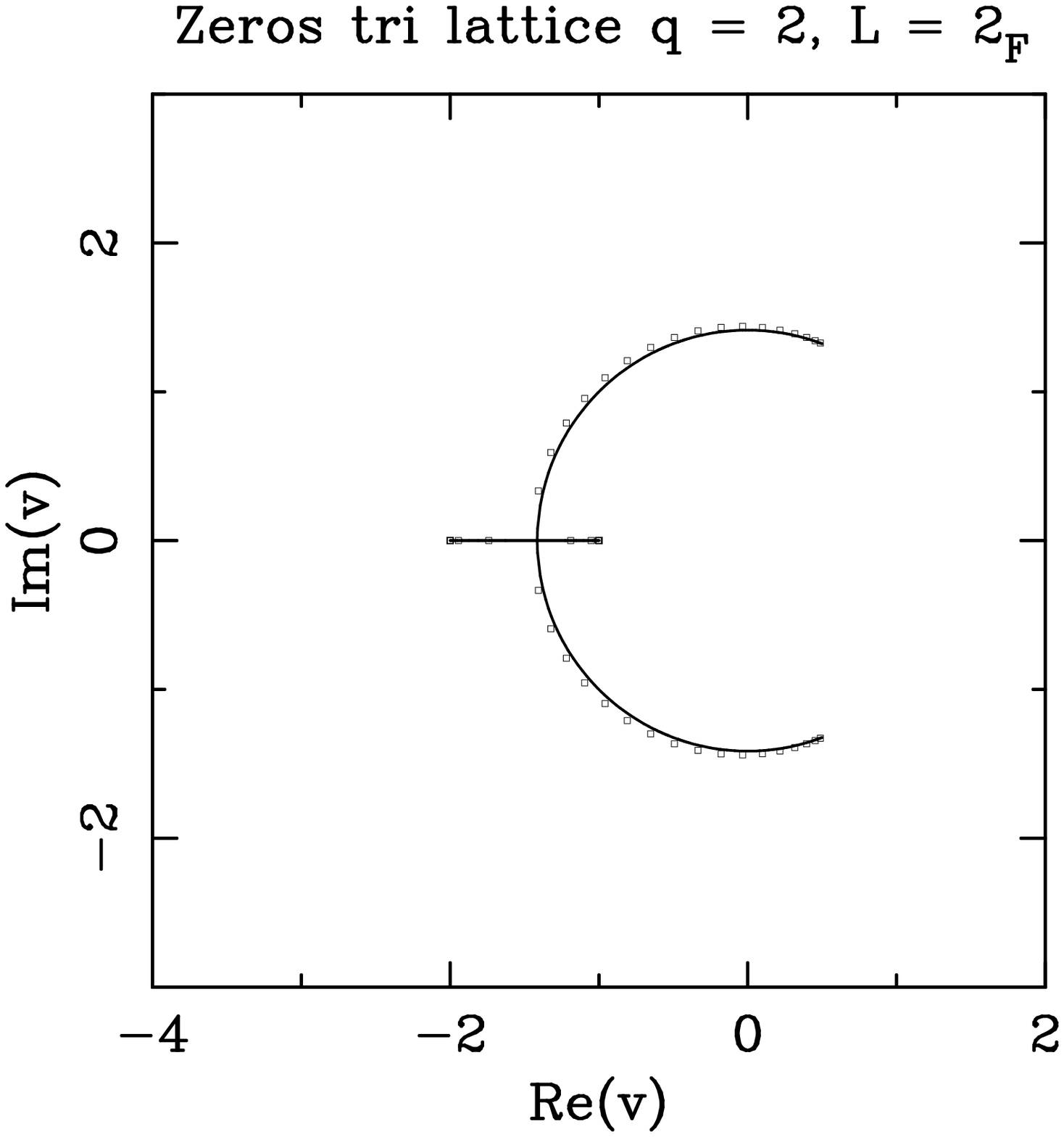} &
   \includegraphics[width=200pt]{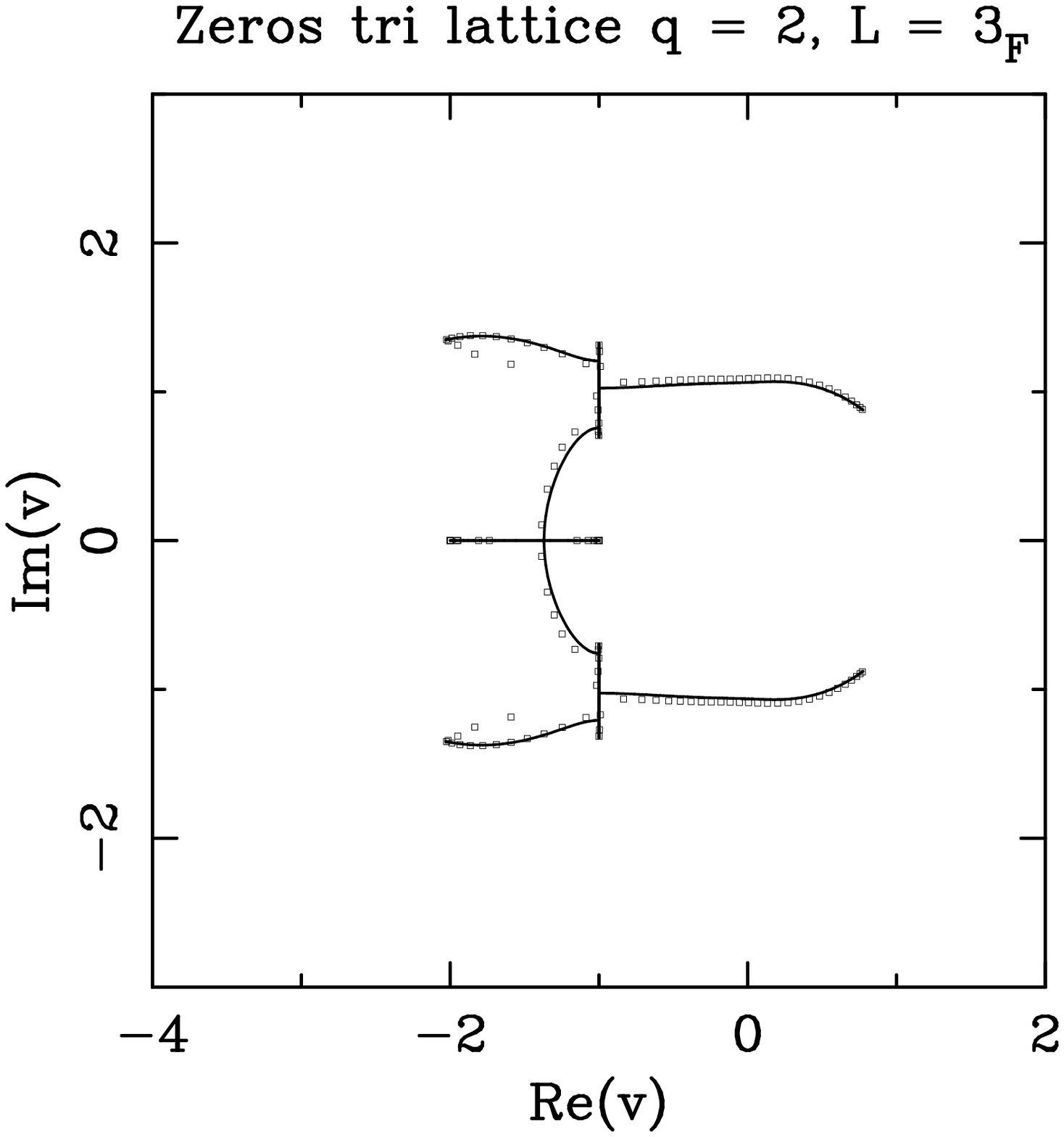} \\[1mm]
   \phantom{(((a)}(a)    & \phantom{(((a)}(b) \\[5mm]
   \includegraphics[width=200pt]{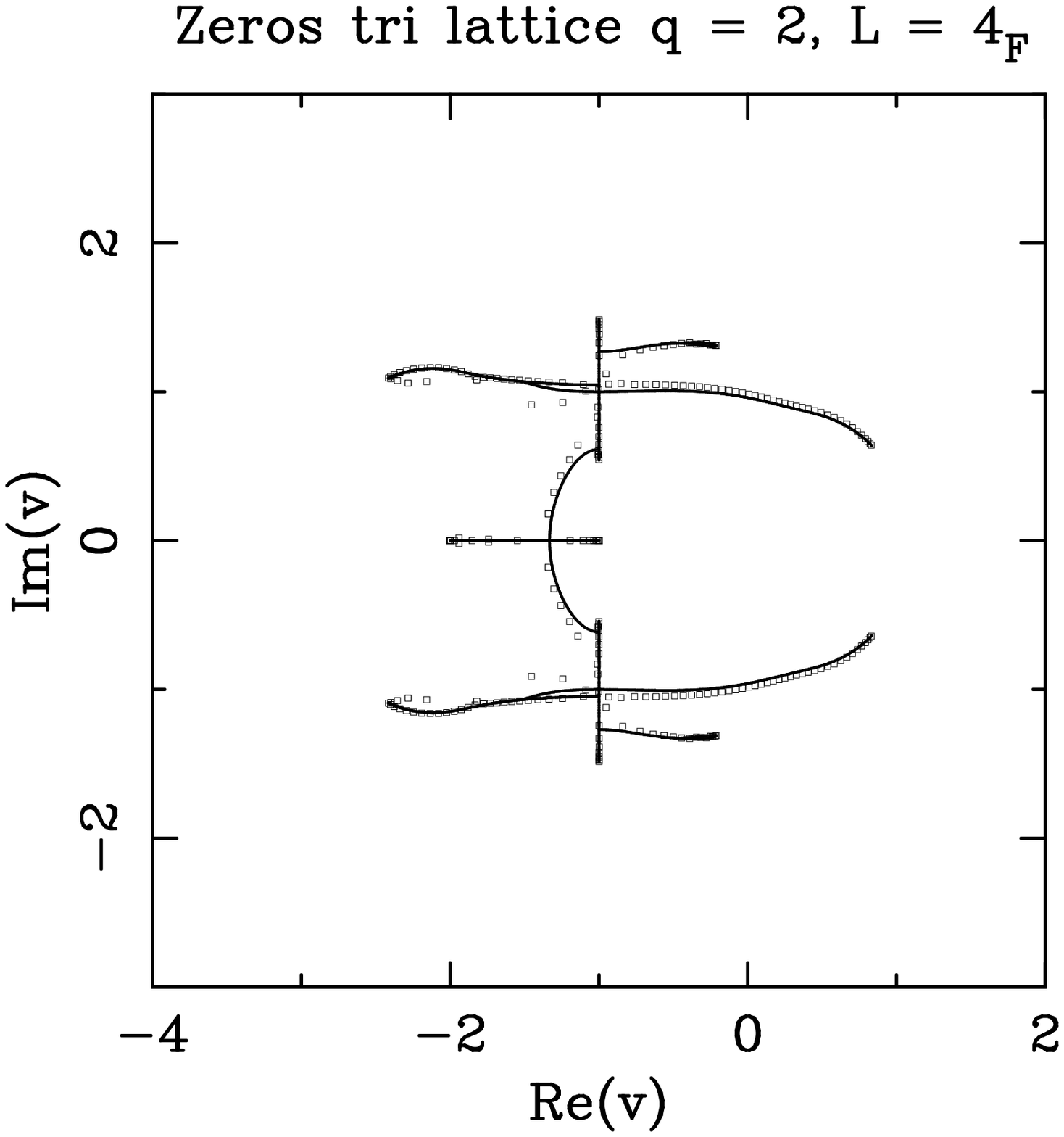} &
   \includegraphics[width=200pt]{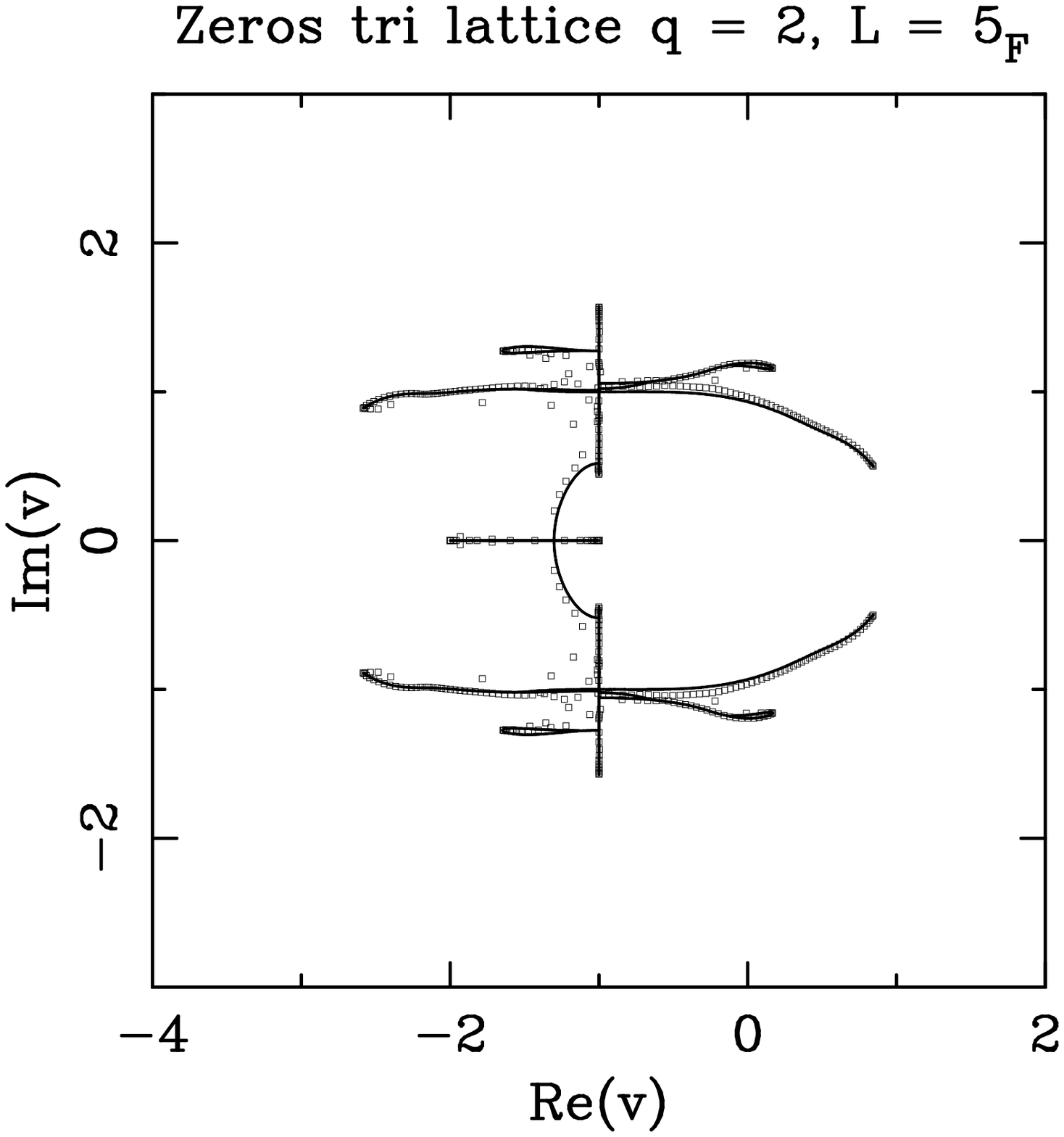} \\[1mm]
   \phantom{(((a)}(c)    & \phantom{(((a)}(d) \\
\end{tabular}
\caption[a]{\protect\label{zeros_tri_allF_q=2} Limiting curves forming the
singular locus ${\cal B}$, in the $v$ plane, for the free energy of the Potts
model for $q=2$ on the $L_{\rm F} \times \infty_{\rm F}$ triangular-lattice
strips with (a) $L=2$, (b) $L=3$, (c) $L=4$, and (d) $L=5$.  We also show the
partition-function zeros corresponding to the strips $L_{\rm F} \times
(10L)_{\rm F}$ for each value of $L$.  Where the results for $2 \le L \le 4$
overlap with those in \cite{ta,ks}, they agree and are included here for
comparison.  }
\end{figure}

\begin{figure}[hbtp]
\centering
\begin{tabular}{cc}
   \includegraphics[width=200pt]{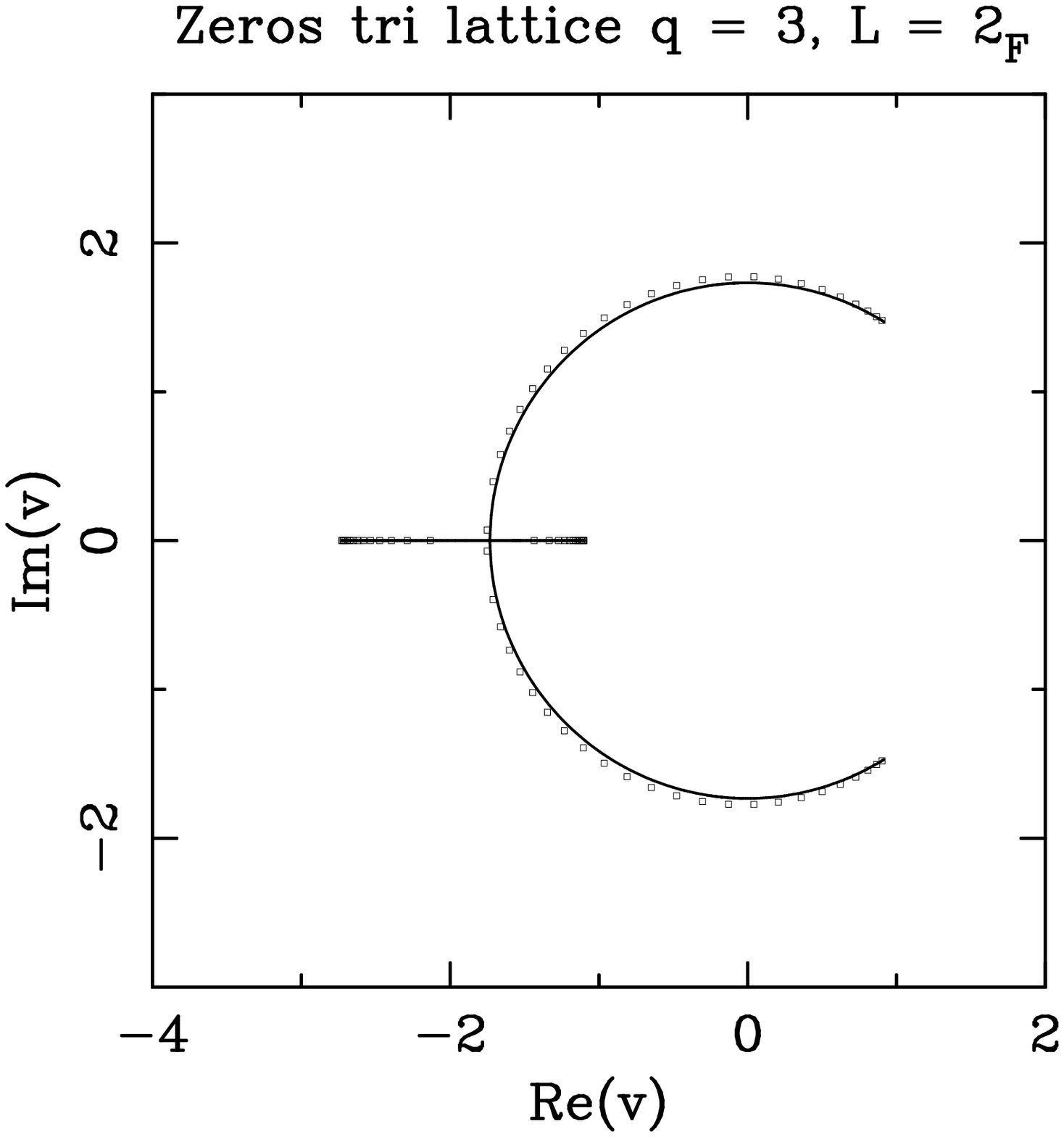} &
   \includegraphics[width=200pt]{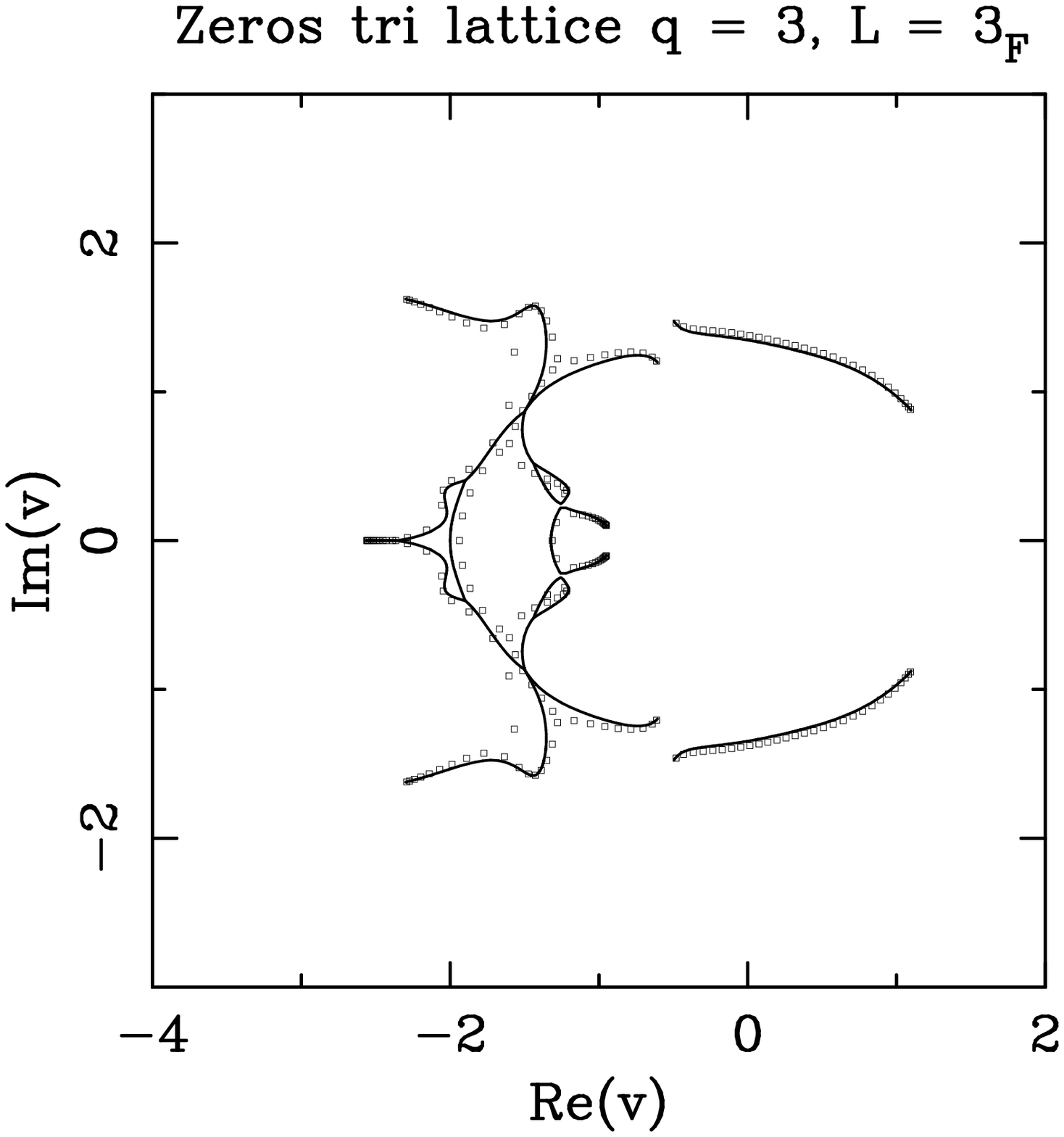} \\[1mm]
   \phantom{(((a)}(a)    & \phantom{(((a)}(b) \\[5mm]
   \includegraphics[width=200pt]{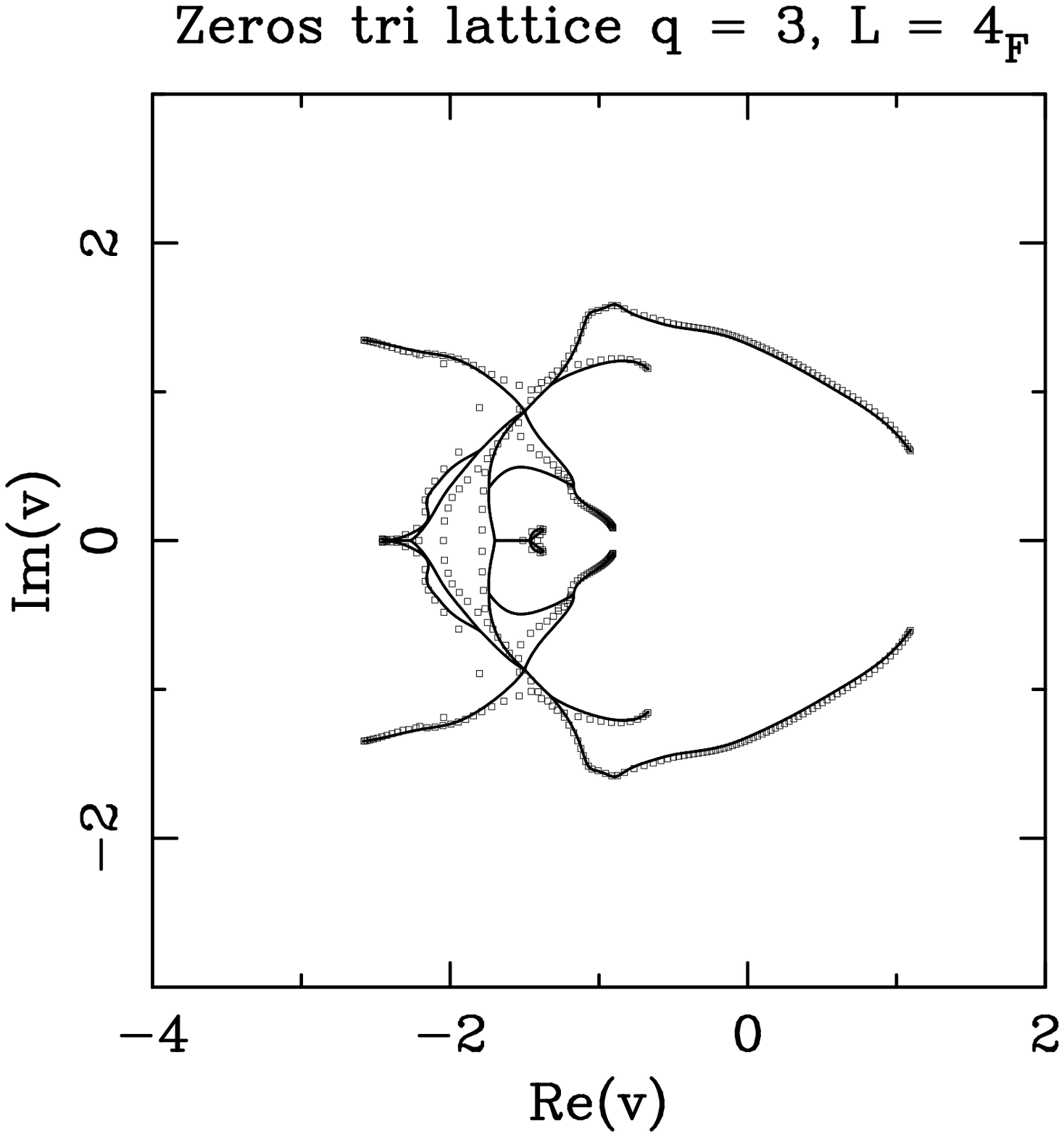} &
   \includegraphics[width=200pt]{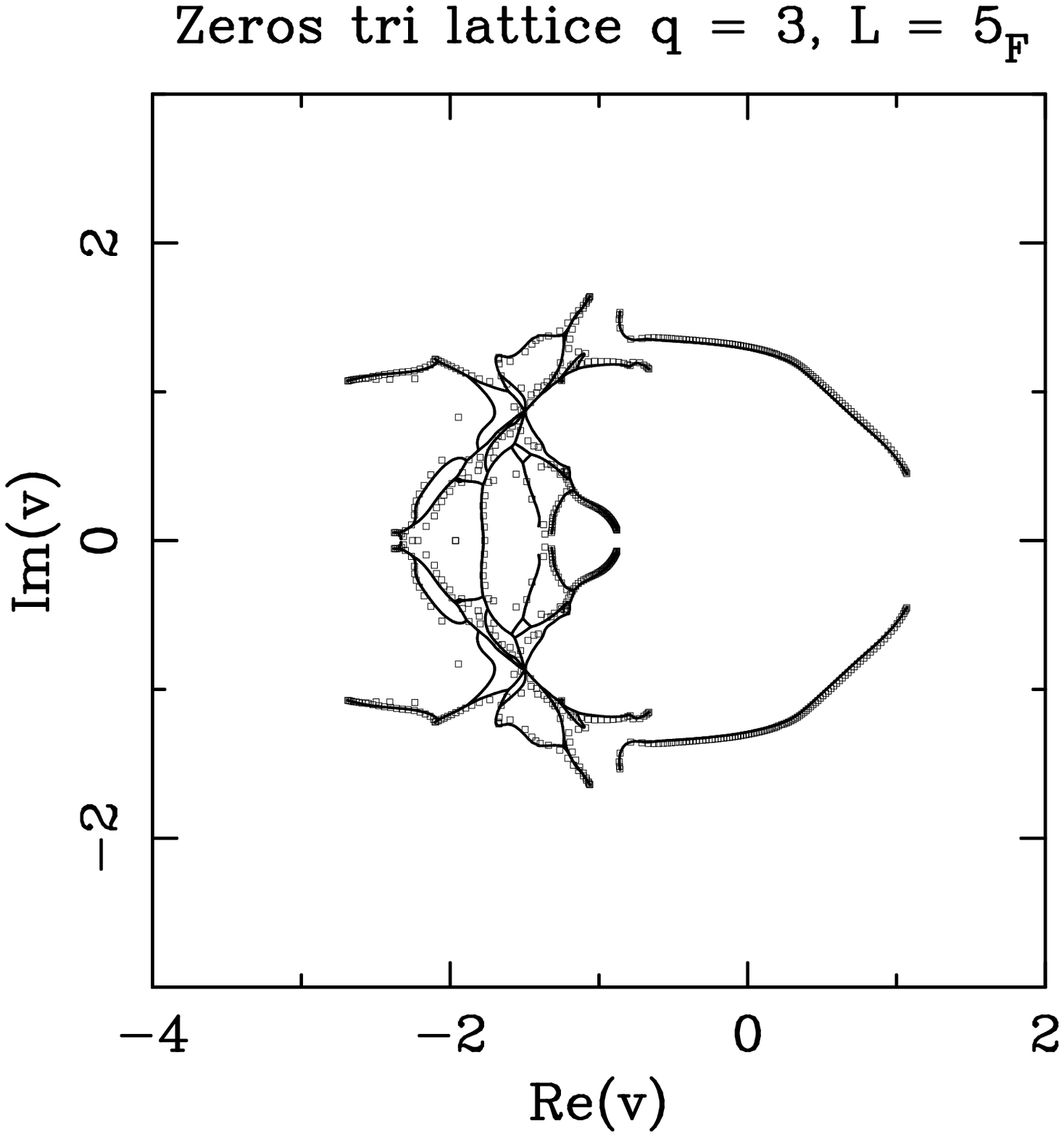} \\[1mm]
   \phantom{(((a)}(c)    & \phantom{(((a)}(d) \\
\end{tabular}
\caption[a]{\protect\label{zeros_tri_allF_q=3}
Limiting curves forming the singular locus ${\cal B}$, in the $v$ plane, for
the free energy of the Potts model for $q=3$ on the
$L_{\rm F} \times \infty_{\rm F}$ triangular-lattice strips with
(a) $L=2$, (b) $L=3$, (c) $L=4$, and (d) $L=5$.
We also show the partition-function zeros corresponding to the strips
$L_{\rm F} \times (10L)_{\rm F}$ for each value of $L$.
Where the results for
$2 \le L \le 4$ overlap with those in \cite{ta,ks}, they agree and are
included here for comparison.

}
\end{figure}

\begin{figure}[hbtp]
\centering
\begin{tabular}{cc}
   \includegraphics[width=200pt]{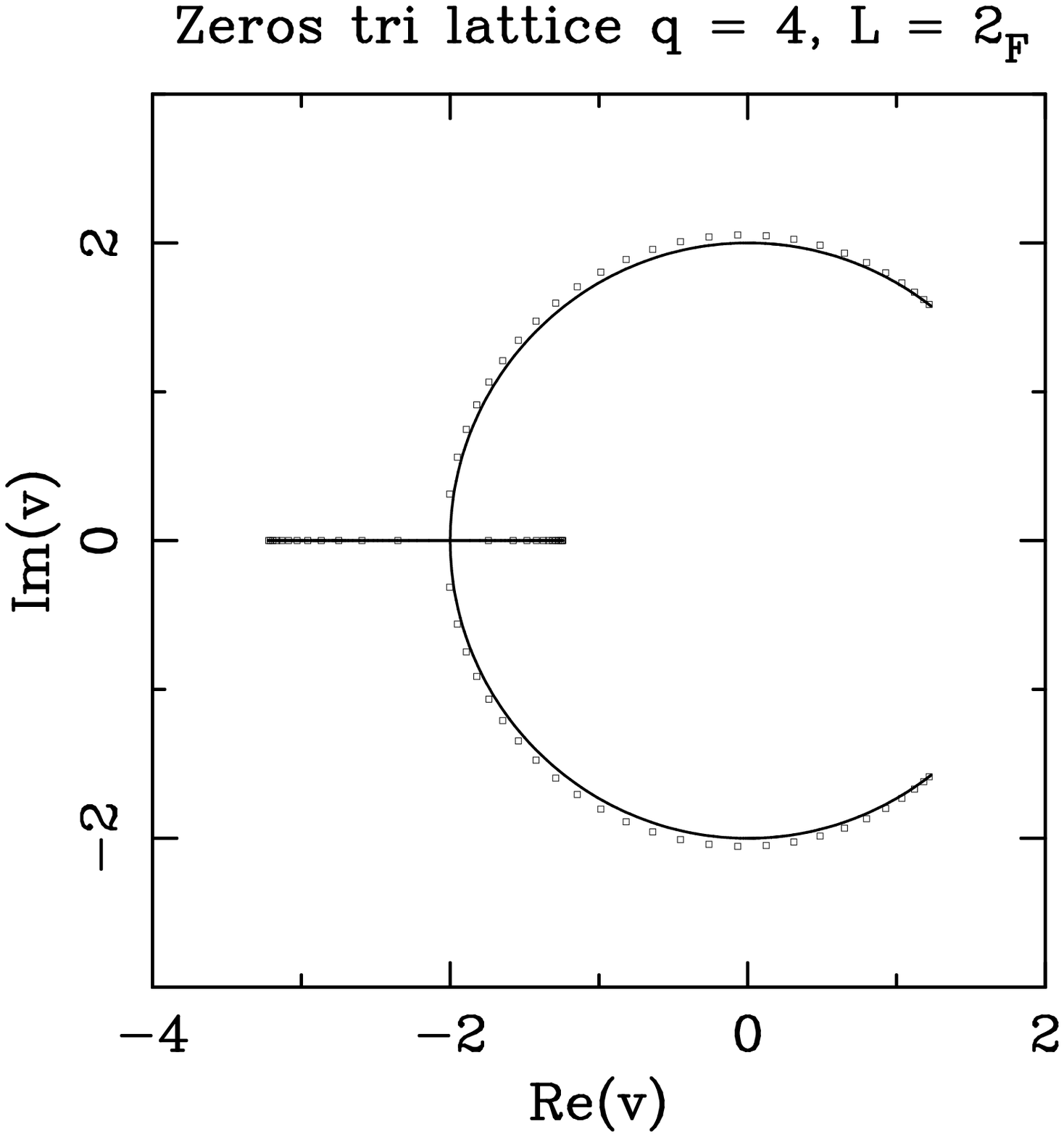} &
   \includegraphics[width=200pt]{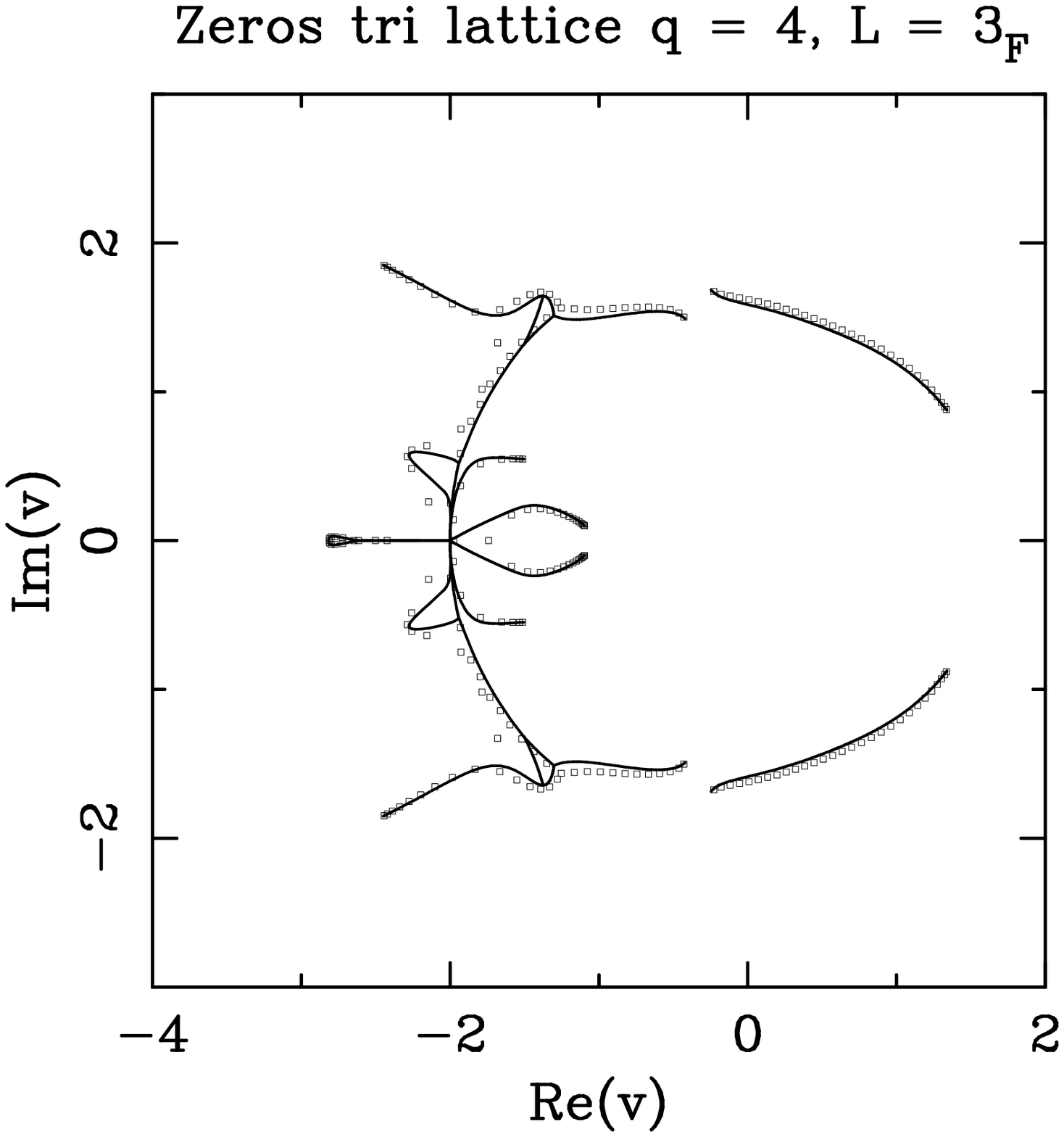} \\[1mm]
   \phantom{(((a)}(a)    & \phantom{(((a)}(b) \\[5mm]
   \includegraphics[width=200pt]{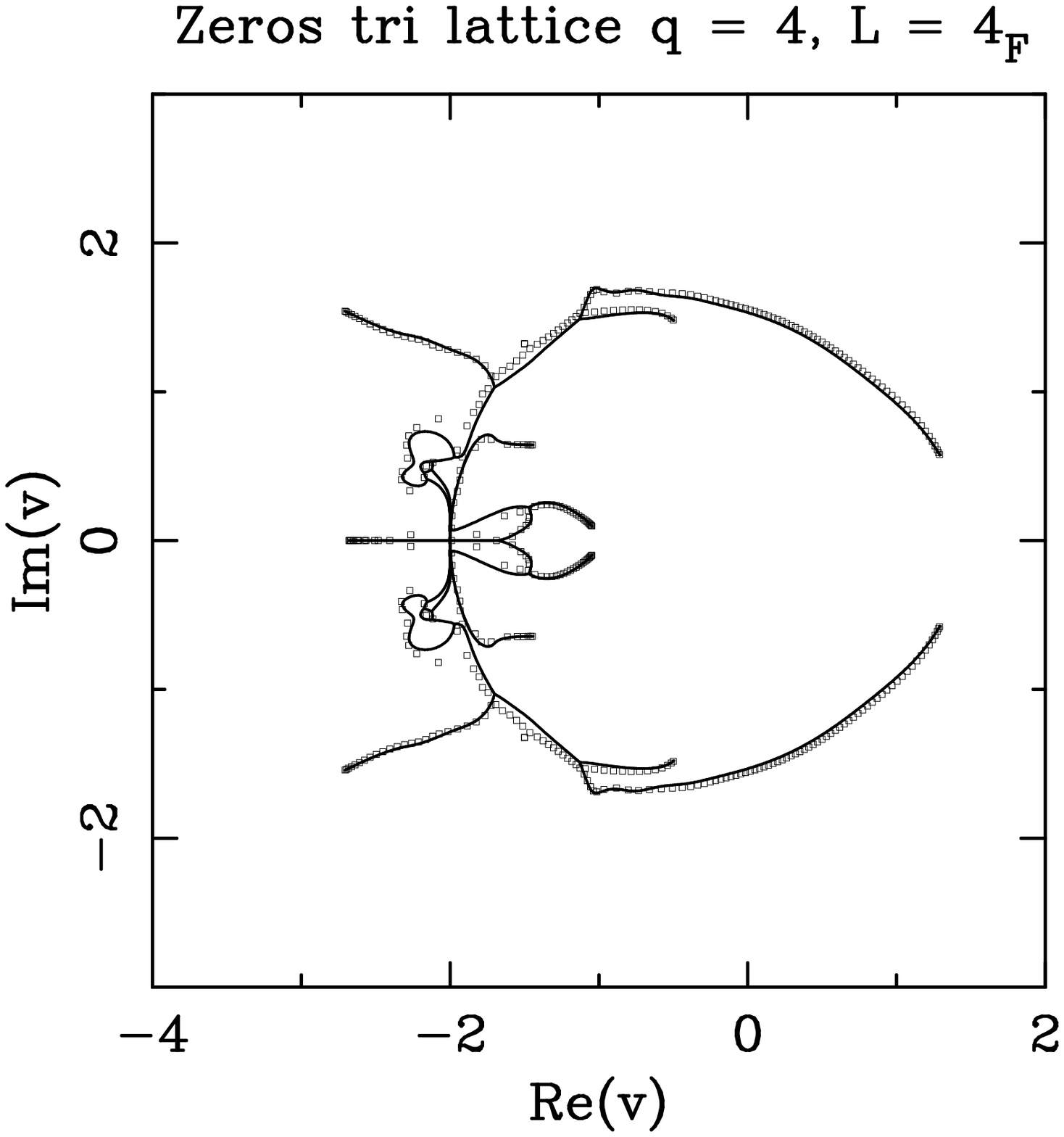} &
   \includegraphics[width=200pt]{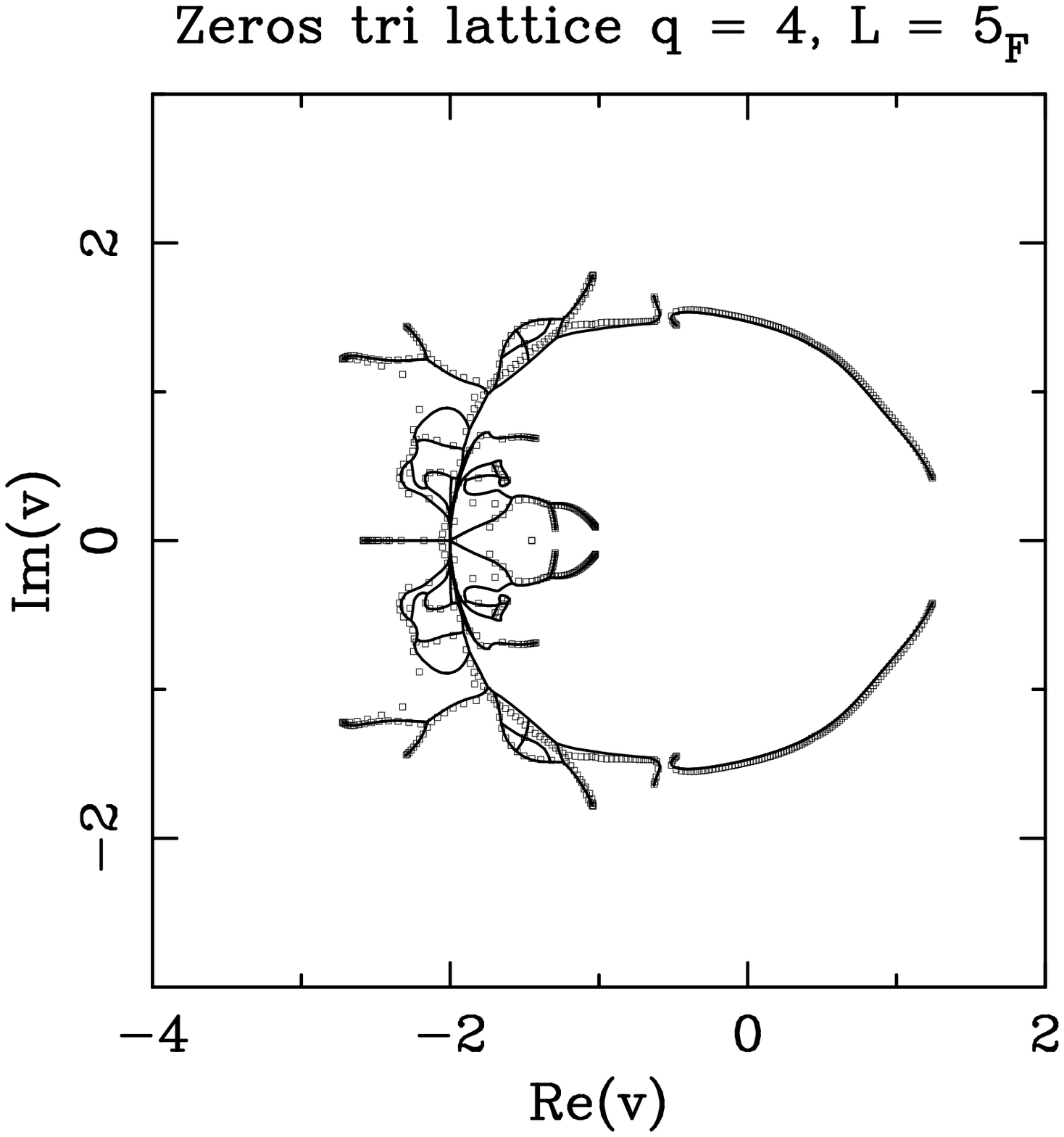} \\[1mm]
   \phantom{(((a)}(c)    & \phantom{(((a)}(d) \\
\end{tabular}
\caption[a]{\protect\label{zeros_tri_allF_q=4}
Limiting curves forming the singular locus ${\cal B}$, in the $v$ plane, for
the free energy of the Potts model for $q=4$ on the
$L_{\rm F} \times \infty_{\rm F}$ triangular-lattice strips with
(a) $L=2$, (b) $L=3$, (c) $L=4$, and (d) $L=5$.
 Where the results for
$2 \le L \le 4$ overlap with those in \cite{ta,ks}, they agree and are
included here for comparison.
We also show the partition-function zeros corresponding to the strips
$L_{\rm F} \times (10L)_{\rm F}$ for each value of $L$.
}
\end{figure}

\begin{figure}[hbtp]
\centering
\begin{tabular}{cc}
   \includegraphics[width=200pt]{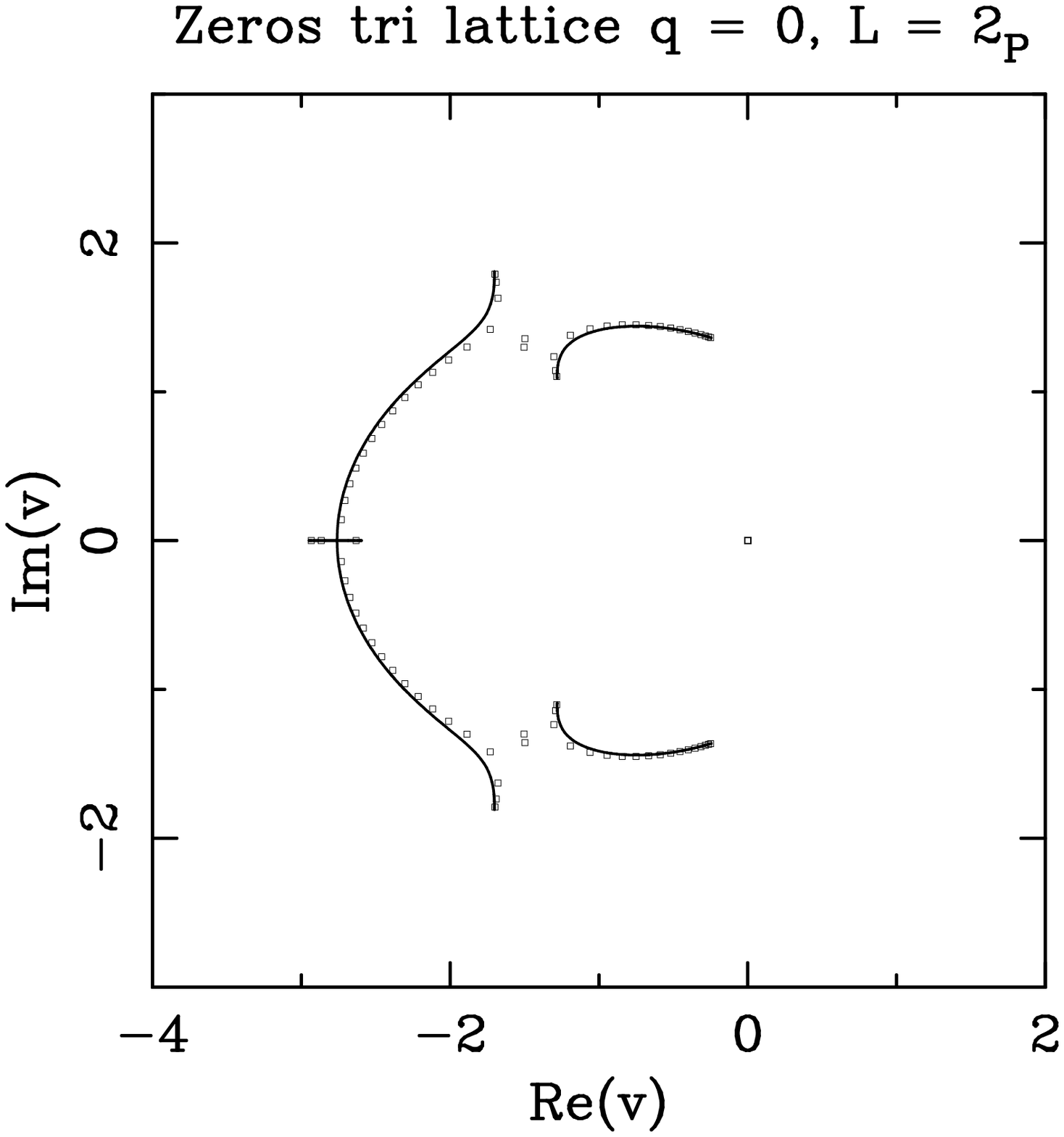} &
   \includegraphics[width=200pt]{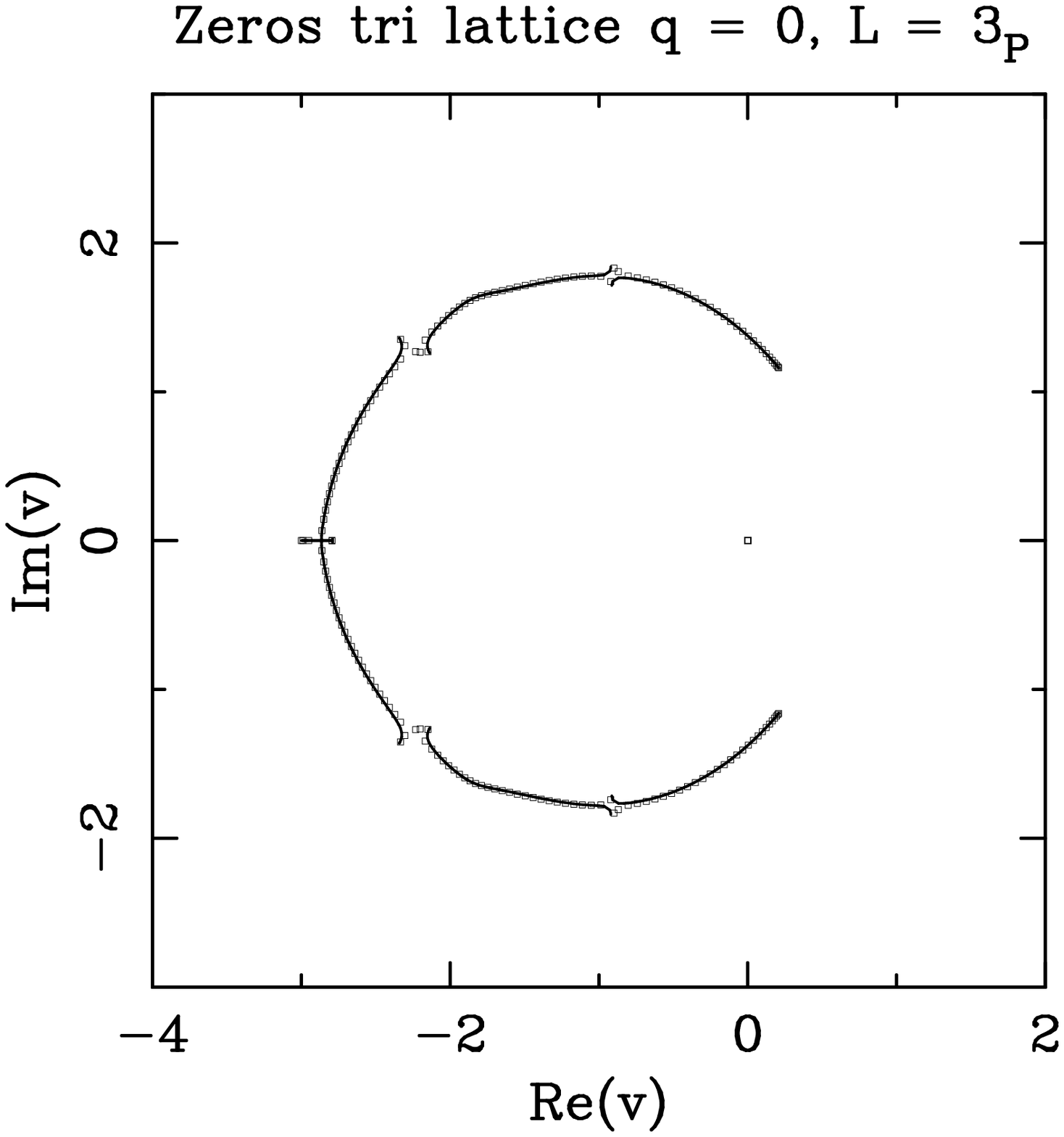} \\[1mm]
   \phantom{(((a)}(a)    & \phantom{(((a)}(b) \\[5mm]
   \includegraphics[width=200pt]{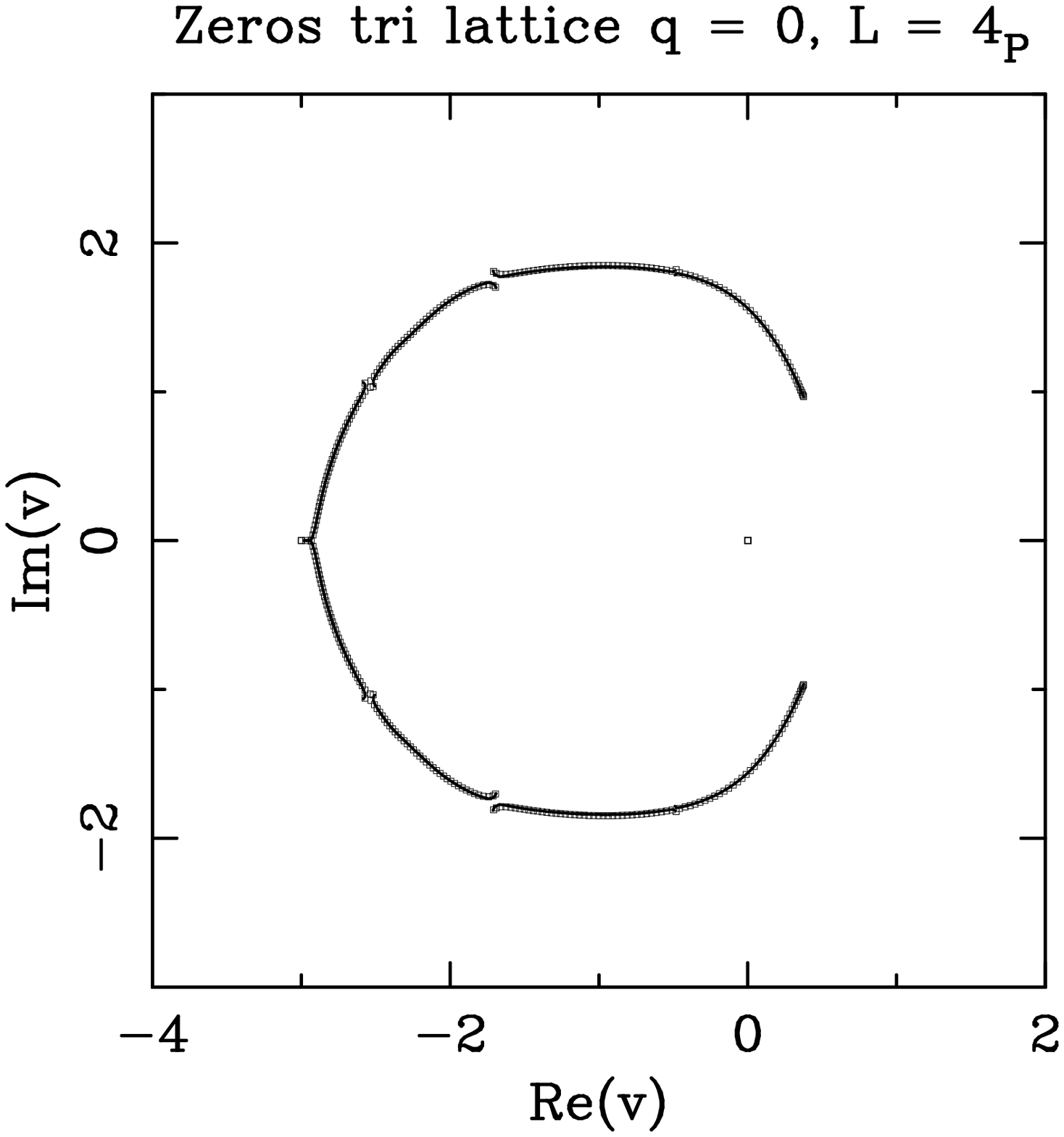} &
   \includegraphics[width=200pt]{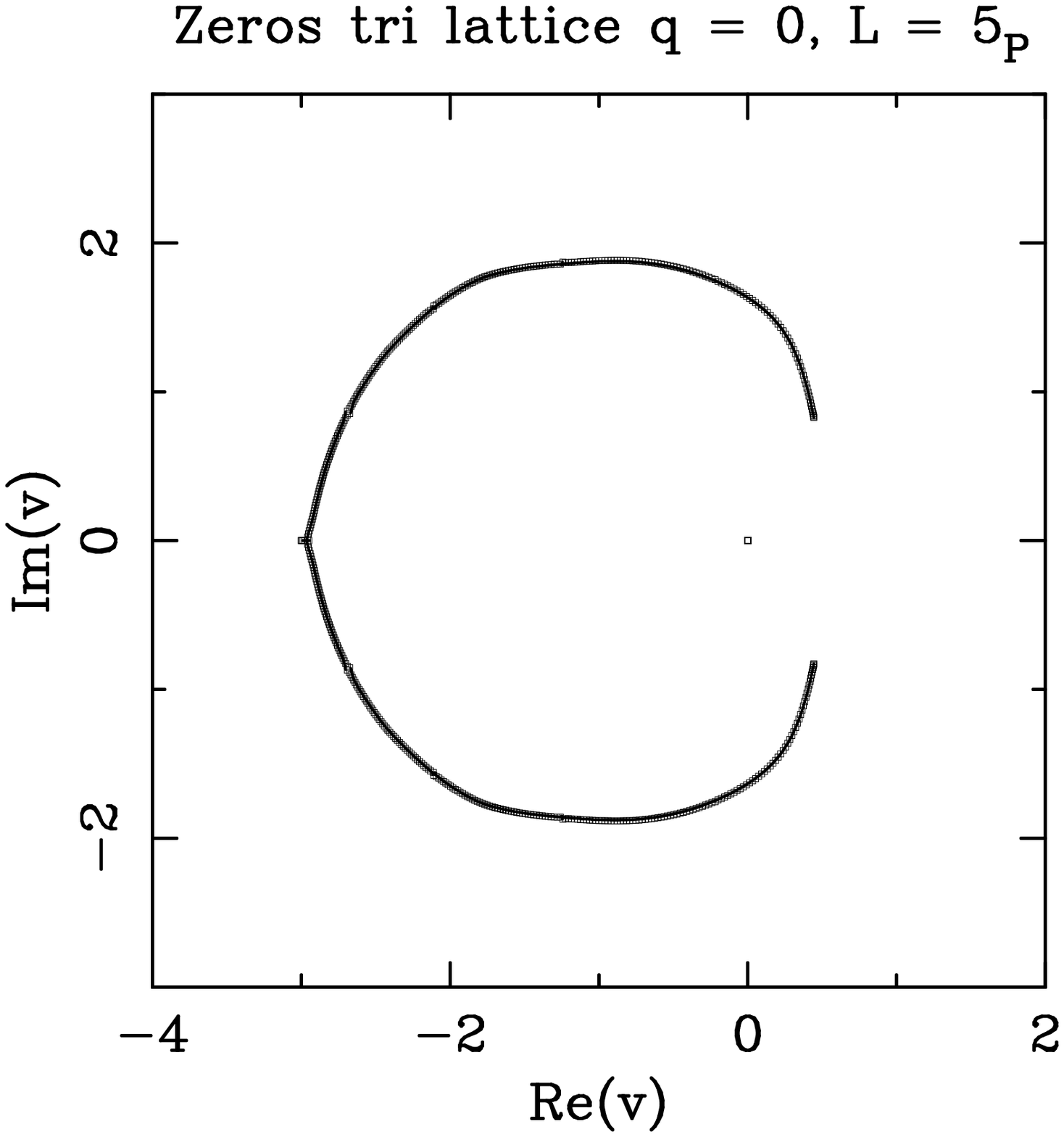} \\[1mm]
   \phantom{(((a)}(c)    & \phantom{(((a)}(d) \\
\end{tabular}
\caption[a]{\protect\label{zeros_tri_allP_q=0} Limiting curves forming the
singular locus ${\cal B}$, in the $v$ plane, for the free energy of the Potts
model, defined with the order $f_{qn}$, for $q=0$ on the $L_{\rm P} \times
\infty_{\rm F}$ triangular-lattice strips with (a) $L=2$, (b) $L=3$, (c) $L=4$,
and (d) $L=5$.  We also show the zeros of $Z(G,0,v)/q$ corresponding to the
strips $L_{\rm P} \times (10L)_{\rm F}$ for each value of $L$.  }
\end{figure}

\begin{figure}[hbtp]
\centering
\begin{tabular}{cc}
   \includegraphics[width=200pt]{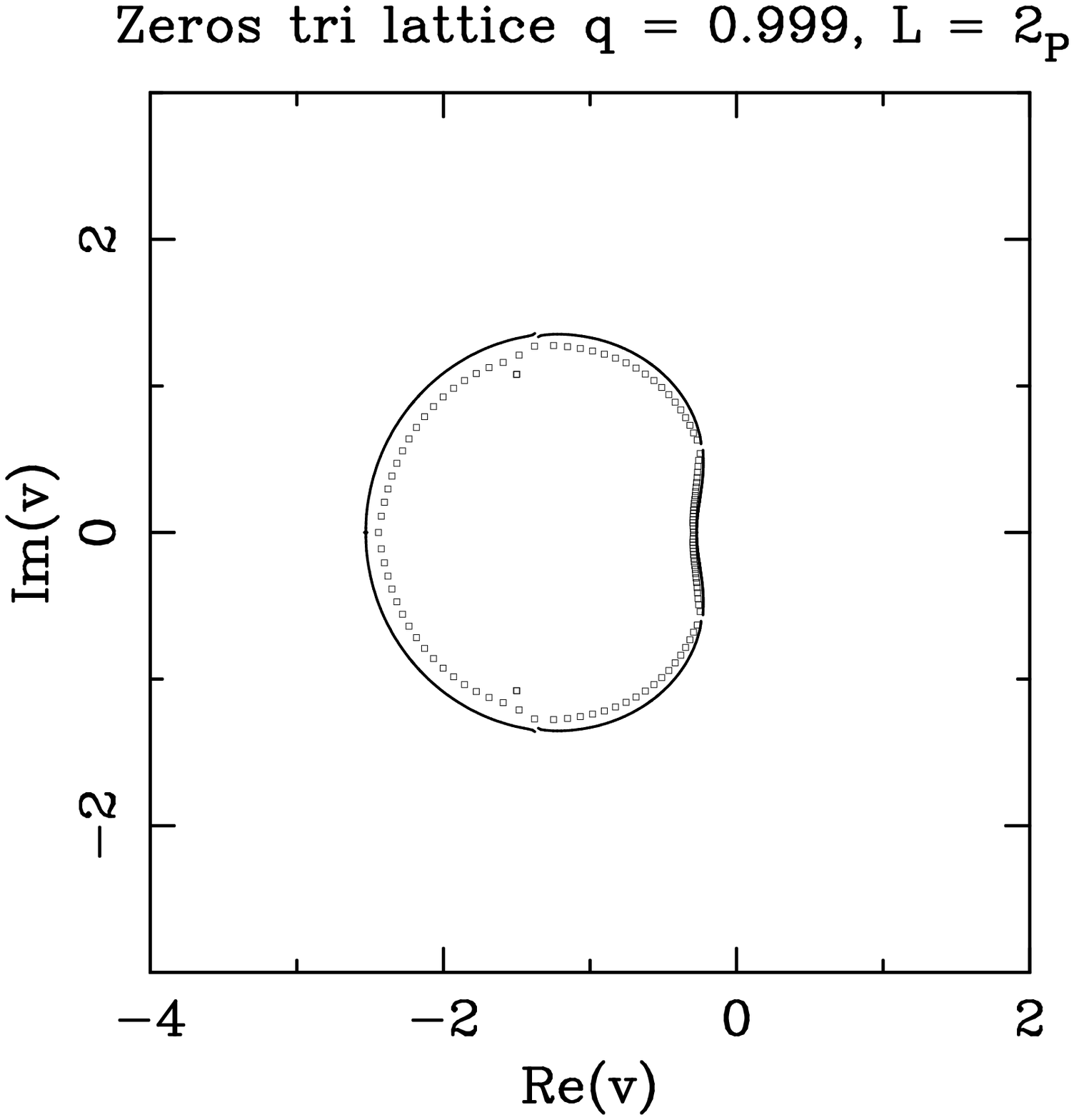} &
   \includegraphics[width=200pt]{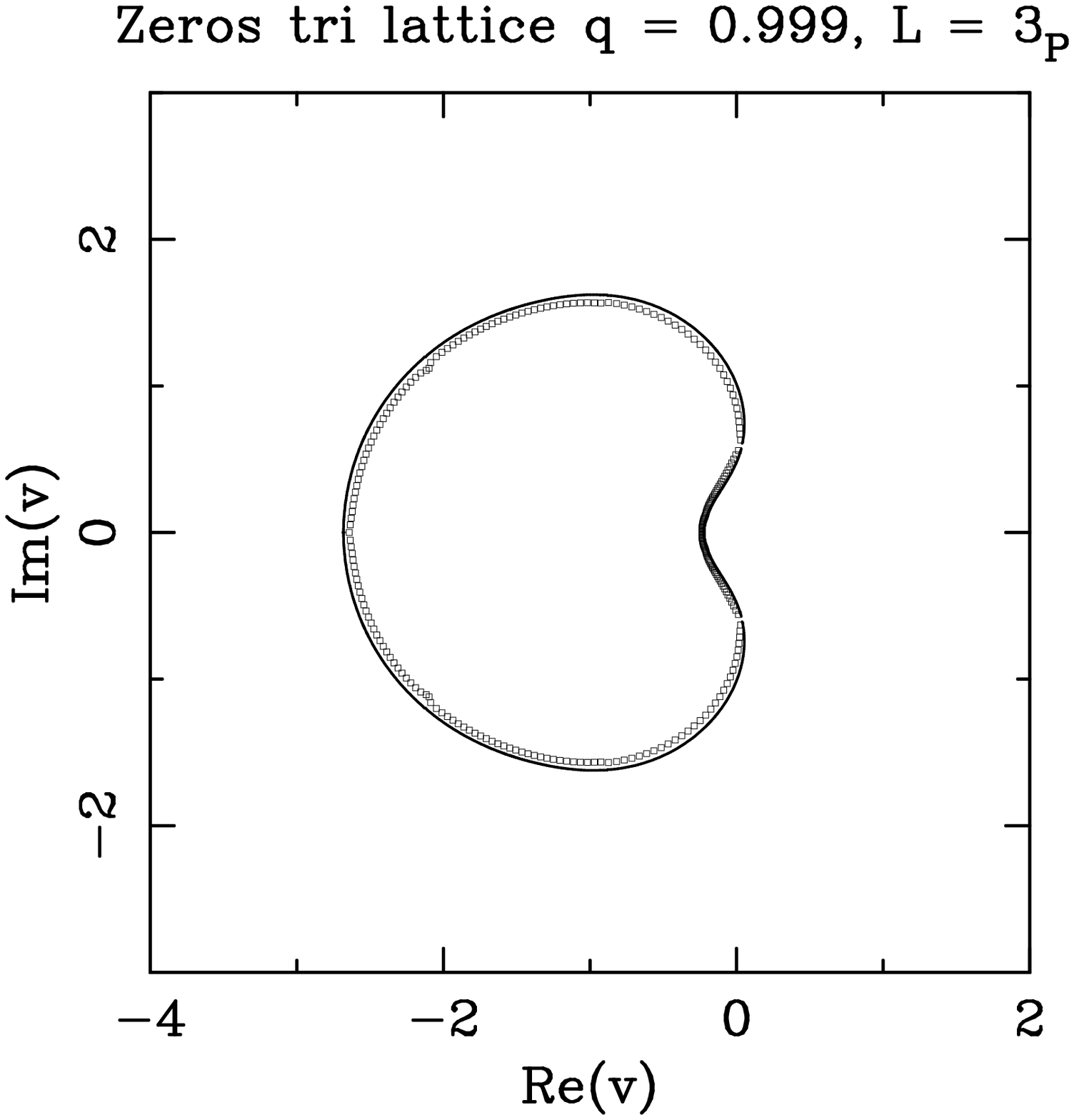} \\[1mm]
   \phantom{(((a)}(a)    & \phantom{(((a)}(b) \\[5mm]
   \includegraphics[width=200pt]{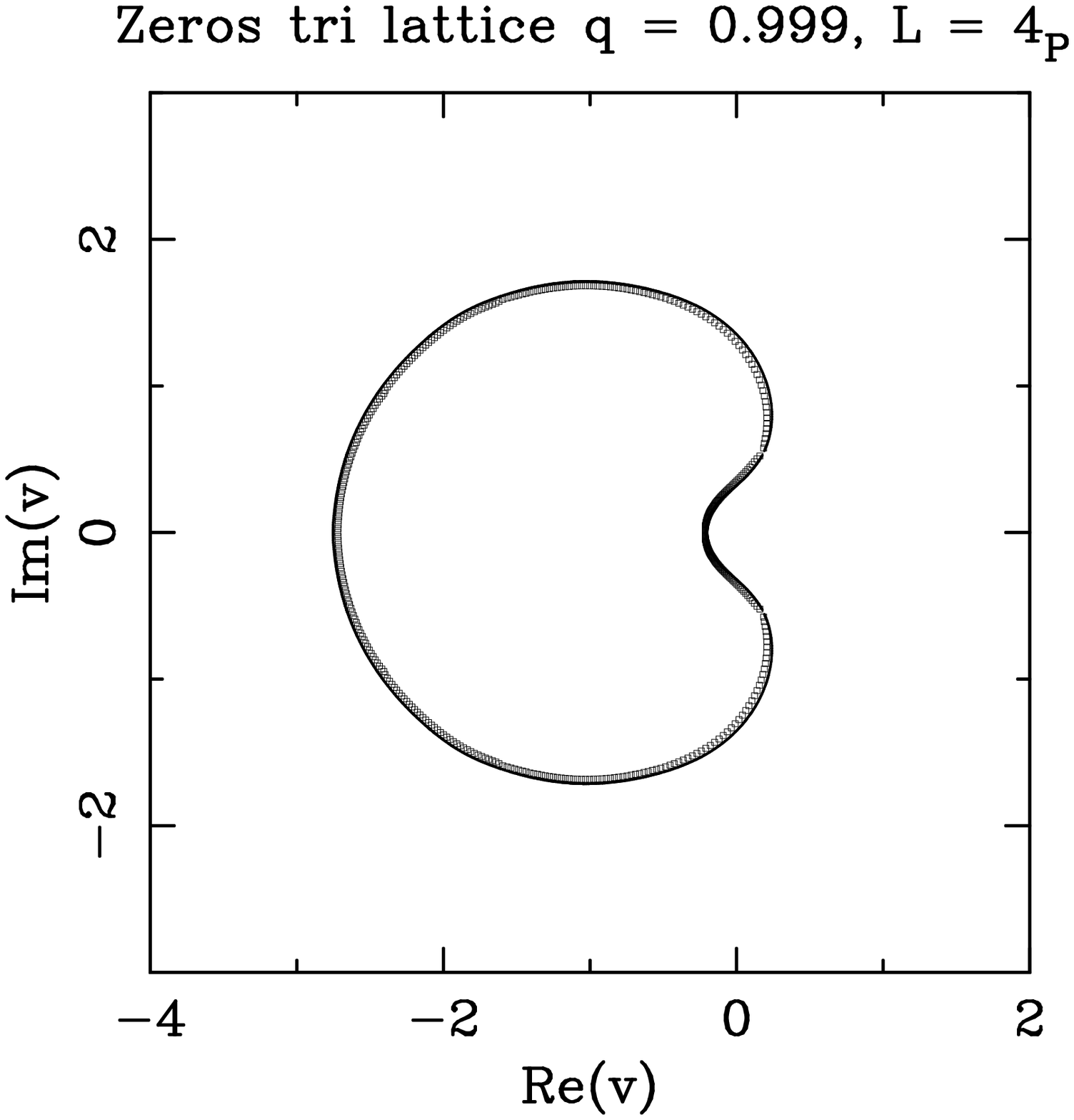} &
   \includegraphics[width=200pt]{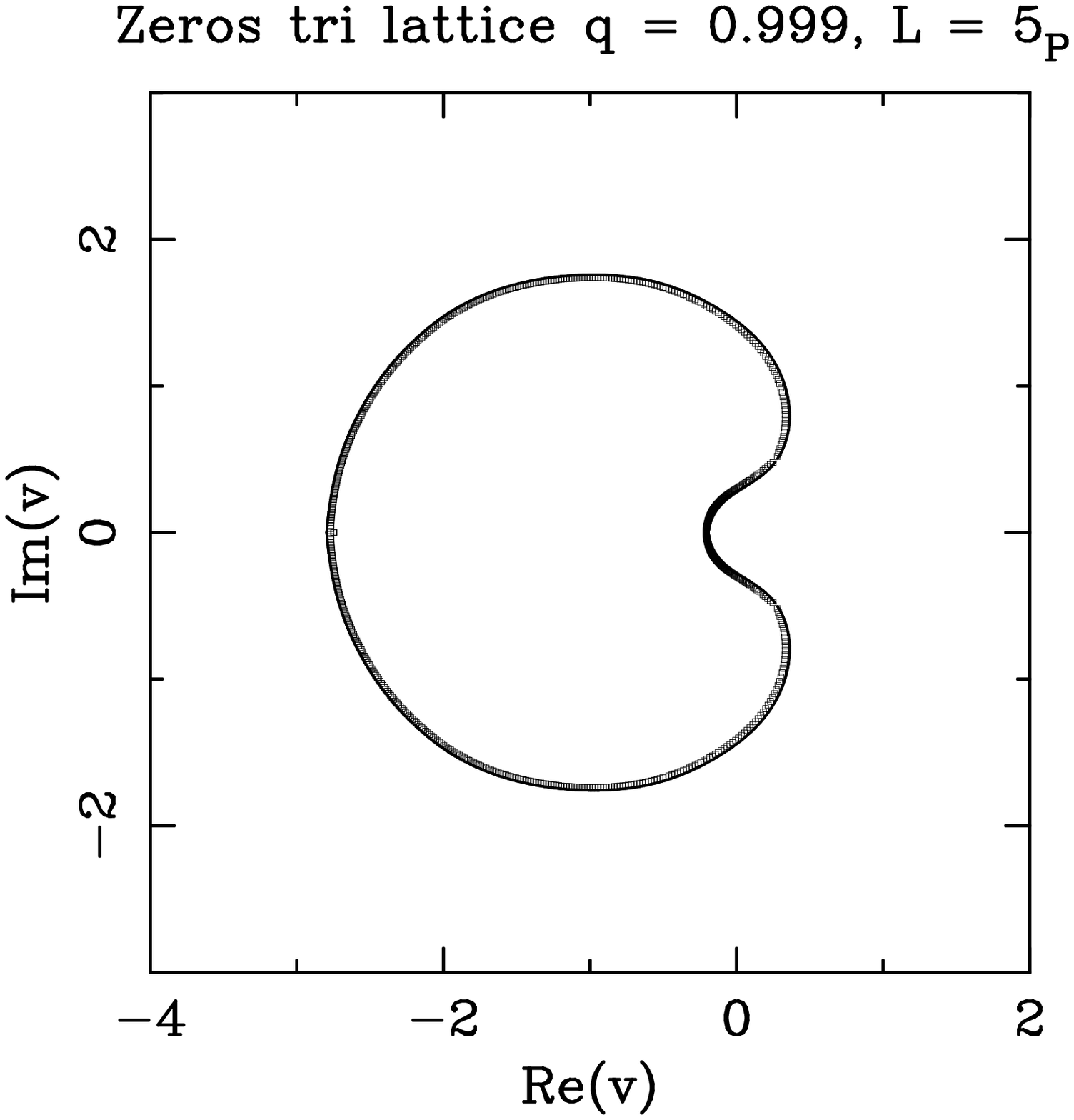} \\[1mm]
   \phantom{(((a)}(c)    & \phantom{(((a)}(d) \\
\end{tabular}
\caption[a]{\protect\label{zeros_tri_allP_q=1}
Limiting curves forming the singular locus ${\cal B}$, in the $v$ plane, for
the free energy of the Potts model for $q=0.999$ on the
$L_{\rm P} \times \infty_{\rm F}$ triangular-lattice strips with
(a) $L=2$, (b) $L=3$, (c) $L=4$, and (d) $L=5$.  These are essentially
equivalent to the limiting curves for $f_{qn}$ at $q=1$.  
We also show the partition-function zeros corresponding to the strips
$L_{\rm P} \times (10L)_{\rm F}$ for each value of $L$.
}
\end{figure}

\begin{figure}[hbtp]
\centering
\begin{tabular}{cc}
   \includegraphics[width=200pt]{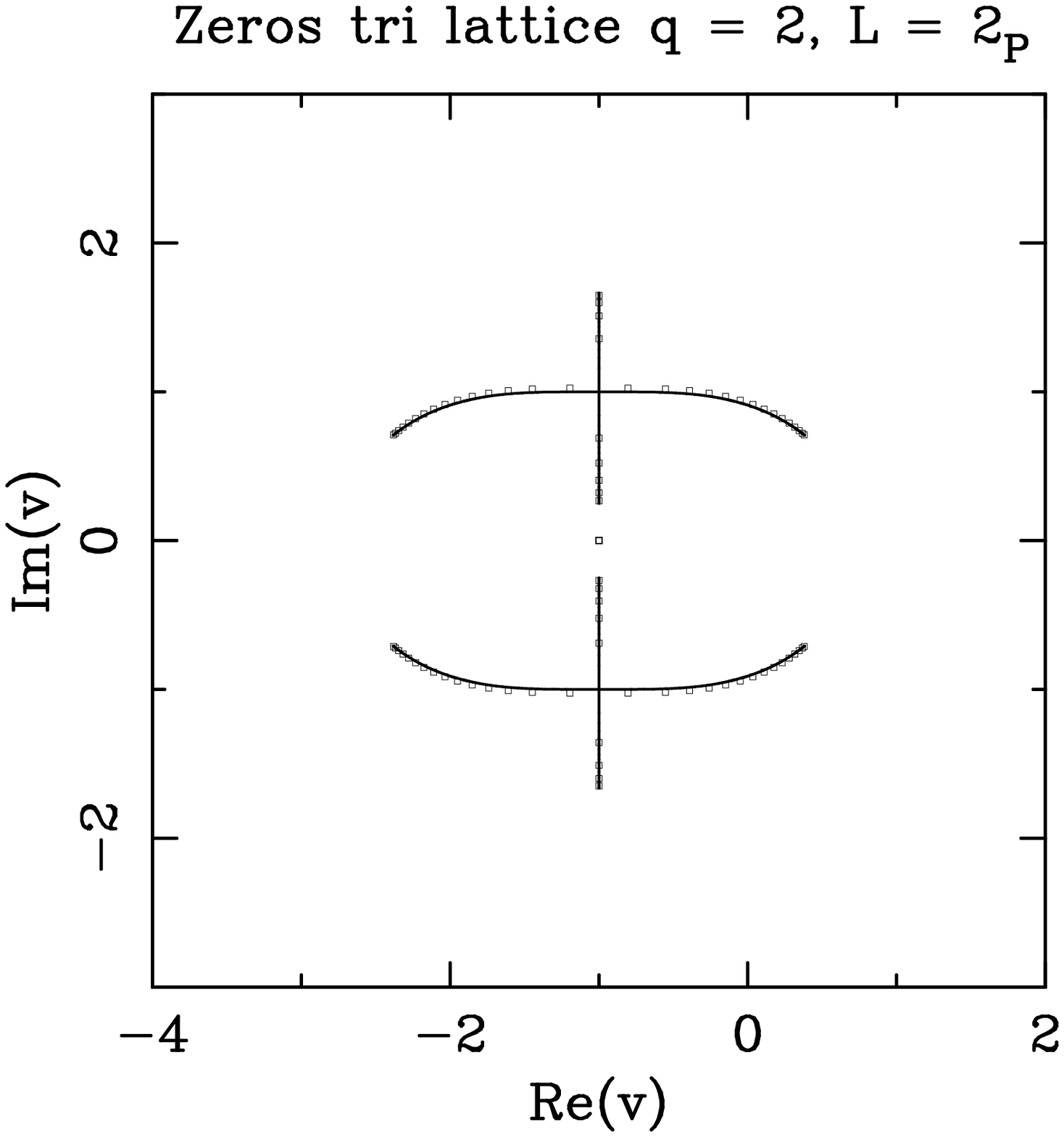} &
   \includegraphics[width=200pt]{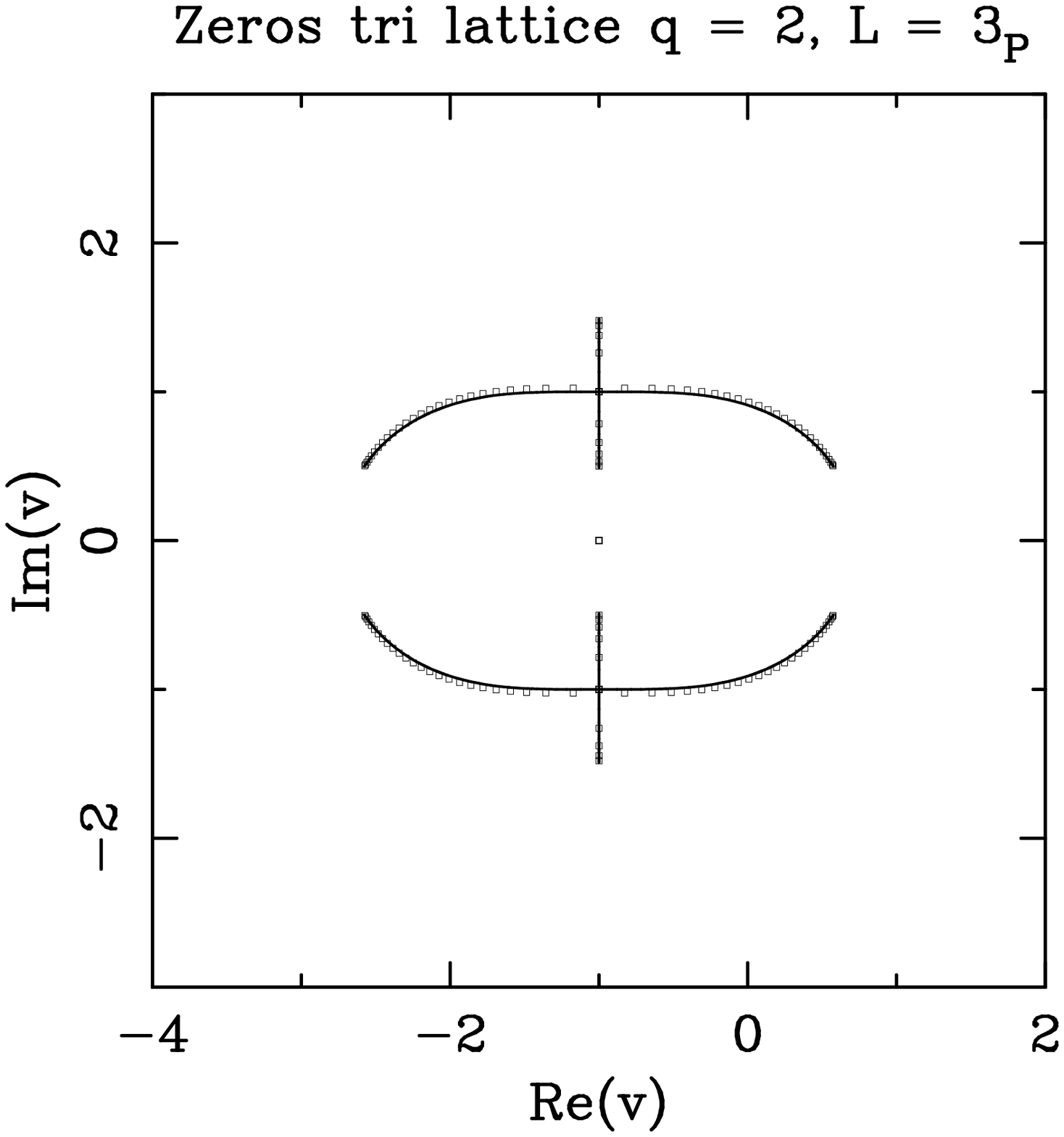} \\[1mm]
   \phantom{(((a)}(a)    & \phantom{(((a)}(b) \\[5mm]
   \includegraphics[width=200pt]{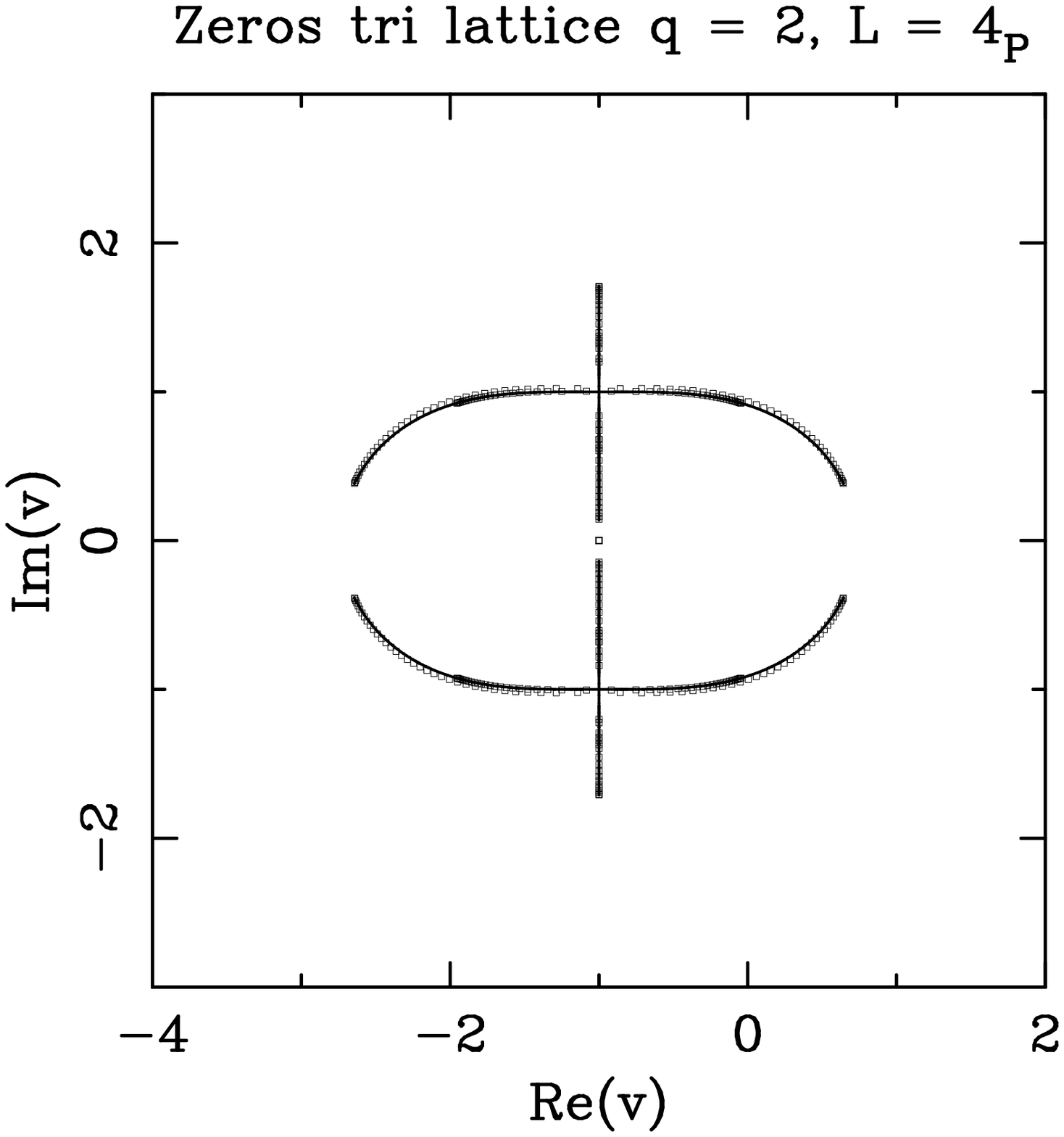} &
   \includegraphics[width=200pt]{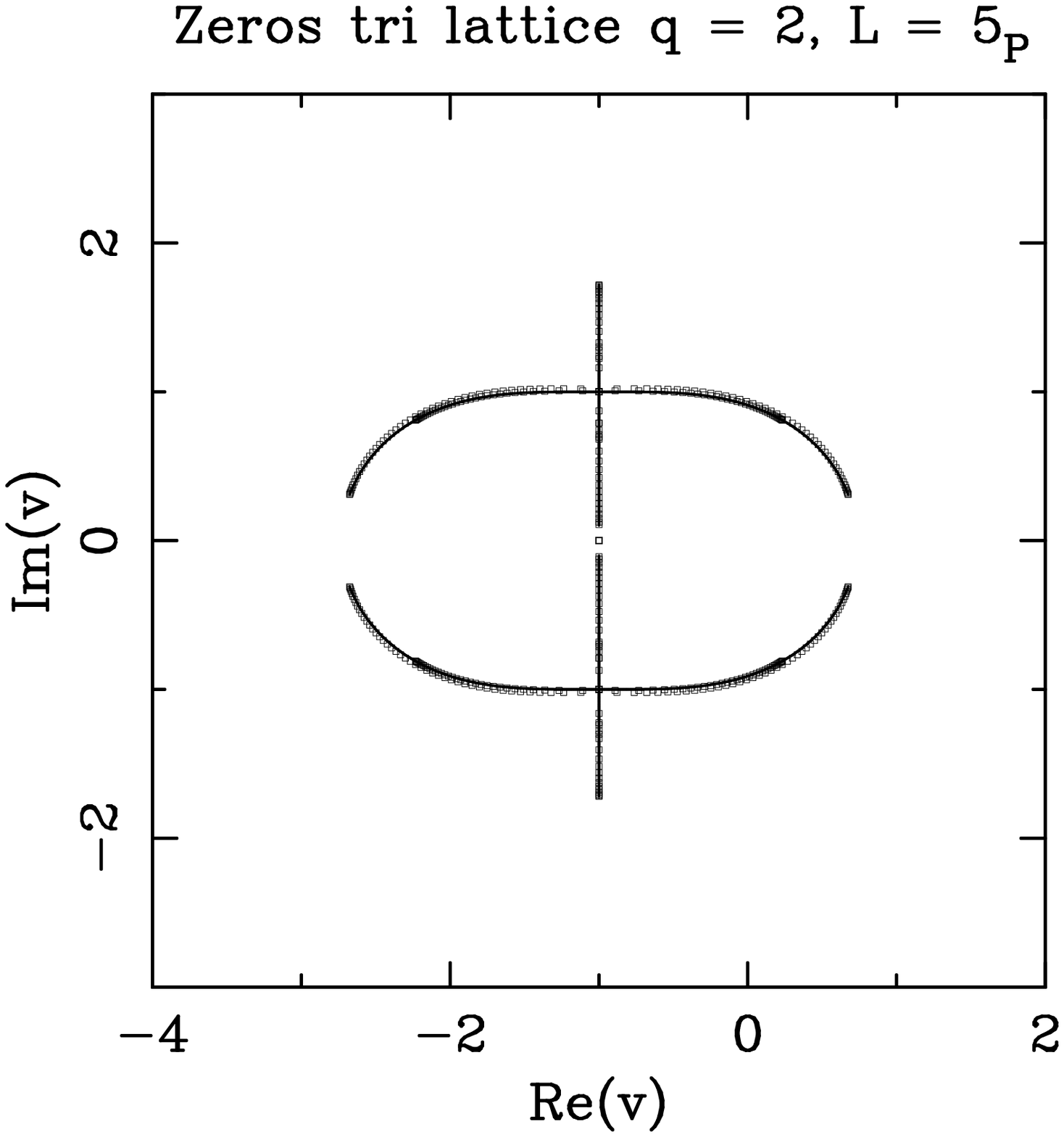} \\[1mm]
   \phantom{(((a)}(c)    & \phantom{(((a)}(d) \\
\end{tabular}
\caption[a]{\protect\label{zeros_tri_allP_q=2}
Limiting curves forming the singular locus ${\cal B}$, in the $v$ plane, for
the free energy of the Potts model for $q=2$ on the
$L_{\rm P} \times \infty_{\rm F}$ triangular-lattice strips with
(a) $L=2$, (b) $L=3$, (c) $L=4$, and (d) $L=5$.
We also show the partition-function zeros corresponding to the strips
$L_{\rm P} \times (10L)_{\rm F}$ for each value of $L$.
}
\end{figure}

\begin{figure}[hbtp]
\centering
\begin{tabular}{cc}
   \includegraphics[width=200pt]{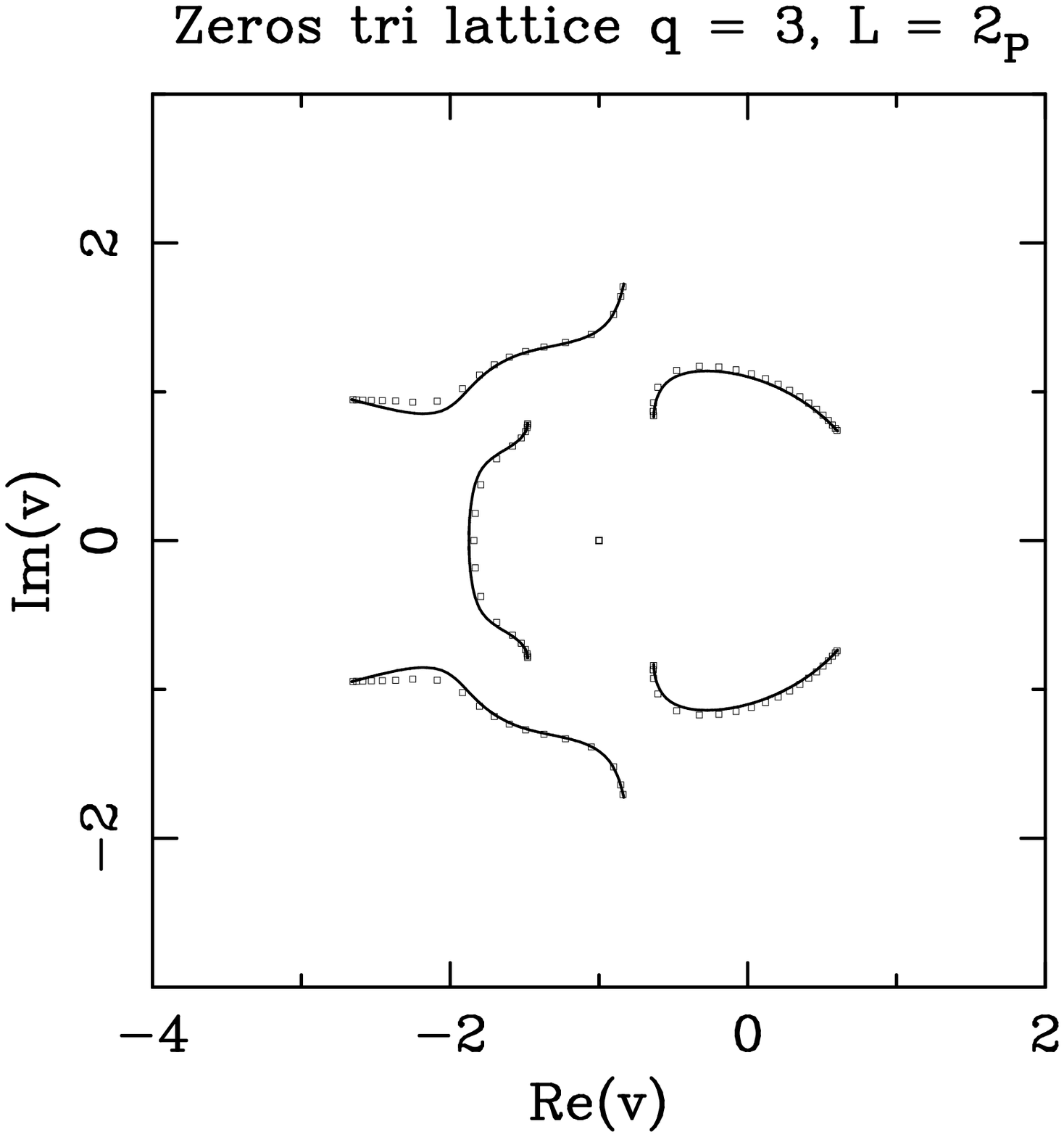} &
   \includegraphics[width=200pt]{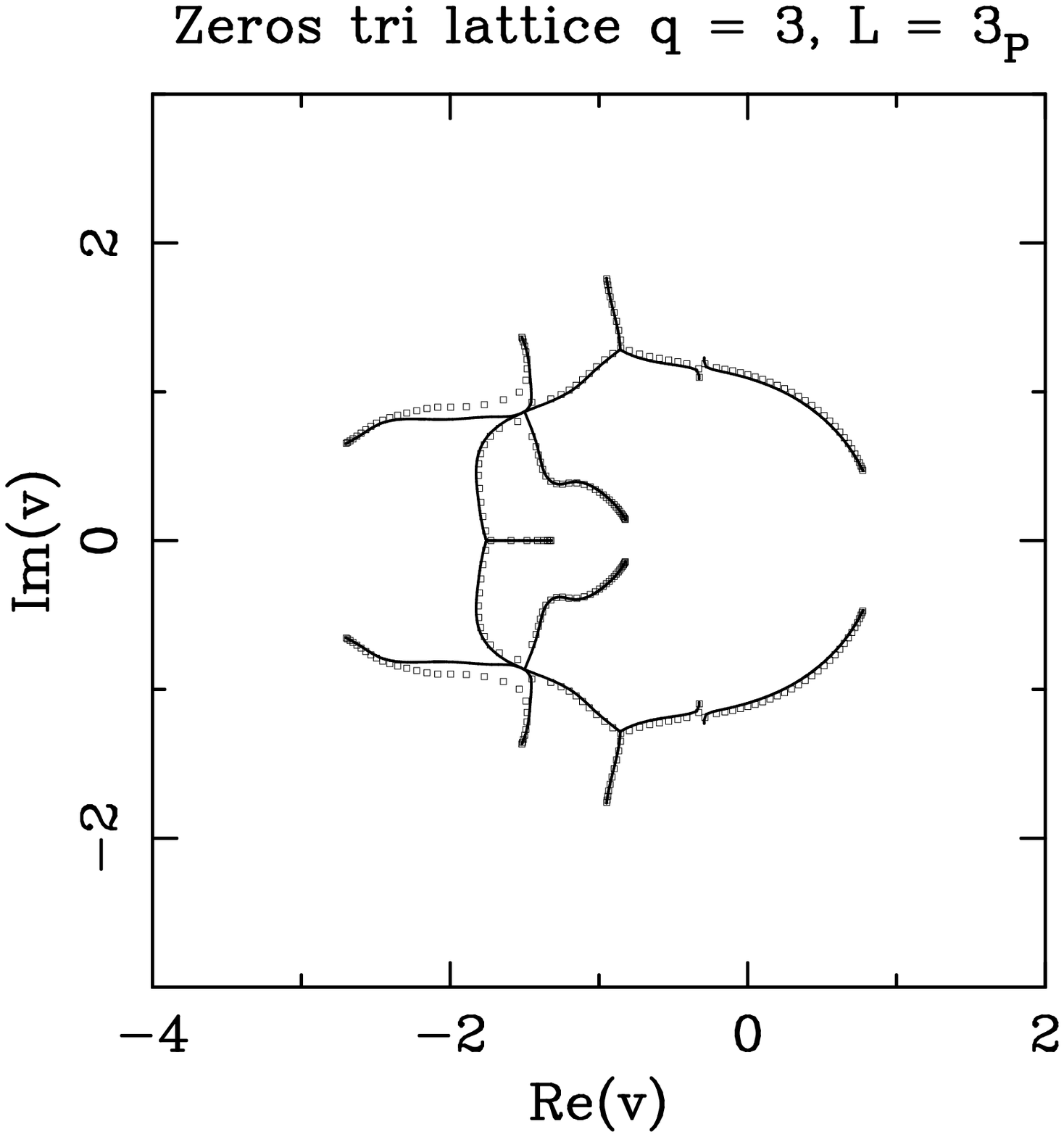} \\[1mm]
   \phantom{(((a)}(a)    & \phantom{(((a)}(b) \\[5mm]
   \includegraphics[width=200pt]{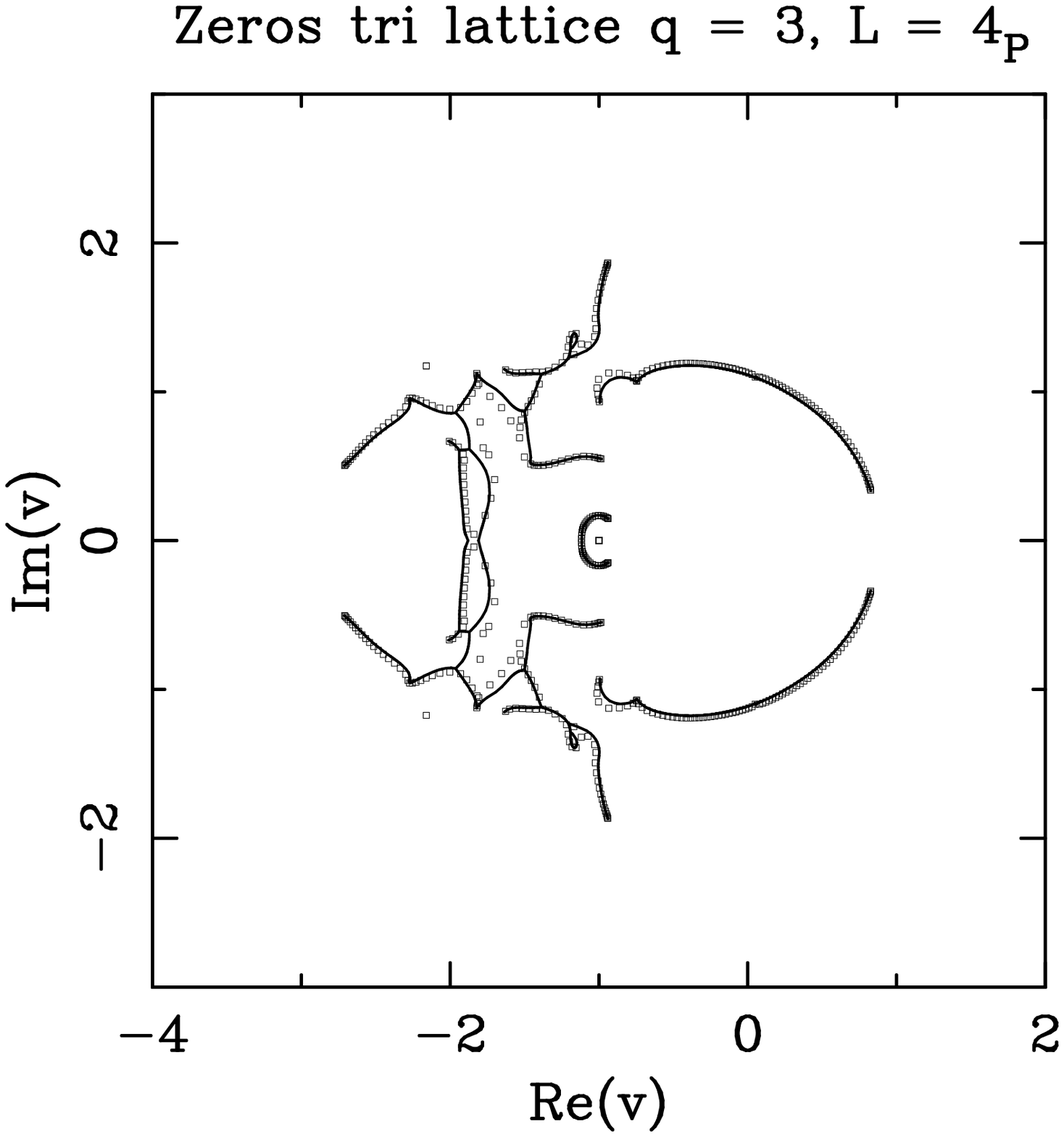} &
   \includegraphics[width=200pt]{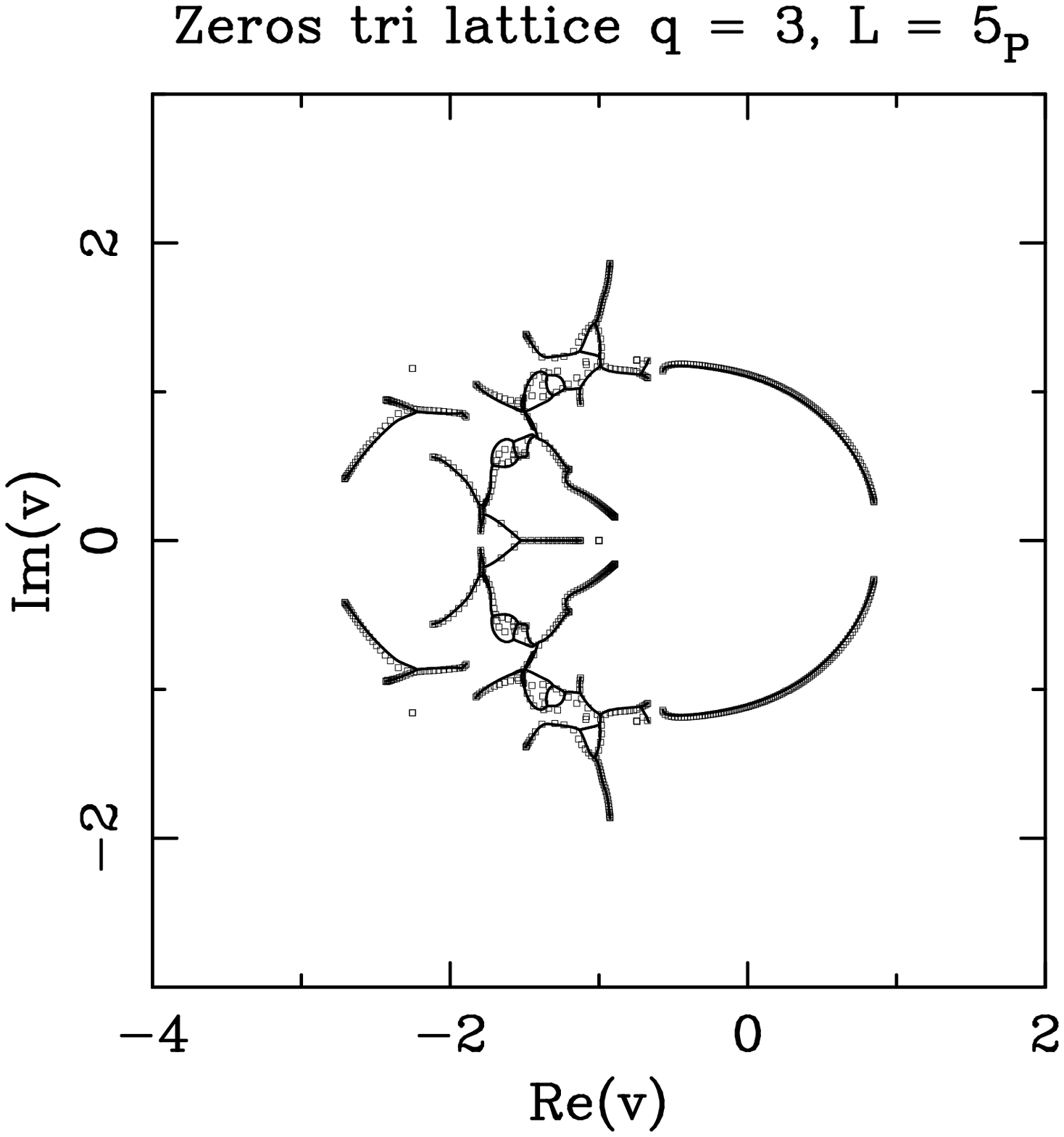} \\[1mm]
   \phantom{(((a)}(c)    & \phantom{(((a)}(d) \\
\end{tabular}
\caption[a]{\protect\label{zeros_tri_allP_q=3}
Limiting curves forming the singular locus ${\cal B}$, in the $v$ plane, for
the free energy of the Potts model for $q=3$ on the
$L_{\rm P} \times \infty_{\rm F}$ triangular-lattice strips with
(a) $L=2$, (b) $L=3$, (c) $L=4$, and (d) $L=5$.
We also show the partition-function zeros corresponding to the strips
$L_{\rm P} \times (10L)_{\rm F}$ for each value of $L$.
}
\end{figure}

\begin{figure}[hbtp]
\centering
\begin{tabular}{cc}
   \includegraphics[width=200pt]{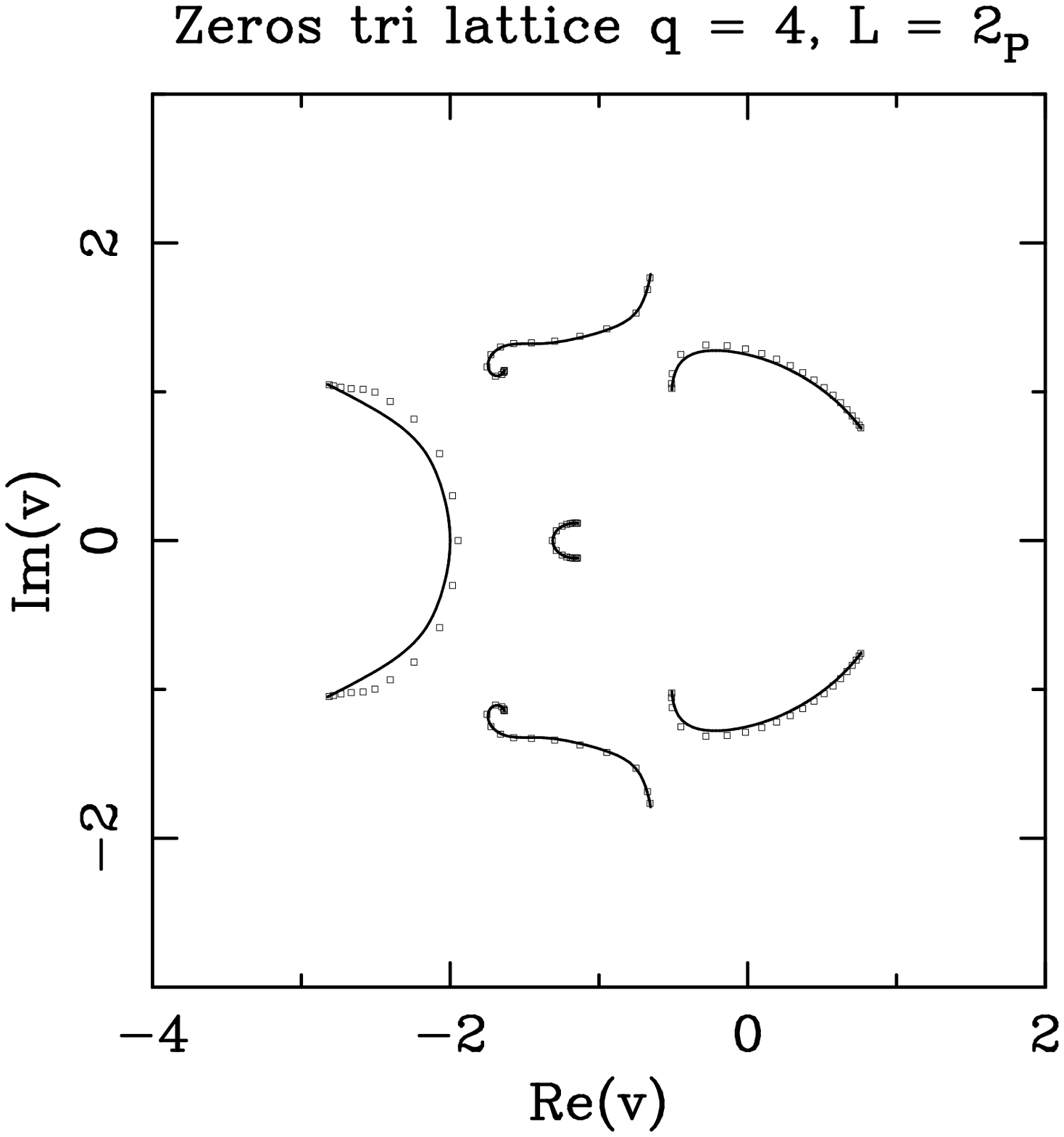} &
   \includegraphics[width=200pt]{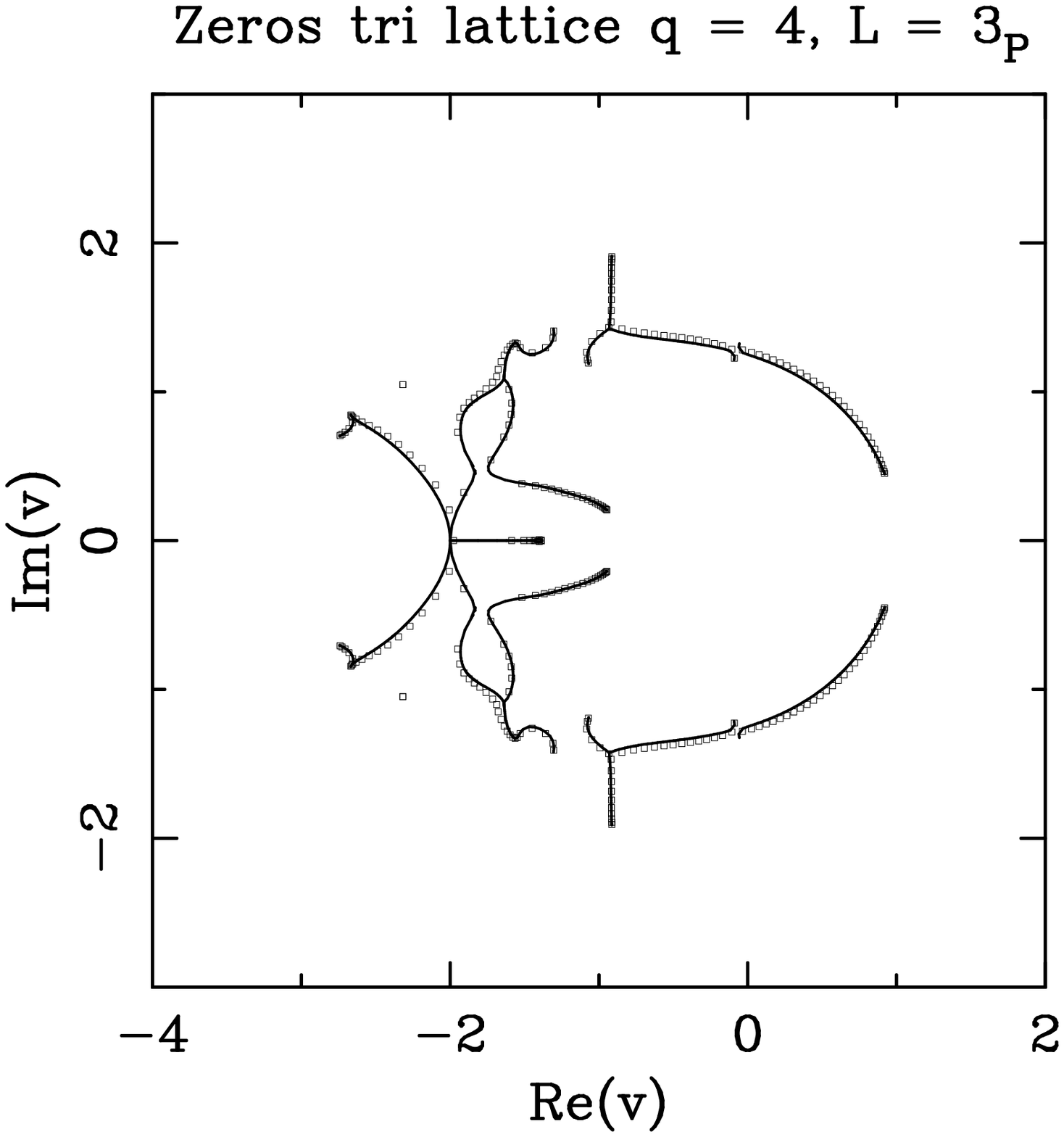} \\[1mm]
   \phantom{(((a)}(a)    & \phantom{(((a)}(b) \\[5mm]
   \includegraphics[width=200pt]{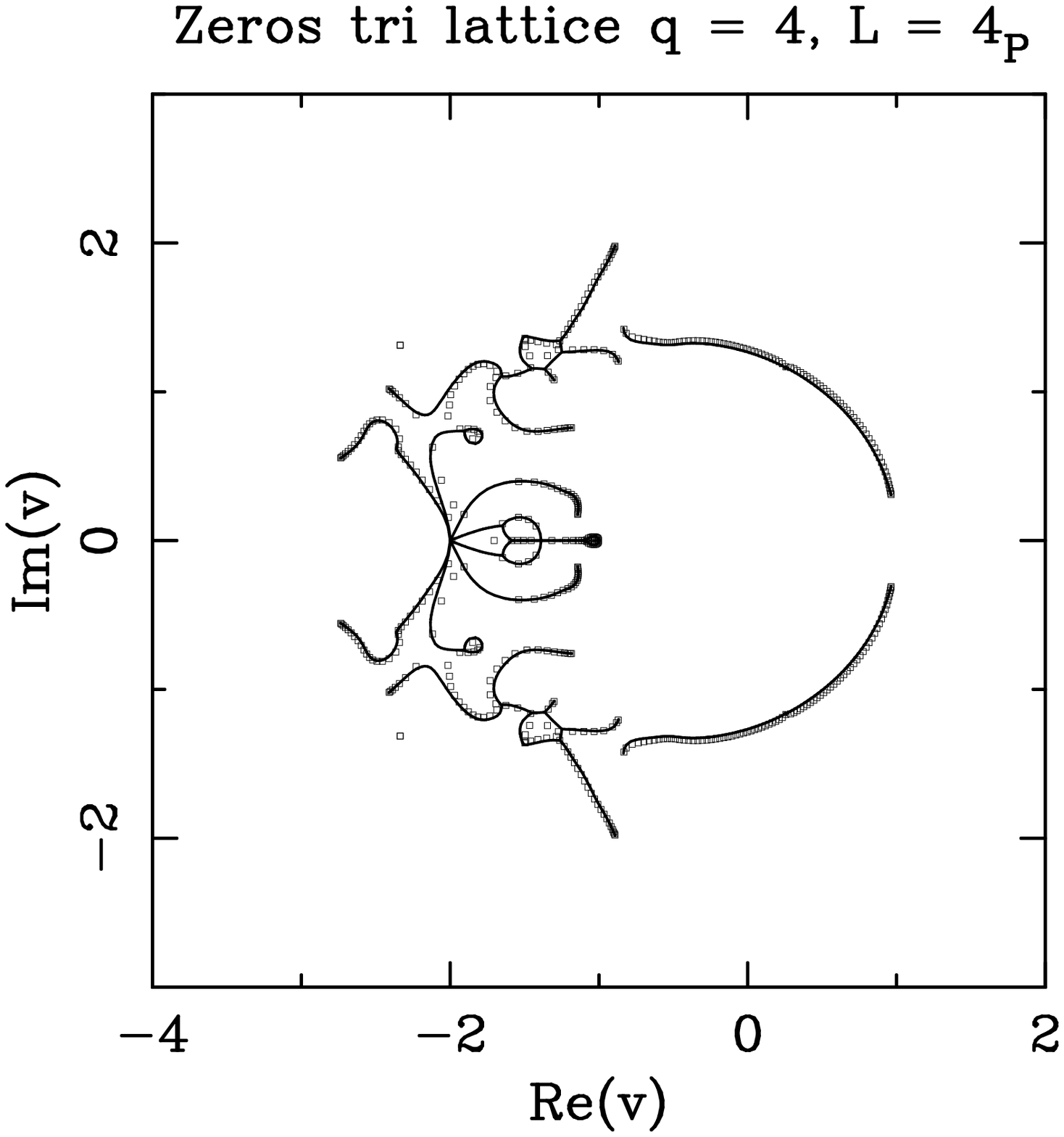} &
   \includegraphics[width=200pt]{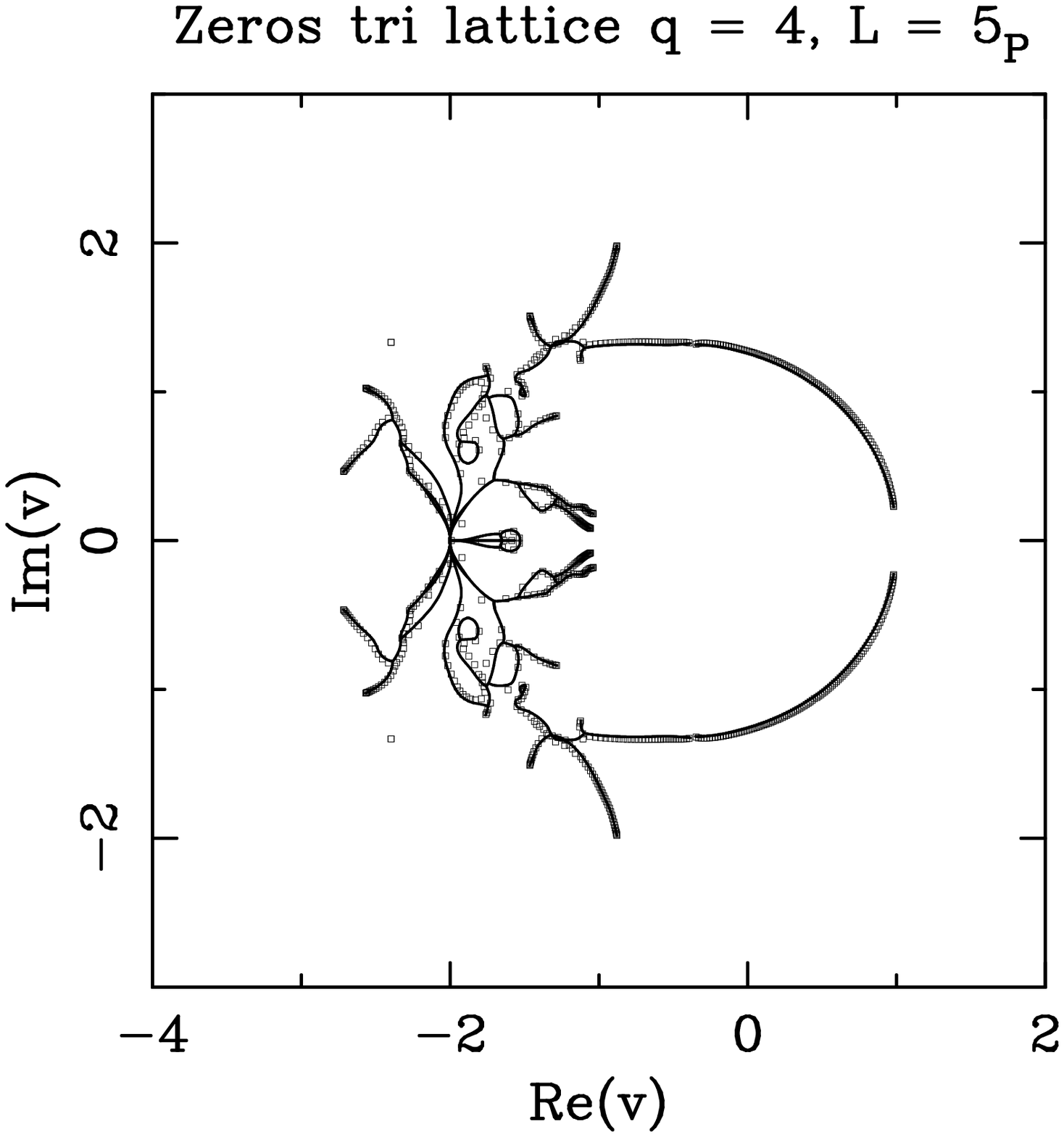} \\[1mm]
   \phantom{(((a)}(c)    & \phantom{(((a)}(d) \\
\end{tabular}
\caption[a]{\protect\label{zeros_tri_allP_q=4}
Limiting curves forming the singular locus ${\cal B}$, in the $v$ plane, for
the free energy of the Potts model for $q=4$ on the
$L_{\rm P} \times \infty_{\rm F}$ triangular-lattice strips with
(a) $L=2$, (b) $L=3$, (c) $L=4$, and (d) $L=5$.
We also show the partition-function zeros corresponding to the strips
$L_{\rm P} \times (10L)_{\rm F}$ for each value of $L$.
}
\end{figure}

The zeros and accumulation sets for $q=2$ are displayed in
Figures~\ref{zeros_tri_allF_q=2} and~\ref{zeros_tri_allP_q=2} for free and
cylindrical boundary conditions. Fig. \ref{zeros_tri_allF_q=2}(a) contains the
same information as Fig. 4 of \cite{ta} (which is plotted in a different
temperature variable, $a^{-1}$).  The finite-size effects for the accumulation
sets ${\cal B}_v(L,q=2)$ are noticeably larger for free, in comparison with
cylindrical, boundary conditions, as expected.  In the latter case, the curves
${\cal B}_v(L,q=2)$ for $L=2,3,4,5$ fall very approximately one on top of the
preceding one. For cylindrical boundary conditions, we see that the curve
${\cal B}_v(L,q=2)$ is symmetric under the replacement $a \to -a$.  The reason
for this is that in this case all of the vertices except for the end-vertices,
which constitute a vanishingly small fraction in the limit of infinite length,
are equivalent (i.e., the graph is $r$-regular) and have even degree $r$.  In
general, this property applies for the complex-temperature phase diagram of the
$q=2$ (Ising) special case of the Potts model for an infinite lattice where the
coordination number is even \cite{chisq,cmo,chitri}.  Our strips with free
transverse boundary conditions are not $r$-regular graphs because the vertices
on the upper and lower sides have a different degree than those in the
interior.
Because of this, the ${\cal B}_v$ in this case does not have the $a \to -a$
symmetry.  From previous work \cite{a,ta} one knows that the loci ${\cal B}_v$
are different for strips with free or periodic transverse boundary conditions
and free longitudinal boundary conditions, on the one hand, and free or
periodic transverse boundary conditions and periodic (or twisted periodic)
longitudinal boundary conditions.  One anticipates, however, that in the limit
of infinite width, the subset of the complex-temperature phase diagram that is
relevant to real physical thermodynamics will be independent of the boundary
conditions used to obtain the 2D thermodynamic limit.

In the 2D thermodynamic limit, one knows the complex-temperature phase diagram
exactly for the $q=2$ (Ising) case.  (This isomorphism involves the
redefinition of the spin-spin exchange constant 
$J_{\rm Potts} = 2J_{\rm Ising}$ and
hence $K_{\rm Potts} = 2K_{\rm Ising}$, where $K_{\rm Potts}$ is denoted 
simply $K$ here.)
The simplest way to portray the complex-temperature phase diagram is in the
$a^2$ or $u=a^{-2}$ plane since this automatically incorporates the $a \to -a$
symmetry noted above.  In the $u$ plane, the complex-temperature phase diagram,
with boundaries given by ${\cal B}_u$, is (see Fig. 1(a) of \cite{chitri} which
is equivalent, by duality to the complex-temperature phase diagram for the
honeycomb lattice given as Fig. 2 in \cite{abe91})
\beq
{\cal B}_u:  \quad \{ |u+\frac{1}{3}| = \frac{2}{3} \} \quad \cup \quad 
\{ -\infty \le u \le -\frac{1}{3} \}
\label{Bqeq2}
\eeq
i.e., the union of a circle centered at $u=-1/3$ with radius 2/3 and the
semi-infinite line segment extending leftward from $u=-1/3$ along the real $u$
axis.  In the $a^{-1}$ plane (Fig. 1(b) of \cite{chitri}, related by duality to
Fig. 3 of \cite{abe91}), ${\cal B}$ is the union of a vertically elongated oval
crossing the real axis at $\pm 1/\sqrt{3}$, the imaginary $a^{-1}$ axis at $\pm
i$, and two semi-infinite line segments extending from $i/\sqrt{3}$ to
$i\infty$ and from $-i/\sqrt{3}$ to $-i\infty$ along the imaginary axis.
Equivalently, in the $a$ plane, ${\cal B}$ is the union of a horizontally
elongated oval crossing the real $a$ axis at $\pm \sqrt{3}$ and the imaginary
$a$ axis at $\pm i$, and a line segment along the imaginary axis extending
between $\sqrt{3}i$ and $-\sqrt{3}i$.  The locus ${\cal B}_v$ in the $v$ plane
is obtained from this by translation by one unit, since $v=a-1$.  This locus
separates the complex $v$ plane into three phases: (i) the paramagnetic phase,
including the infinite-temperature point $v=0$, where the $S_q$ symmetry is
realized explicitly ($S_q$ being the symmetric group on $q$ numbers, the
symmetry group of the Hamiltonian), (ii) the ferromagnetic phase, including the
real interval $v_c(q=2) \le v \le \infty$ where the $S_q$ symmetry is
spontaneously broken by the existence of a nonzero magnetization, and (iii) an
unphysical phase (denoted ``O'' for ``other'' in \cite{chitri}) including the
point $v=-2$.  Here 
\beq
a_c(q=2)=v_c(q=2)+1 = \sqrt{3}
\label{acq2}
\eeq
is the physical critical point separating the PM and FM phases (for a review of
the Ising model on the triangular lattice, see, e.g., \cite{domb}) .  These
physical PM and FM phases have complex-temperature extensions off the real $v$
axis.  The PM and O phases are separated by the subset of the vertical line
segment extending between $a=i$ and $a=-i$; this line segment terminates at the
points $a=\pm \sqrt{3}i$.  Because of the maximal frustration, there is no
antiferromagnetic phase at finite temperature.  The presence of a
zero-temperature critical point in the 2D Ising antiferromagnet
\cite{stephenson64} is manifested by the fact that ${\cal B}_v$ passes through
$v=-1$, i.e., $a=0$ (as part of the above-mentioned vertical line segment).
The complex-temperature phase boundary ${\cal B}_v$ crosses the real $v$ axis
at $v=\sqrt{3}-1$, separating the FM and PM phases, at $v=-1$, separating the
PM and O phases, and at $v=-1-\sqrt{3}$, separating the O and
(complex-temperature analytic continuation of the) FM phases.  In \cite{ta} the
${\cal B}$ for an infinite-length free or cyclic strip with width $L=2$ were
compared with this 2D phase diagram.  These three points, $v=-1, -1 \pm
\sqrt{3}$, are the three roots of the $q=2$ special case of eq.~(\ref{tri_eq}).

Using our exact results, we can compare our loci ${\cal B}_v$ for a wide
variety of widths and either free or periodic transverse boundary conditions
with the known complex-temperature phase diagram for the Ising model on the
infinite 2D triangular lattice.  This comparison is simplest for the case of
cylindrical boundary conditions, so we concentrate on these results.  For the
finite values of $L$ that we have considered, ${\cal B}_v$ has the form of two
complex-conjugate arcs that cross two complex-conjugate line segments on the
imaginary axis at $v=-1 \pm i$.  One sees that as $L$ increases, the endpoints
of the arcs move down toward the real axis, as do the endpoints of the line
segments.  As $L \to \infty$, we expect that these arc endpoints will close,
forming the above-mentioned horizontally elongated oval and vertical line
segment extending from $v=-1+\sqrt{3}i$ to $v=-1-\sqrt{3}i$ that constitute the
complex-temperature phase boundaries ${\cal B}$ for the Ising model on the
infinite triangular lattice. 

\subsection{$q=3$} 

In contrast to the $q=2$ case, the free energy of the $q$-state Potts model has
not been calculated exactly for $q \ge 3$ on any 2D (or higher-dimensional)
lattice and hence the complex-temperature phase diagrams are not known exactly.
The $q=3$ special case of eq.~(\ref{tri_eq}) has the root 
\beq
v_{PM-FM,q=3} \equiv v_c(q=3) = -1+\cos(2\pi/9)+\sqrt{3}\sin(2\pi/9) 
=0.879385...
\label{vcq3}
\eeq
corresponding to the physical PM-FM phase transition point, and two other roots
at the complex-temperature values 
\beq
v=-1+\cos(2\pi/9)-\sqrt{3}\sin(2\pi/9)  = -1.347296...
\label{vq3point2}
\eeq
and
\beq 
v=-1-2\cos(2\pi/9) = -2.532089...
\label{vq3point3}
\eeq
Discussions of the complex-temperature solutions of eq.~(\ref{tri_eq}) and
their connections with the complex-temperature phase diagram have been given in
\cite{mm,hcl,p,p2}.  A number of studies involving exact calculation of the
partition function for various $q$ values on large sections of the triangular
lattice have been performed \cite{mm,hcl,p,p2}.  (There have also been many
studies calculating zeros for the Potts model with $q \ge 3$ on the square
lattice; see \cite{ts} for references to these works.) 

The zeros and accumulation sets for $q=3$ are displayed in
Figures~\ref{zeros_tri_allF_q=3} and~\ref{zeros_tri_allP_q=3} for free and
cylindrical boundary conditions. We expect that the pair of complex-conjugate
endpoints in this regime will eventually converge to the ferromagnetic critical
point $v_c(q=3)$ as $L \to \infty$.  However, obviously, an infinite-length
strip of finite width $L$ is a quasi-one-dimensional system, so the Potts model
has no physical finite-temperature phase transition on such a strip for any
finite $L$. 

In the antiferromagnetic regime $-1 \leq v < 0$, we observe noticeable
finite-size effects even with cylindrical boundary conditions.  In this regime,
we also observe a complex-conjugate pair of endpoints with small value of
$\imag(v)$ that, as $L \to \infty$, are expected to approach the real $v$ axis
at the transition point separating the paramagnetic and antiferromagnetic (AFM)
phases of the $q=3$ Potts antiferromagnet on the infinite triangular lattice.
Monte Carlo and series analyses \cite{grest,saito,entingwu,adler_93} 
have yielded the
conclusion that the PM-AFM transition in the $q=3$ Potts antiferromagnet on the
triangular lattice is weakly first-order.  A high-accuracy determination of the
location of the PM-AFM transition temperature $T$ was obtained in
\cite{adler_93} by means of Monte Carlo simulations: $T = 0.62731 \pm 0.00006$,
or equivalently
\beq
v_{PM-AFM,q=3} = -0.79691 \pm 0.00003 
\label{estimate_vc_q=3}
\eeq
We shall improve this estimate below.  

Finally, in the complex-temperature interval $v < -1$, the finite-size and
boundary condition effects are evidently very strong.  Because of this, in
previous work, a combination of partition-function zeros and analyses of
low-temperature series expansions was used \cite{p2}; these enable one at least
to locate some points on the complex-temperature phase boundary.  As regards
the infinite 2D triangular lattice, because of a duality relation, the complete
physical temperature interval $0 \le T \le \infty$, i.e., $0 \le a \le 1$ of
the $q$-state Potts antiferromagnet on the honeycomb lattice is mapped to the
complex-temperature interval $-\infty \le v \le -q)$ on the triangular lattice
(and vice versa) \cite{hcl}.  As was noted in \cite{hcl}, it follows that
because the $q=3$ Potts antiferromagnet on the honeycomb lattice is disordered
for all temperatures, including $T=0$, the free energy for this model on the
triangular lattice is analytic in the interval $-\infty < v \le -3$, and hence
no part of the complex-temperature phase boundary ${\cal B}_v$ can cross the
negative real axis in this interval.  In particular, one anticipates that as $L
\to \infty$ for the infinite-length, width-$L$ strips, the left-most arcs on
${\cal B}_v$ will not close and pinch the negative real axis in this interval.
Our calculations of ${\cal B}_v$ are consistent with the inference that as $L
\to \infty$, this locus crosses the real axis at the points (\ref{vq3point2})
and (\ref{vq3point3}), although there are significant differences between the
loci obtained with free and cylindrical boundary conditions.  

We also observe certain line segments on the real $v$ axis in the
complex-temperature region.  We note that massless phases with algebraic decays
of correlation functions have been suggested for the Potts model on the
(infinite) square lattice at real values of $v$ and $q$ in the intervals
$-2-\sqrt{4-q} \le v \le -2 + \sqrt{4-q}$ with $q \in (0,4)$ and $q \ne B_r =
4\cos^2(\pi/r)$, and it was conjectured that these might also occur for other
2D lattices \cite{saleur}.  However, the correspondence of these suggestions
with our results is not clear; for example, the above interval suggested in
\cite{saleur} shrinks to zero as $q \to 4$, but we observe clear line segments
on the real $v$ axis for $q=4$ for both free and cylindrical boundary
conditions (see Figs.  \ref{zeros_tri_allF_q=4} and \ref{zeros_tri_allP_q=4}).
A possible physical subset of the above range of $v$ given in \cite{saleur}
would be the antiferromagnet interval $-1 \le v < 0$.  However, the condition
that $q \ne B_r$ excludes all of the integral values of $q$ in the indicated
range (recall that $B_2=0$, $B_3=1$, $B_4=2$, $B_6=3$, and
$B_1=\lim_{n\to\infty}B_n=4$).  The claim in \cite{saleur} is thus complicated
by the fact that although it is possible formally to define the Potts model
partition function $Z(G,q,v)$ using (\ref{cluster}) for real positive
non-integral $q$ for the antiferromagnetic case, $-1 \le v < 0$, here the model
does not satisfy the usual statistical mechanical requirement that the
partition function is positive, and hence does not, in general, admit a Gibbs
measure \cite{ss97,a}.  This leads to pathologies that preclude a physical
interpretation, such as negative partition function, negative specific heat,
and non-existence of a $|V| \to \infty$ limit for thermodynamic functions that
is independent of boundary conditions \cite{ss97,a,ta}.  As regards the
connection with the locus ${\cal B}$, a signal of a massless phase would be a
line segment on ${\cal B}$ on the real $v$ axis for fixed $q$ or the real $q$
axis for fixed $v$.  For the zero-temperature Potts antiferromagnet, i.e.,
chromatic polynomial, $v=-1$, these phases would thus occur in the intervals
between the Beraha numbers, $0 < q < 1$, $1 < q < 2$, $2 < q <
(1/2)(3+\sqrt{5})$, and so forth.  However, it has been proved that there are
no real zeros of a chromatic polynomial in the intervals $-\infty < q < 0$, $0
< q < 1$, and $1 < q \le 32/27$ \cite{jackson,thomassen}.  Since ${\cal B}$
forms as an accumulation set of zeros, this makes it difficult to see how there
could be a line segments in these intervals, in particular, the intervals $0 <
q < 1$ and $1 < q \le 32/27$.  Again, it is not clear how to reconcile the
results of these theorems with a suggestion that there would be massless phases
with associated real line segments on ${\cal B}$ in these intervals.

\subsection{$q=4$}

For $q=4$, eq.~(\ref{tri_eq}) has the physical root 
\beq
v_{PM-FM,q=4} \equiv v_c(q=4) = 1
\label{vcq4}
\eeq
corresponding to the PM-FM phase transition point and a double root at the 
complex-temperature point
\beq
v=-2
\label{vq4point23}
\eeq
The zeros and accumulation sets for $q=4$ are displayed in
Figures~\ref{zeros_tri_allF_q=4} and~\ref{zeros_tri_allP_q=4}.  One observes
the approach of the right-most complex-conjugate arcs to the real axis as $L$
increases, i.e. the approach to the PM-FM critical point in this case.  For a
given $L$, the approach to the exactly known value $v_{PM-FM,q=4}=1$ in
eq.~(\ref{vcq4}) is closer for cylindrical versus free boundary conditions, as
is anticipated since the former minimize boundary effects.  The $q=4$ Potts
antiferromagnet on the triangular lattice has a zero-temperature critical
point, so that $v=-1$ is on ${\cal B}_v$ \cite{baxter_70} (this is not a root
of eq.~(\ref{tri_eq})). For $L \ge 2$ for free boundary conditions and for $L
\ge 2$ for cylindrical boundary conditions, we see how a pair of
complex-conjugate arc endpoints approaches the real axis as $L$ increases,
consistent with the inference that these would pinch at $v=-1$ in the $L \to
\infty$ limit.  For both free and cylindrical boundary conditions and various
values of $L$, one sees that ${\cal B}_v$ contains an intersection point at the
complex-temperature value $v=-2$, in agreement with the expectation from
eq.~(\ref{vq4point23}).  The fact that the $q=4$ Potts antiferromagnet is
disordered on the honeycomb lattice for all temperatures $T$ including $T=0$
implies that ${\cal B}_v$ does not cross the negative real axis in the interval
$-\infty < v \le -4$ \cite{hcl}.  In particular, this implies that as $L \to
\infty$, the leftmost arc endpoints on ${\cal B}_v$ in the figures do not move
down to pinch the negative real axis in this interval $-\infty < v \le -4$,
provided that the limit $L \to \infty$ of these infinite-length, width-$L$
strips commutes with the 2D thermodynamic limit for the triangular lattice as
regards this aspect of the complex-temperature phase diagram.

We do not show plots for $q \ge 5$, but recall that the Potts antiferromagnet
is expected to be disordered (with exponential decay of spin-spin correlation
functions) even at $T=0$ on the triangular lattice.  This can be proved
rigorously for $q \ge 11$ as a slight improvement of the result that $q$-state
Potts antiferromagnet is disordered at all temperatures on a lattice with
coordination number $r$ if $q > 2r$ \cite{ss97}. The property that the Potts
antiferromagnet is disordered at all $T$ on the triangular lattice for $q \ge
5$ is reflected in the property that ${\cal B}_v$ does not pass through $v=-1$.
 
%
%
\section{Internal Energy and Specific Heat} 

The partition function \reff{zfun} can be used to derive the free-energy
density $f(G,q,v)$ 
\be
f(G,q,v) = {1\over |V|} \log Z(G,q,v) 
\label{def_free_energy}
\ee
for finite $|V|$, with the $V \to \infty$ limit having been defined in
eq.~(\ref{ef}) above. 
The internal energy $E$ and the specific heat $C$ are derived in the usual way
from the free energy as 
\beq
E(G,q,v)=-J\frac{\partial f}{\partial K} = -J(v+1)\frac{\partial f}{\partial v}
\slabel{def_E}
\eeq
and
\beq
C = \frac{\partial E}{\partial T} = k_B K^2(v+1)\left [
\frac{\partial f}{\partial v} + (v+1)\frac{\partial^2 f}{\partial v ^2}
\right ] \equiv k_B K^2 C_H 
\label{def_CH}
\eeq
Henceforth, for convenience, we shall use a definition of $E$ without the
factor $-J$ in (\ref{def_E}), and we shall use the dimensionless function $C_H$
in discussions of the specific heat.  Let us suppose that $G$ is a
triangular-lattice strip graph of size $L\times m$.  In the limit $m \to
\infty$, since only the dominant eigenvalue $\lambda_d(q,v)$ of the transfer
matrix contributes to the free energy, one has
\begin{subeqnarray}
f(q,v;L)   &=& {1 \over L} \log \lambda_d \\
E(q,v;L)   &=& {1 \over L} \Lambda_1 \\ 
C_H(q,v;L) &=& {1 \over L} \left[ \Lambda_2 + \Lambda_1 - \Lambda_1^2 \right]  
\end{subeqnarray}
where the $\Lambda_i(q,v;L)$ are defined by 
\be
\Lambda_i(q,v;L) = {(1+v)^i \over \lambda_d(q,v;L)} 
                   {\partial^i \lambda_d \over \partial v^i } 
\label{def_Lambdas}
\ee

The infinite-length limits of the triangular-strips considered here
are quasi-one-dimensional systems with analytic free energies at all
temperatures. Hence the dominant eigenvalue $\lambda_d$ is the
same on the whole semi-axis $\imag(v) = 0$, $\real(v) \geq -1$.
Furthermore, as discussed in \cite{a}, the free energy and its derivatives
with respect to the temperature are independent of the longitudinal boundary  
conditions in the limit $m\to\infty$ 
(although they depend on the transverse boundary conditions).

\begin{figure}[hbtp]
\centering
\begin{tabular}{cc}
   \includegraphics[width=200pt]{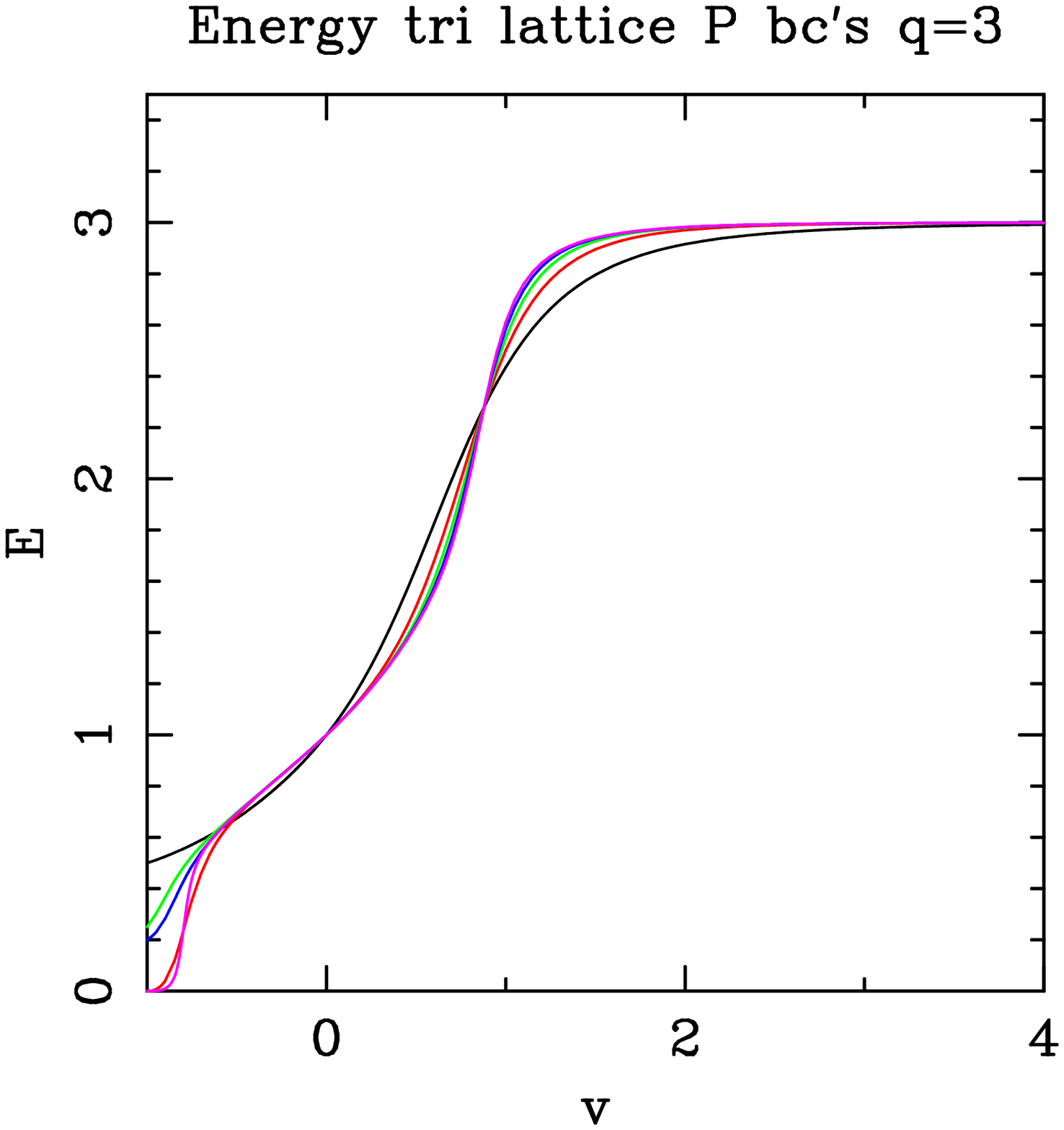} &
   \includegraphics[width=200pt]{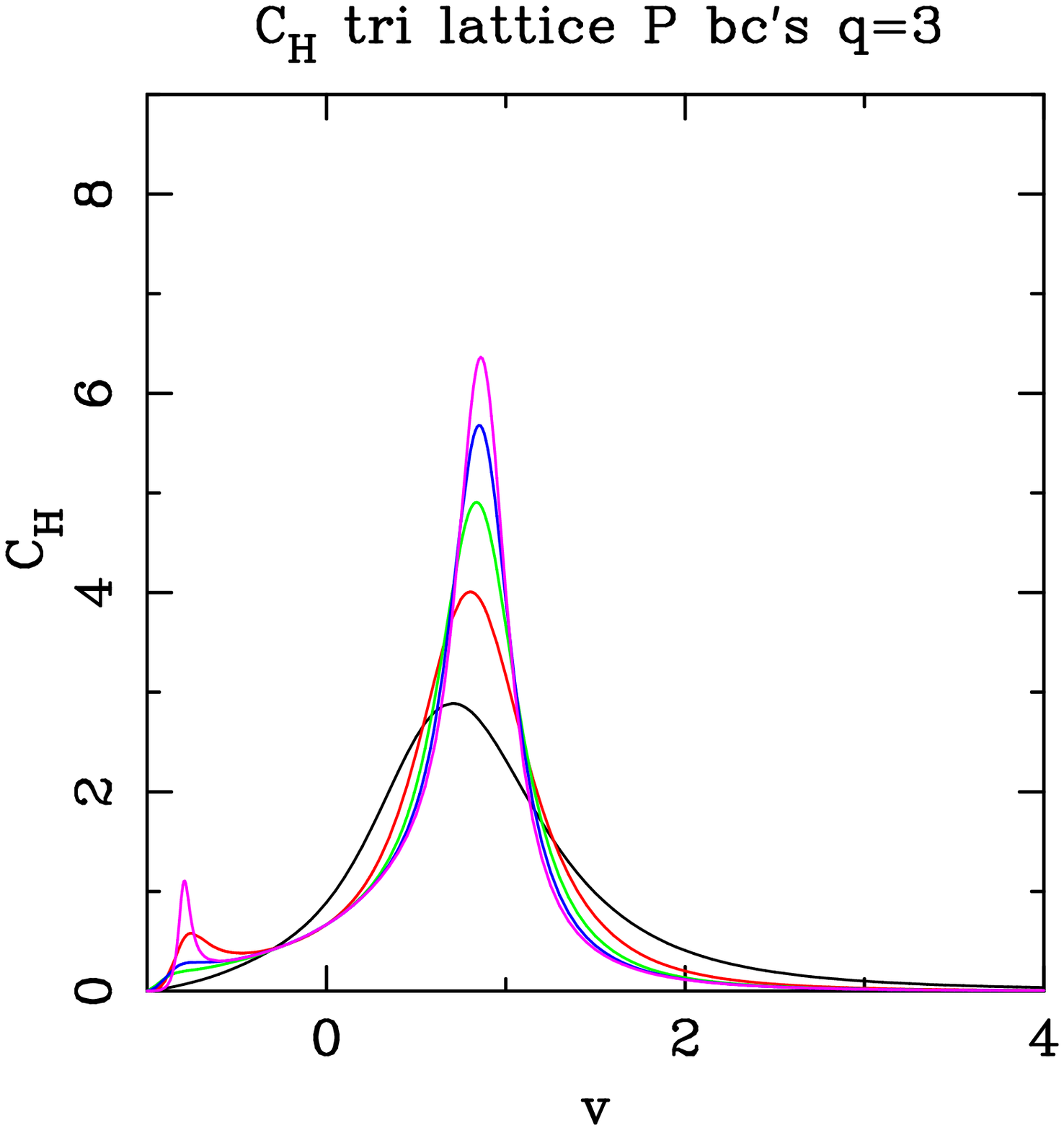} \\[1mm]
   \phantom{(((a)}(a)    & \phantom{(((a)}(b) \\
   \multicolumn{2}{c}{ \includegraphics[width=200pt]{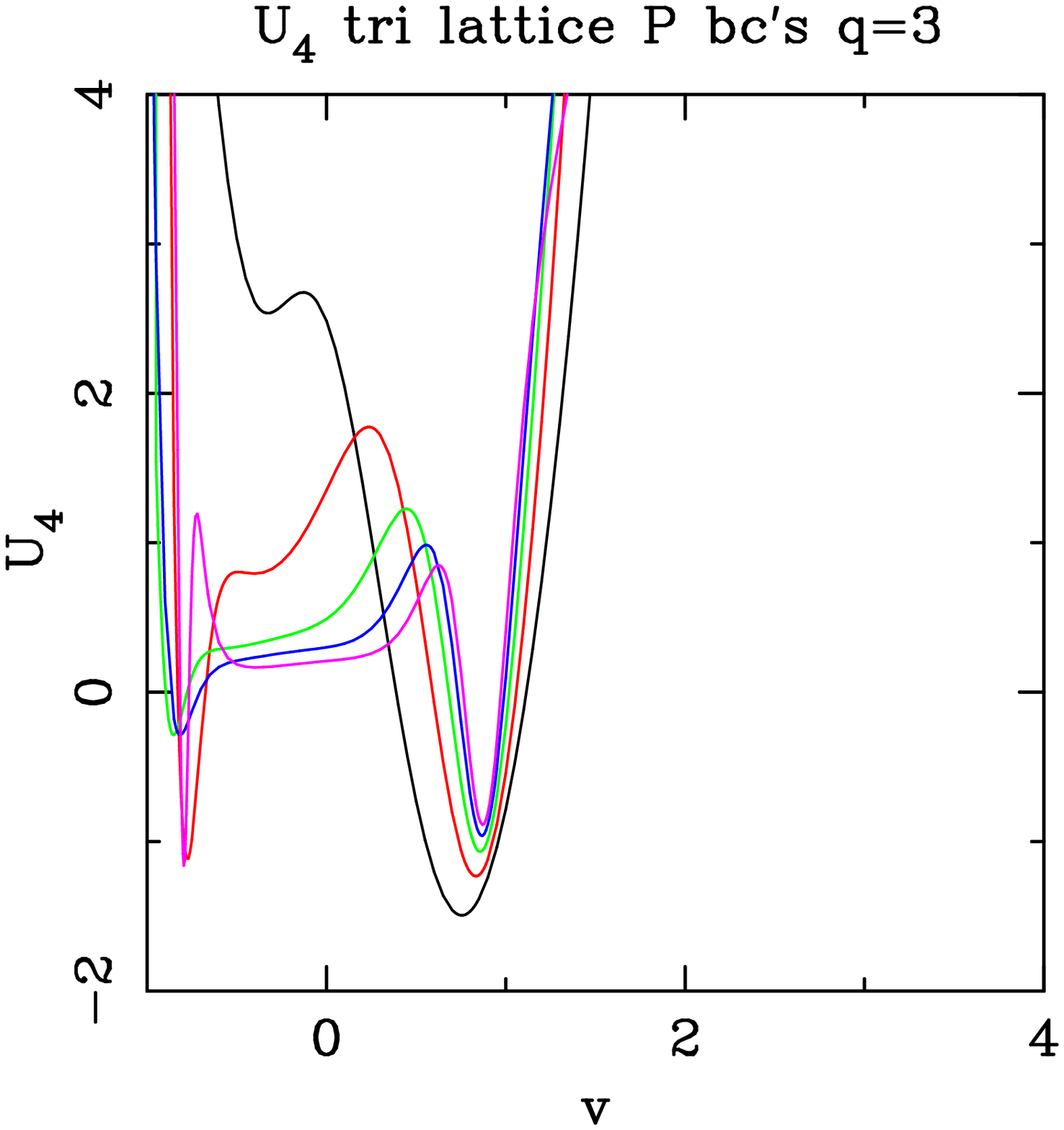} }
   \\[1mm]
   \multicolumn{2}{c}{ \phantom{(((a)}(c) }
\end{tabular}
\caption[a]{\protect\label{observables_P_q=3}
Thermodynamic observables for the 3-state Potts model on triangular-lattice 
strips of sizes $L_{\rm P} \times \infty_{\rm F}$. 
We show (a) the energy density $E$, (b) the specific heat
$C_H$, and (c) the Binder cumulant $U_4$ as a function of the temperature-like
parameter $v$ for several strip widths $L$: 2 (black), 3 (red), 4 (green), 
5 (blue), and 6 (pink). $-1\leq v < 0$ corresponds to the antiferromagnetic 
regime, while $v > 0$  to the ferromagnetic one. 
}
\end{figure}

In Figure~\ref{observables_P_q=3} we have plotted the internal energy $E$
\reff{def_E}, the specific heat $C_H$ \reff{def_CH}, and the Binder cumulant
$U_4$ \reff{def_U4} (see below) for $q=3$ on a triangular-lattice strip of
width up to $L=6$ and infinite length with cylindrical boundary conditions.

The behavior of the energy for the triangular-lattice strips with cylindrical
boundary conditions is interesting: the curves cross each others close to the
critical value $v_c$ in the ferromagnetic regime.  In particular, for $q=2$ we
find that all curves cross at $v_c(2) = \sqrt{3}-1 \approx 0.7320508...$.
For $q=3,4$ we find that the crossings are
close to the respective PM-FM critical points $v_c(3)$ given exactly in
eq.~(\ref{vcq3}) ($=0.8793852...$) and $v_c(4)=1$, but they do not coincide
precisely with these critical points. 
The reason for the above behavior is the following: the triangular-lattice 
Potts model on a triangular lattice at a given value of the temperature
Boltzmann variable $v$ is related by 
duality to the hexagonal-lattice Potts model at a different temperature
variable 
\beq
v' = \frac{q}{v}
\label{vdual}
\eeq
This relation is exact when the original triangular-lattice is defined on an
infinitely long cylinder of width $L$.  In the Ising case $q=2$, there is an
additional transformation (namely, the star-triangle transformation) that maps
back the hexagonal-lattice Potts model onto a triangular-lattice Potts model at
temperature variable $v''(v)$. This transformation allows us to compute the
critical temperature $v_c$ (i.e., $v_c$ is the unique fixed point of the
equation $v''(v) = v$) and the critical energy $E_c = E(q,v_c;L)$.  When we
perform the computation of $E_c$ in the limit $m\to\infty$, all finite-size
corrections disappear, so that \cite{Salas_02}
\be
E_c(q=2;L) \equiv E(q=2,v_c;L) = E_c(q=2;\infty) = {5\over 2} 
\ee
However, for $q\neq 2$ there is no star-triangle transformation, 
and this implies the existence of corrections to scaling: 
\be
E_c(q;L) \equiv E(q,v_c;L) = E_c(q;\infty) + 
           \sum\limits_{k=1}^\infty {A_k \over L^{\omega_k} }\ ,
\qquad q \neq 2 
\ee
where the parameters $\{\omega_k(q)\}$ are correction-to-scaling exponents 
that depend in general on the value of $q$. 

One can obtain a pseudo-critical temperature $v_E \equiv v_E(q,L,L')$ by 
solving the equation
\be
E(q,v_E,L) = E(q,v_E,L')
\label{crossing_eq}
\ee
When $L,L'\to\infty$, we expect that this quantity will converge to the true
critical value $v_c(q)$. This method has been employed in the the literature 
to locate critical points for several statistical-mechanical systems
\cite[and references therein]{yurishchev_00,yurishchev_02}.

Another pseudo-critical temperature can be obtained by looking at the point
$v_C \equiv v_C(q,L)$ where the specific heat $C(q,v;L)$ attains a 
maximum value. This value differs from the bulk critical value 
$v_c(q)$ by finite-size-corrections of order $\sim L^{-1/\nu}$ 
\cite{Barber_83}.  

We can also consider higher derivatives of the free energy with respect to 
$K$. In particular, the quantity $\bar{u}_4$ is the fourth derivative of the 
free energy with respect to $K$ and, in the limit $m\to\infty$ it can be
written as 
\beq
\bar{u}_4(q,v;L)  = \frac{\partial^4 f}{\partial K^4}
\label{def_ubar4}
\eeq
Instead of $\bar{u}_4$, it is more useful to deal with the phenomenological 
quantity $U_4$ (also called the Binder cumulant \cite{Binder_86}) defined as 
\be
U_4(q,v;L) = {1\over L^2} {\bar{u}_4 \over C^2 }
\label{def_U4}
\ee
(Since this involves the energy instead of the magnetization, one could also
consider a third cumulant.) A plot of this quantity for $q=3$ is given in 
Figure~\ref{observables_P_q=3}(c).
A third pseudo-critical temperature $v_U \equiv v_U(q,L)$ can be defined as the
value at which the Binder cumulant \reff{def_U4} attains a minimum value.
Again, this estimate is expect to differ from the bulk critical temperature
$v_c(q)$ by terms of order $L^{-1/\nu}$ \cite{Barber_83}.

The computation of the transfer matrices for triangular-lattice strips
of width $L$ allows us to compute these three pseudo-critical temperatures
(namely, $v_E(q,L,L')$, $v_C(q,L)$, and $v_U(q,L)$). We have computed these
estimates for several values of $1/2 \leq q \leq 4$ in the ferromagnetic
regime for strips with cylindrical boundary conditions. 
We expect a finite-size behavior of these three pseudo-critical temperatures 
of the type
\be
v_Q(q;L) = v_c(q) + \sum\limits_{k=1}^\infty {A_{Q,k} \over L^{\omega_k}}
\label{def_Ansatz}
\ee
where $Q = E,C,U$, and the $\{A_{Q,k},\omega_k\}$ are correction-to-scaling
amplitudes and exponents, respectively. These quantities depend in general
on $q$. We can try to estimate the critical temperature $v_c(q)$ by fitting
the data to the above Ansatz \reff{def_Ansatz} with only the leading
correction-to-scaling term included
\be
v_Q(q;L) = v_c(q) + A_Q L^{-\Delta_Q}
\label{def_Ansatz_bis}
\ee

Although the exact phase transition temperature for the $q$-state Potts
ferromagnet is known, it is useful to compare these pseudo-critical temperature
with the exactly known values for the infinite lattice, since we do not know,
{\em a priori}, which estimate $v_Q$ will give the most accurate results. It 
is important to perform a comprehensive check of the method before
studying a phase transition whose critical temperature is not known (for
instance, the 3-state triangular-lattice Potts antiferromagnet; see below).  We
have done this and have found that the estimates coming from the
pseudo-critical temperature $v_E$ are better by far than the other two; and the
estimate $v_U$ is more accurate than $v_C$. More precisely, the difference
between the extrapolated value for $v_c$ and the exact value is $\sim 10^{-7}$
for $v_E$ for all the $1.5 \leq q \leq 4$ values considered. However, for $v_U$
the discrepancy is of order $\sim 10^{-5}$ for $q\gtapprox 1.5$ and $\sim
10^{-2}$ to $10^{-4}$ for $q \ltapprox 1$. Finally, for $v_C$ the discrepancy
is $\sim 10^{-4}$ for $q\gtapprox 2$, and $\sim 10^{-3}$ for $1\ltapprox q
\ltapprox 1.5$. In conclusion, we can establish the position of the
ferromagnetic critical temperature $v_c(q)$ (over the whole range of values of
$q$) by using the pseudo-critical temperature $v_E(L,L')$.  The results for
$v_C$ and $v_U$ are at least two orders of magnitudes worse, and the accuracy
also depends on the value of $q$: it worsens for $q\ltapprox 2$ for $v_C$, and
for $q\ltapprox 1.5$ for $v_U$.

One can try to extend the previous analysis to the antiferromagnetic regime.
In Table~\ref{table_crossings_P_AF} we show the estimates for the critical
temperature of the 3-state Potts antiferromagnet using strips with cylindrical
boundary conditions and widths that are multiples of 3.  This constraint is due
to the $(\mod 3)$-oscillations that appear in antiferromagnets: in
Figure~\ref{observables_P_q=3} we clearly observe such oscillations in the
antiferromagnetic regime. Thus, we keep only the data with $L=3,6,9,12$ that
is expected to be closer to the thermodynamic limit.\footnote{
 To compute the estimates $v_E(q=3,L,L^\prime)$ we need the transfer matrices
 $\T(L_{\rm P},q=3)$. For $L_{\rm P}\leq 9$, we have used the symbolic
 transfer matrices of section~4 evaluated at $q=3$. For larger
 widths (i.e., $L=12,15$), we have computed numerically the corresponding
 transfer matrices at the particular value $q=3$. 
} 
(When $L$ is not a
multiple of 3, the corresponding triangular-lattice strip with cylindrical
boundary conditions is not tripartite, unlike those strips with $L$ a multiple
of 3 or the infinite triangular lattice; recall our earlier discussion of the
chromatic number for these strips.)  The value quoted on the row labelled "MC"
comes from the Monte-Carlo study by Adler {\em et al}.\ \cite{adler_93}.
For $L=12$ we just list the estimate from the energy crossing since this is 
superior to the other two estimates, and this gives the value
\be
v_c(q=3) = -0.796927(20)
\label{estimate_vc_q=3_crossE}
\ee
The error bar quoted in \protect\reff{estimate_vc_q=3_crossE} was roughly
estimated by comparing the above result to the value of $v_c(q=3)$ obtained by
fitting the data points $v_E(L,L')$ with $L=3,6,9$ and $L'=L+3$, namely,
$v_c(q=3) = -0.796907$. This is indeed a very conservative estimate for this
error bar. Our results in (\ref{estimate_vc_q=3_crossE}) 
is in agreement with, and more accurate than, the estimate from 
\cite{adler_93} listed above in eq.~(\ref{estimate_vc_q=3}).

%
%
\subsection*{Acknowledgment} 
This research of was partially supported by NSF grant PHY-0098527 (R.S.) and 
PHY-0116590 (J.S.).  Computations were performed on several computers including
those at NYU and YITP, Stony Brook.

\clearpage
%
%
\begin{table}
\centering

\begin{tabular}{rrrrr}
\hline\hline
\multicolumn{1}{c}{$L$}          & 
\multicolumn{1}{c}{$N_{Z,{\rm tri},{\rm FF},L}$} & 
\multicolumn{1}{c}{$C_{L+1}$}    & 
\multicolumn{1}{c}{$N_{Z,{\rm tri},{\rm PF},L}$} &
\multicolumn{1}{c}{$N_{Z,{\rm sq},{\rm PF},L}$}  \\ 
\hline 
1   &      1   &       2  &     1 &    1\\
2   &      2   &       5  &     2 &    2\\
3   &      5   &      14  &     3 &    3\\
4   &     14   &      42  &     6 &    6\\
5   &     42   &     132  &    10 &   10\\
6   &    132   &     429  &    28 &   24\\
7   &    429   &    1430  &    63 &   49\\
8   &   1430   &    4862  &   190 &  130\\
9   &   4862   &   16796  &   546 &  336\\
10  &  16796   &   58786  &  1708 &  980\\
11  &  58786   &  208012  &  5346 & 2904\\
12  & 208012   &  742900  & 17428 & 9176\\
\hline\hline
\end{tabular}
\caption{\protect\label{Table_dimensions}
Dimensions of the transfer matrix for triangular-lattice strips.
For each strip width $L$ we give the dimension of the transfer matrix
for free boundary conditions $N_{Z,{\rm tri},{\rm FF},L}$ 
(which is equal to the Catalan number $C_L$), the dimension of the full 
transfer matrix for cylindrical boundary conditions (which is $C_{L+1}$), 
the dimension for cylindrical boundary conditions when translational 
symmetry is taken into account $N_{Z,{\rm tri},{\rm PF},L}$, and the 
dimension when we project onto the subspace of reflection-invariant 
connectivities $N_{Z,{\rm sq},{\rm PF},L}$. 
}
\end{table}

\clearpage

%
%
\begin{table}[thb]
\centering
\begin{tabular}{|c|rll|rl|}
\hline
\multicolumn{1}{|c|}{$q$} & \multicolumn{1}{|c}{$L$} &
                               \multicolumn{1}{c}{$v_{C}$} &
                               \multicolumn{1}{c|}{$v_{U}$} &
                               \multicolumn{1}{|c}{$L'$} &
                               \multicolumn{1}{c|}{$v_{E}$} \\
\hline\hline
 3   &    3     & -0.7537129688 & -0.7746989054 &    6 & -0.7984897326 \\
     &    6     & -0.7916555922 & -0.7942230532 &    9 & -0.7971540641 \\
     &    9     & -0.7952008646 & -0.7959971452 &   12 & -0.7969905288 \\
     &   12     &               &               &   15 & -0.7969527708 \\
\cline{2-6}
     & $\infty$ & -0.79660      & -0.79668      &      & -0.796927(20) \\
\cline{2-6}
     & MC       & -0.79691(3)   & -0.79691(3)   &      & -0.79691(3)  \\
\hline\hline
\end{tabular}
\caption{\protect\label{table_crossings_P_AF}
Pseudo-critical temperatures for the 3-state Potts model in the
antiferromagnetic regime. For each value of the strip width $L=3k$,
we show the pseudo-critical temperatures $v_C$, $v_U$, and $v_E$ computed
on strips with cylindrical boundary conditions.
The row labelled ``MC'' shows the Monte Carlo estimate for $v_c(q)$.
}
\end{table}

\clearpage
%
%

\clearpage
%
%

\clearpage
%
%
\clearpage

\clearpage
%
%
\clearpage

\clearpage
%
%
\clearpage

\clearpage
%
%
\clearpage

\clearpage
%
%
\clearpage

\clearpage
%
%
\clearpage

\clearpage
%
%
\clearpage

\clearpage
%
%
\clearpage

\clearpage
%
%
\clearpage

\clearpage
%
%
\clearpage

\clearpage
%
%
\clearpage

\vfill
\eject
\end{document}